\documentclass[conference]{IEEEtran}
\IEEEoverridecommandlockouts
\usepackage{cite,amsmath,amssymb,amsfonts}
\usepackage{algorithmic,graphicx,textcomp,xcolor,tabularx,multirow,array}
\usepackage{caption}
\usepackage{subcaption}
\usepackage[utf8]{inputenc}
\usepackage[english]{babel}
\usepackage{url}
\renewcommand{\arraystretch}{1.5}

\def\BibTeX{{\rm B\kern-.05em{\sc i\kern-.025em b}\kern-.08em
    T\kern-.1667em\lower.7ex\hbox{E}\kern-.125emX}}
    
\begin{document}

\title{Network Traffic Characteristics of IoT Devices \\in Smart Homes}

\author{\IEEEauthorblockN{Md Mainuddin, Zhenhai Duan}
\IEEEauthorblockA{Department of Computer Science\\
Florida State University\\
Tallahassee, FL 32306, USA\\
\textit{\{mainuddi, duan\}@cs.fsu.edu}%
}
\and
\IEEEauthorblockN{Yingfei Dong\\}
\IEEEauthorblockA{Department of Electrical Engineering\\
University of Hawaii\\
Honolulu, HI 96822 USA\\
\textit{yingfei@hawaii.edu}%
}
}

\maketitle

\begin{abstract}
Understanding network traffic characteristics of IoT devices plays a critical role in improving both the performance and security of IoT devices, including IoT device identification, classification, and anomaly detection. Although a number of existing research efforts have developed machine-learning based algorithms to help address the challenges in improving the security of IoT devices, none of them have provided detailed studies on the network traffic characteristics of IoT devices. In this paper we collect and analyze the network traffic generated in a typical smart homes environment consisting of a set of common IoT (and non-IoT) devices. We analyze the network traffic characteristics of IoT devices from three complementary aspects: remote network servers and port numbers that IoT devices connect to, flow-level traffic characteristics such as flow duration, and packet-level traffic characteristics such as packet inter-arrival time. Our study provides critical insights into the operational and behavioral characteristics of IoT devices, which can help develop more effective security and performance algorithms for IoT devices.
\end{abstract}


\section{Introduction}\label{sec:introduction}
Internet of Things (IoT) devices have been increasingly adopted and deployed in diverse environments including both home and enterprise networks to provide a variety of data collection, monitoring, and control functionalities. For example, the number of IoT devices surpassed the number of non-IOT devices in 2020 for the first time, and it was estimated that there would be more than 30 billions of IoT devices in 2025, which would be three times more than non-IoT devices at that time\cite{numberofiot}.



On the other hand, the unprecedented growth of IoT devices poses a significant security threat on the Internet. Typical IoT devices possess low processing capabilities, limited memory and storage, as well as minimal network protocol support, to both prolong the battery life and to lower the prices of IoT devices in an increasingly competitive IoT  market~\cite{challengessecurigiot,whymanuinsecure}. As a consequence, IoT devices in general lack strong security measures to protect themselves from security attacks. Furthermore, security patches might not be updated regularly due to irregular software release or a lack of awareness and expertise of the IoT device users~\cite{iotwildlyinsecure}. These factors ultimately create and keep a large number of IoT devices vulnerable on the Internet~\cite{hadar2017lightweight}.

IoT devices and systems are deployed in public, private, and corporate spaces, and a significant amount of information is being collected from these environments to make intelligent decisions. This vast amount of information, along with the vulnerabilities of the IoT devices, attracts the attackers who wish to exploit the IoT systems. As an example, in 2016 the massive DDoS attack on Dyn data centers made its DNS service unreachable to its users~\cite{dynddos}. The attack was performed using the Mirai botnet on IoT devices, and it took down more than $1200$ domains that Dyn was supporting, including Amazon, Twitter, Shopify, and many others. 

A thorough investigation of network traffic characteristics of IoT devices is critical in our understanding of the network operation and behavior of IoT devices, and in developing effective IoT device identification, classification, and anomaly detection mechanisms.
Given the significance of the IoT device security issue, a number of studies have developed machine-learning based algorithms to identify and classify IoT devices, and to detect compromised IoT devices (see Section~\ref{sec:related-work} on related work), relying on various network traffic characteristics (or features) of IoT devices. However, none of them have provided comprehensive studies on the network traffic characteristics of IoT devices. 

In this paper we investigate the network traffic characteristics of IoT devices (and non-IoT devices) in a typical smart home environment consisting of $10$ common smart home IoT devices and $5$ non-IoT devices. In this paper we adopt a commonly accepted definition of IoT devices. An IoT device performs a specific functionality, and can operate autonomously without directly human controls. After the network traffic is collected for the devices, we perform a thorough investigation of the network traffic characteristics from three complementary aspects: remote network servers and port numbers that IoT devices connect to, flow-level traffic characteristics such as flow duration, and packet-level characteristics such as packet inter-arrival time. 

Based on our investigation of the network traffic characteristics of both IoT and non-IoT devices, we make a few promising observations that can provide insights into developing effective IoT security mechanisms, including IoT device identification, classification, and anomaly detection. In particular, IoT devices behave very differently from non-IoT devices in a number of aspects, for example, in terms of number of network domains they communicate with, and in terms of both control traffic and data traffic. Second, IoT devices are also more consistent and stable in certain network traffic that they generate. We will report the details of our studies in Section~\ref{sec:characteristics}.

The remainder of the paper is organized as follows. In Section~\ref{sec:related-work}, we discuss related work. In Section~\ref{sec:setup} we describe the set-up of the smart home environment on which the network traffic used in this study is collected, and tools for data pre-processing. We perform the detailed study of the network traffic characteristics of IoT devices in Section~\ref{sec:characteristics}. We summarize our study and discuss the future work in Section~\ref{sec:summary}.






\section{Related Work}\label{sec:related-work}
In this section we discuss a number of research efforts that are most relevant to our work. We first note that the existing works focused on developing machine-learning based mechanisms for IoT identification, classification, and anomaly detection. Although they also explored network traffic characteristics (or features) in the development of their machine-learning based algorithms, none of them have provided a thorough study on network traffic characteristics of IoT devices, which is the objective of this work.

Mazhar and Shafiq~\cite{mazhar2020characterizing} performed a study on IoT traffic analysis in a home network environment, based on flow-level summary information obtained through a home gateway management company. They made a few interesting observations based on this flow-level summary data. For example, they found that the majority of IoT traffics were related to communications with Google Cloud and amazon AWS. In addition, they also noted that most of IoT devices used hard-coded Google DNS servers for various reasons. However, the researchers only had access to the flow-level summary data; and as a consequence, they cannot perform packet level studies of IoT traffic. 

Meidan {\it et al.} used IoT traffic data to detect and white list IoT devices connected to the current network~\cite{meidan2017detection}. They used selected features from TCP sessions information to train their machine learning model. Doshi {\it et al.} presented a model to distinguish normal and attack IoT traffic by using machine learning techniques on different network traffic features~\cite{doshi2018machine}. Their study relied on simulated IoT traffic data.

Shahid {\it et al.} developed a few machine learning algorithms to recognize IoT devices using the sizes and inter-arrival times of the first N packets of TCP flows of IoT devices~\cite{shahid2018recognition}. Their study used only four IoT devices with no non-IoT devices.
Sivanathan {\it et al.} developed a machine-learning algorithm to classify IoT devices based on statistical attributes of their network activities~\cite{sivanathan2018classifying}, including features such as activity cycles, remote servers and ports, signaling patterns, and cipher suites. The authors investigated the characteristics of these features. Our work complements their studies and provides a more thorough investigation of the network traffic characteristics of IoT devices.

 Miettinen {\it et al.} worked on IoT device fingerprinting and identification that utilized the network traffic during the setup phase~\cite{miettinen2017iot}. Ammar {\it et al.} presented a different machine-learning based approach to identifying IoT devices~\cite{ammar2019network}. Their work used service discovery and DHCP protocols to collect information about the IoT devices during their setup phases.


\section{Network Setup and Traffic Collection} \label{sec:setup}
\subsection{Devices and Network Setup}
\begin{figure}[h!]
    \centering
    \includegraphics[width=0.45\textwidth]{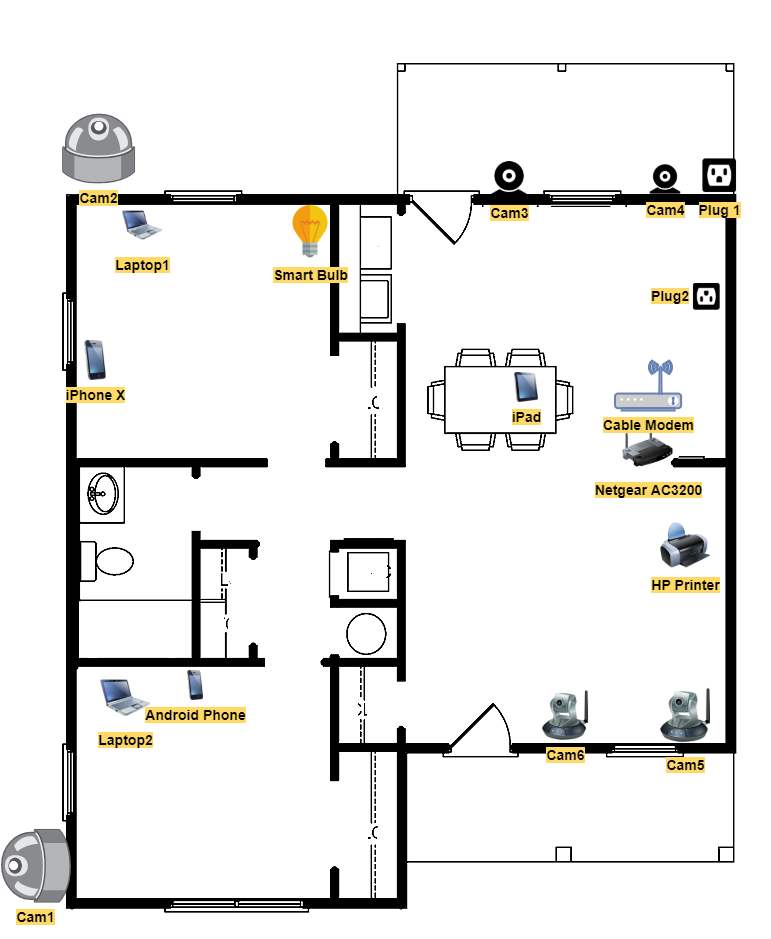}
    \caption{Smart home network setup}
    \label{fig:network1}
\end{figure}

Figure~\ref{fig:network1} illustrates the setup of the network testbed in a home environment, from which we collect the network traffic used in this study. This network comprises a variety of $10$ IoT devices as well as a number of non-IoT devices (see Table~\ref{table:1}), which reflects a typical network setup in a smart home environment. All devices are connected to a Netgear AC3200 Wi-Fi router. The Netgear router is connected to a Technicolor DOCSIS $3$ cable modem, which is also a Wi-Fi router, to connect the home network to the Internet. The Netgear router and the Technicolor router assign IP addresses in two different private IP address ranges.


This home network contains three pairs of security cameras. A pair of Logitech Circle-2 cameras are placed at different outdoor locations; two indoor cameras (Wyze and Eufy Indoor) are placed at the same location at the front porch; and another pair of indoor cameras (Eufy pan/tilt and LittleElf) are placed inside the living room. This combination of similar cameras at different locations and different cameras at the same location helps us to observe the traffic behavior of these security cameras from different perspectives. This home network also contains two smart plugs from two different manufacturers, a smart bulb, and an HP printer. Two laptops, two smart phones and an iPad, representing non-IoT devices, are also connected with the Netgear Wi-Fi router.

All devices are assigned with static IP addresses on the Netgear router so that we can keep track of their traffic with IP addresses in addition to their MAC addresses. Table~\ref{table:1} shows these IoT and non-IoT devices used in this study, including their corresponding MAC and IP addresses (assigned by the Netgear router). In the table, the column  "Device" indicates the manufacturer and type of the device. For each device, we also provide an "Alias Name", which is used to refer to the device throughout the paper to simplify our exposition. We also note that none of the IoT and non-IoT devices have been intentionally shut down for the studies reported in this paper. It could be interesting to allow users to use their (non-IoT) devices in the normal way, for example, to shut down a laptop or PC when they do not use it. This may provide us with additional behavioral and operational differences between IoT and non-IoT devices. However, in this study we focus on the differences between these two types of devices without considering the impact of this operational differences between IoT and non-IoT devices. We plan to investigate this difference in our future work.  

The Netgear Wi-Fi router is flushed with OpenWrt firmware~\cite{openwrt} to ease the collection of network traffic on the home network. A few software packages are installed on the router to help capture network traffic from all the connected devices. A Unix cron job runs the tcpdump tool on the router, which captures the traffic from all the devices and saves the captured traffic to an attached 512GB USB flash drive. The cron job runs tcpdump for each device separately and restarts the tcpdump processes at midnight. In this way we obtain a separate trace file every $24$ hours for each device while keeping the file sizes within a reasonable limit.

\begin{table*}[htbp]
\renewcommand{\arraystretch}{1.5}
\caption{Connected Devices in Smart Home Network}
\label{table:1}
\centering
\begin{tabular}{|l|l|l|l|l|}
  \hline
  Device & Alias Name & Category & MAC Address & IP Address \\
  \hline
  \multirow{2}{*}{Logitech Circle-2} & Cam1 & \multirow{10}{*}{IoT} & 44:73:D6:0C:36:AD & 192.168.1.158 \\
  \cline{2-2} \cline{4-5}
  & Cam2 & & 44:73:D6:09:BD:C9 & 192.168.1.186 \\
  \cline{1-2} \cline{4-5}
  Wyze Cam & Cam3 & & 2C:AA:8E:95:F3:18 & 192.168.1.228 \\
  \cline{1-2} \cline{4-5}
  Eufy Indoor Cam & Cam4 & & 8C:85:80:38:98:AF & 192.168.1.182 \\
  \cline{1-2} \cline{4-5}
  Eufy Pan and Tilt & Cam5 & & 8C:85:80:3A:12:B4 & 192.168.1.131 \\
  \cline{1-2} \cline{4-5}
  LittleElf Cam & Cam6 & & 0C:8C:24:61:50:29 & 192.168.1.168 \\
  \cline{1-2} \cline{4-5}
  Epicka Smart Plug & Plug1 & & DC:4F:22:0E:C6:36 & 192.168.1.127 \\
  \cline{1-2} \cline{4-5}
  Amazon Smart Plug & Plug2 & & F8:54:B8:25:AA:C9 & 192.168.1.204 \\
  \cline{1-2} \cline{4-5}
  Smart Bulb & Bulb & & 84:0D:8E:7F:4B:B4 & 192.168.1.207 \\
  \cline{1-2} \cline{4-5}
  HP Envy Printer & Printer & & 94:57:A5:0C:5B:66 & 192.168.1.248 \\
  \hline
  HP Elitebook  & Laptop1 & \multirow{5}{*}{non-IoT} & AC:FD:CE:01:7C:9B & 192.168.1.105 \\
  \cline{1-2} \cline{4-5}
  HP ZBook & Laptop2 & & CC:2F:71:3B:0E:DE & 192.168.1.247 \\
  \cline{1-2} \cline{4-5}
  Apple iPhone X & iPhone & & 34:08:BC:DE:E9:7E & 192.168.1.203 \\
  \cline{1-2} \cline{4-5}
  Apple iPad & iPad & & E8:8D:28:14:82:30 & 192.168.1.125 \\
  \cline{1-2} \cline{4-5}
  Samsung S20 & Android & & 16:05:DD:78:5F:20 & 192.168.1.215 \\
  \hline
\end{tabular}
\end{table*}


The motion detection feature has been enabled on all security cameras. When any motion is detected, a camera will upload the footage to the remote server and send a notification to the companion mobile app. Also, all the cameras have a live view feature, which enables a user to view the live video through the mobile app. 

We note that we sometimes directly connect our Android phone to the Technicolor router (instead of the Netgear Wi-Fi router) in order to examine the communication behavior of IoT devices when an IoT device and the corresponding companion mobile phone are on different LANs. In the following we highlight some observations we made on the behaviors of different IoT devices to illustrate the diversity and complexity of the IoT device operations. At a high level, the behavior of the same type of IoT devices can be very different depending on the manufacturers. In addition, they can also operate differently depending on the location of the corresponding companion mobile apps. 

For example, Logitech cameras (Cam1 and Cam2) always send the data to their corresponding remote servers whenever an event occurs (for example, a motion is detected), or when we watch the live stream through the mobile app. In contrast, Wyze (Cam3), Eufy (Cam4 and Cam5), and LittleElf (Cam6) cameras send the traffic to their corresponding remote servers when there is an event, but they may send the traffic directly to the mobile device (where mobile app is installed) for a live stream under certain scenarios, for example, when the mobile device is connected locally. 
Similar behavior is observed for smart plugs and bulb. When the mobile phone is connected on the local network, the app sends the control instruction to the IoT device directly; otherwise, the control instruction is delivered from the app to the remote server, from where it is in turn delivered to the device.


\subsection{Data Pre-processing} \label{subsection_data_2}
The USB flash drive has different folders for each device and daily traffic is stored in separate pcap files in corresponding folders. We download those pcap files from the router to a laptop for data processing. In our study we use two tools---CICFlowMeter\cite{cicflowmeter} and tshark\cite{tshark}---and a number of home-made Python scripts to process the collected data traces. In particular, we use CICFlowMeter to extract the flow-level information of the data traces (and also the packet inter-arrival times in a flow), and we use tshark to extract the other network traffic characteristics. 

We updated the source code of CICFlowMeter to reflect a more commonly used definition of TCP and UDP flows. A TCP or UDP flow is defined by a 5-tuple of source IP address, source port number, destination IP address, destination port number, and protocol. In the original CICFlowMeter, a flow is identified by a pair of SYN and FIN packets or a {\em duration} timeout threshold for TCP, and a duration timeout threshold for UDP. For example, with a duration timeout threshold of $120$ seconds, CICFlowMeter will conclude a flow when the next packet is received $120$ seconds {\em after the flow starts}. As a consequence, the maximum flow duration reported in CICFlowMeter is $120$ seconds. Instead of a duration timeout threshold, we update it to a {\em packet inter-arrival} timeout threshold. That is, we only conclude a flow when a packet inter-arrival time is greater than the specified timeout threshold (if we have not received FIN packet for TCP flows). We use a packet inter-arrival timeout threshold of $120$ seconds for extracting flows in this study~\cite{adaptive-timeout}. Both TCP and UDP flows are bidirectional, containing both outgoing and incoming packets.




\section{Network Traffic Characteristics}\label{sec:characteristics}
In this section we analyze the network traffic characteristics of the IoT devices based on the data collected on the home network, and compare with those of non-IoT devices. We investigate the characteristics from three complementary perspectives: the properties related to the remote servers that IoT devices connect to, the flow-level traffic properties, and the packet-level traffic properties. Although our data collection spans more than two months, our investigation will be focused on the data trace collected on one day, which is somewhat randomly chosen to be a Monday. For a number of studies, we also investigate the network traffic characteristics of data traffic in a week, containing the one-day mentioned above. The results obtained are representative of the general network traffic characteristics. 


\subsection{Remote Servers and Services}\label{sec:remote-servers}

\begin{figure}[ht!]
\centering
\subcaptionbox{Number of remote network domains \label{fig:outgoing_hosts}}[0.48\linewidth]
    {\includegraphics[width=0.24\textwidth]{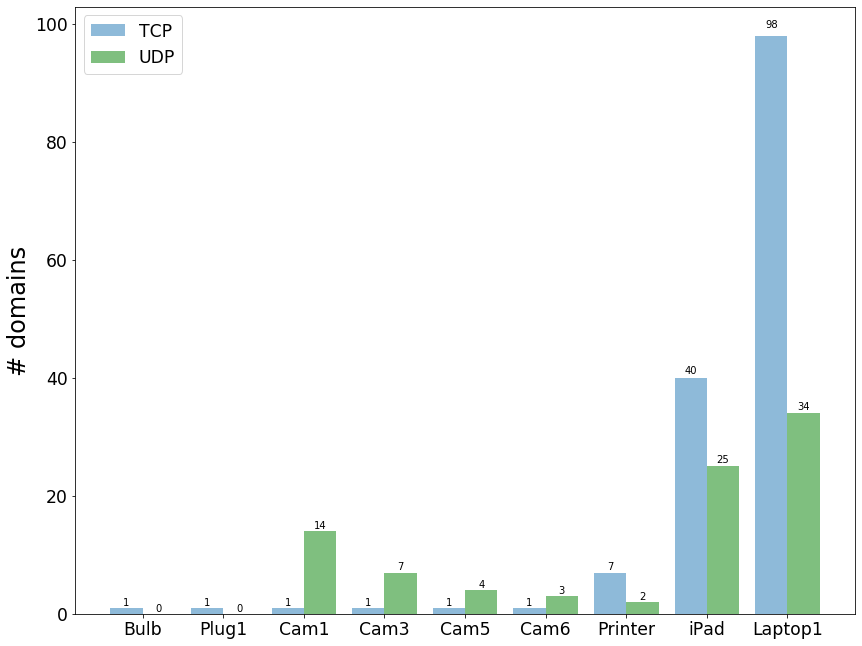}}
\hfill
\subcaptionbox{Number of remote port numbers \label{fig:outgoing_ports}}[0.48\linewidth]
    {\includegraphics[width=0.24\textwidth]{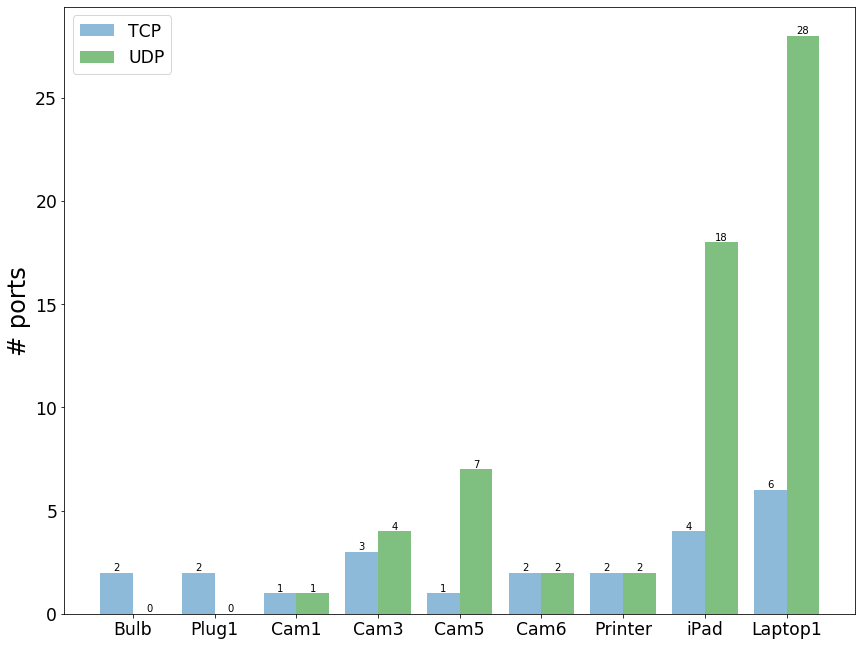}}
\caption{Number of remote network domains and port numbers.}
\label{fig:outgoing_hosts_and_ports}
\end{figure}




\begin{table*}[!ht]
\centering
\caption{Remote Servers and Port Numbers of IoT Devices}
\label{table:outgoing_host_port}
\begin{tabular}{|c|c|c|c|c|}
  \hline
   
  Device & TCP Ports & UDP Ports  &  TCP Servers & UDP Servers  \\
  & &   (Up to 4 most frequent) & & (Up to 4 most frequent) \\
  \hline
  \multirow{2}{*}{Cam1} & 443 & 123 & amazonaws.com & mattnordhoff.net, chaosfire.net \\
  & &  & & internap.com, latt.net \\

 \hline
  \multirow{2}{*}{Cam2} & 443 & 123 & amazonaws.com & ardoin.me, cloudflare.com \\
  & &  & & comcastbusiness.net, ctyme.com \\
  
  \hline
  \multirow{2}{*}{Cam3} & 8443, 443, 8883 & 53, 1001, 10240, 123 & amazonaws.com & dns.google, homeassurednow.com \\
  &  &  &  &   ip-147-135-36.us, nist.gov \\

  \hline
  \multirow{2}{*}{Cam4} & 443 & 32100, 53, 8006, 123 & amazonaws.com & amazonaws.com, dns.google \\
  &  &  &  & 224.0.0.251, flashdance.cx \\

  \hline
  \multirow{2}{*}{Cam5} & 443 & 32100, 53, 26807, 25249 & amazonaws.com & amazonaws.com, dns.google \\
  &  &  &  & 114dns.com, comcast.net \\

  \hline
  \multirow{2}{*}{Cam6} & 19000, 19001 & 53, 123 & amazonaws.com & dns.google, nist.gov \\
  &  &  &  & colorado.edu \\

  \hline
  \multirow{1}{*}{Plug1} & 80, 1883 & - & amazonaws.com & - \\

  \hline
  \multirow{1}{*}{Plug2} & 443 & 123 & amazonaws.com & amazonaws.com \\
  
  \hline
  \multirow{1}{*}{Bulb} & 443, 8886 & - & amazonaws.com & - \\

  \hline
 \multirow{2}{*}{Printer} & 5222, 443 & 930, 5353 & google.com, 1e100.net & amazonaws.com, 224.0.0.251 \\
  &  &  & amazonaws.com &  \\

  \hline
  \multicolumn{5}{|c|}{\scriptsize{Some common ports: 443 = HTTPS, 8443 = SSL, 80 = HTTP, 1883/8883 = MQTT, 5222 = XMPP, 123 = NTP, 53 = DNS, 5353 = mDNS}} \\
  \hline
\end{tabular}
\end{table*}

In this subsection we will study the behavior of IoT devices in terms of both remote servers and services an IoT device communicates with. For the remote servers, we will aggregate at the network domain level instead of individual server machines, given that it is common for an IoT device to communicate with multiple servers in the same domain for performance, reliability, or other purposes. For this reason, we use the terms {\em remote server} and {\em remote network domain} interchangeably. The remote services are represented by the corresponding TCP or UDP port numbers. We will first study the number of remote network domains and port numbers that a device communicate with in one day, and then we study the corresponding behavior on a daily basis for one week. To make the figures more legible, in the figures we only show the results of seven IoT devices and two non-IoT devices. We omit the results of other devices. They have similar results as the ones reported here.

\subsubsection{Remote Servers and Services in One Day}
Fig.~\ref{fig:outgoing_hosts} shows the number of remote network domains that the devices communicate with in the chosen day. We report the results for the TCP and UDP traffic separately to better illustrate the communication patterns of the devices. As we can see from the figure, IoT devices only communicate with a small number of remote network domains. For example, most of the IoT devices only communicate with one network domain for TCP traffic, except the Printer, which communicates with seven remote network domains for TCP. The IoT devices in general communicate with a slightly higher number of remote network domains for UDP traffic. The UDP is normally used for control traffic such as DNS and NTP to support the autonomous operations of the IoT devices. However, we note that in general, the number of UDP network domains is still a relatively small number. In contrast, the non-IoT devices communicate with a much higher number of remote network domains for TCP traffic. Of course, these are normally affected by the specific usage of the device by the users. Non-IoT devices also have a smaller number of remote network domains for UDP traffic. This is understandable as UDP is mostly used for control traffic such as DNS and NTP etc. Users of these devices may not initiate any UDP traffic.

These observations are consistent with what we have expected. As IoT devices are connected to their corresponding remote server for their status update and other operational activities, they regularly send traffic to their remote servers and maintain longer TCP sessions (see Subsection~\ref{sec:flow} for more studies on flow-level characteristics of IoT devices). Smart bulb and Plug1 do not have any external UDP connections while cameras maintain some UDP connections with different time servers.

Figure~\ref{fig:outgoing_ports} shows the number of remote port numbers that devices communicate with. In general, IoT devices communicate with a small number of remote port numbers, except Cam5.
We inspected the traffic of Cam5 manually, and it revealed the following observation. Cam5 was connected very often for a live view from iPhone, which resulted in communications with a number of different UDP ports. 
However, it still holds that IoT devices in general communicate with a limited number of remote port numbers. In contrast, non-IoT devices in general communicate with a larger number of remote port numbers. The specifics of the remote port numbers of non-IoT devices may depend on the specific usage of these devices by the users. For example, the non-IoT devices iPad and Laptop1 communicated with various port numbers on websites such as Youtube and Facebook. 

Table~\ref{table:outgoing_host_port} shows the most frequently used remote network domains and port numbers of TCP and UDP traffic for IoT devices. We see that IoT devices mainly use port $443$ for their TCP traffic, which is an official port number for HTTPS. IoT devices rarely use port number $80$ for the HTTP traffic. Some IoT devices use port numbers other than $443$ for HTTPS traffic. For instance, Cam6 uses port numbers $19000$ and $19001$ to communicate with their server instead of the more commonly used $443$ port number. The remote network domains and port numbers of non-IoT devices are much more diverse compared to IoT devices and unpredictable, because their behavior is largely determined by the users who use these devices.

We also note from Table~\ref{table:outgoing_host_port} that most of the IoT devices use Google DNS server as their DNS provider, which is also supported by other studies~\cite{mazhar2020characterizing}, while some others use the local router for this purpose (not shown in table). On the other hand, all laptops, iPhone, and iPad use the local router as their DNS provider. The Android phone uses the router as well as Google DNS as its DNS provider.

\subsubsection{Daily Remote Network Domains and Port Numbers in One Week}


\begin{figure}[!h]
\begin{subfigure}[b]{0.24\textwidth}
    \includegraphics[width=\textwidth]{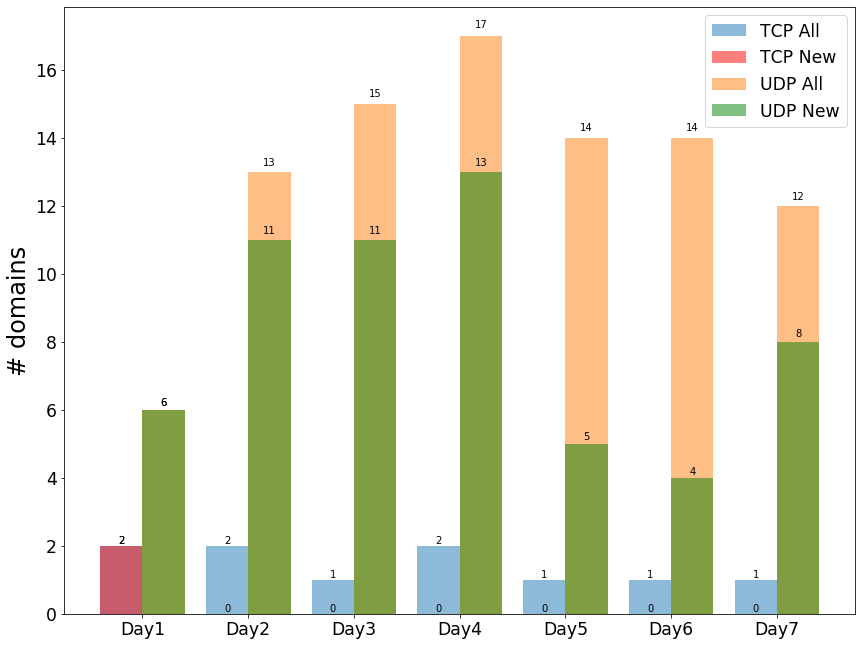}
    \caption{Domains - IoT Device Cam1}
    \label{fig:outgoing_hosts_cam1}
\end{subfigure}
\begin{subfigure}[b]{0.24\textwidth}
    \includegraphics[width=\textwidth]{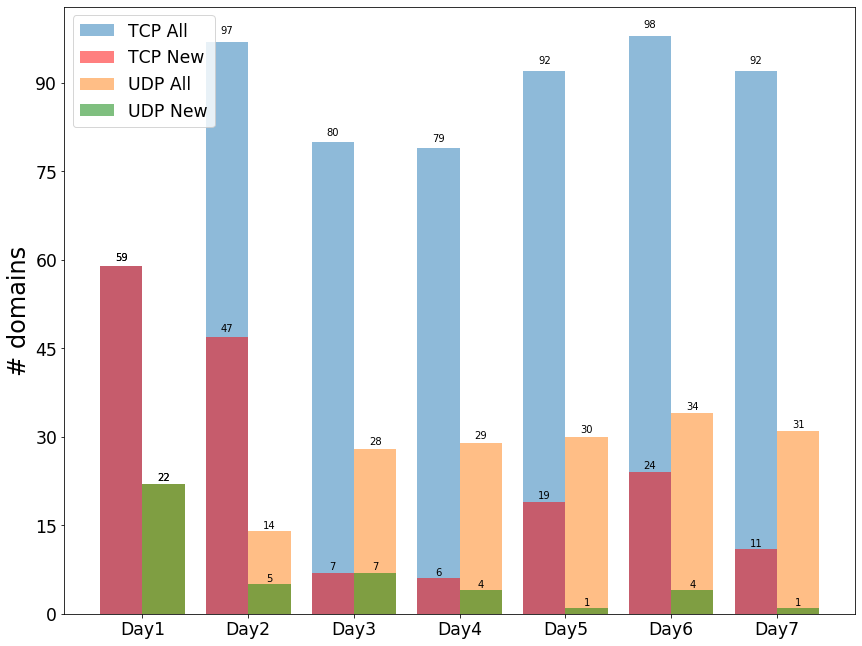}
    \caption{Domains - Non-IoT Device Laptop1}
    \label{fig:outgoing_hosts_laptop1}
\end{subfigure}

\begin{subfigure}[b]{0.24\textwidth}
    \includegraphics[width=\textwidth]{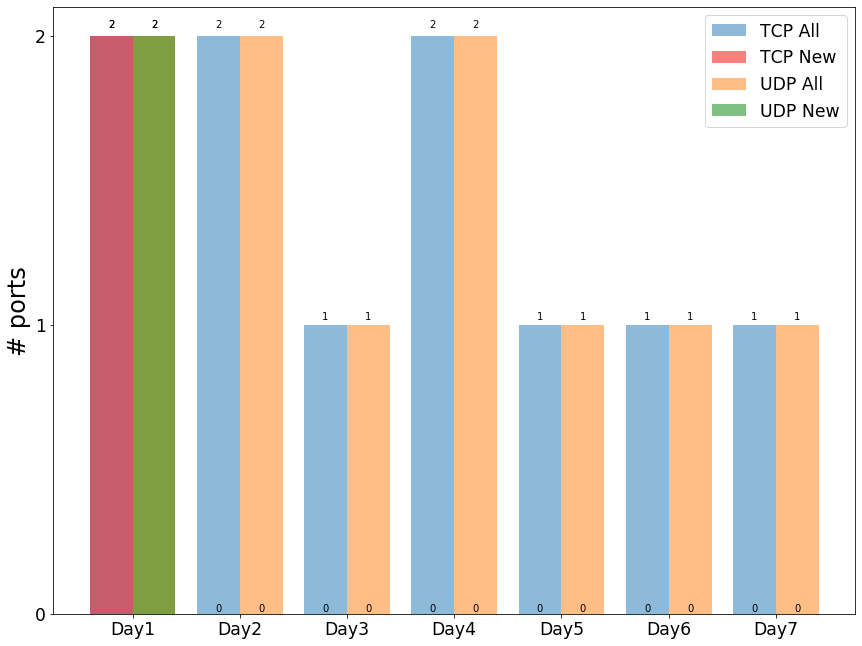}
    \caption{Ports - IoT Device Cam1}
    \label{fig:outgoing_ports_cam1}
\end{subfigure}
\begin{subfigure}[b]{0.24\textwidth}
    \includegraphics[width=\textwidth]{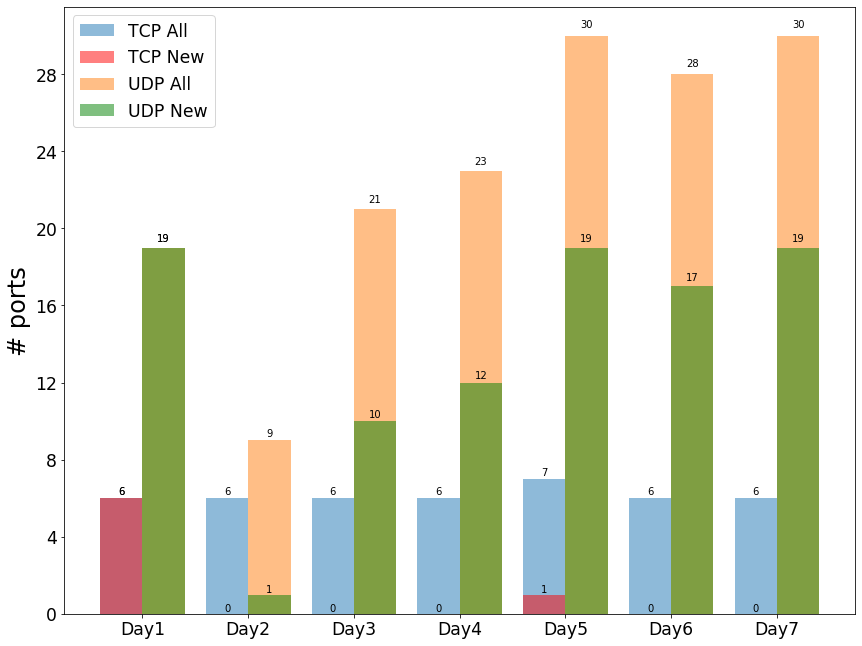}
    \caption{Ports - Non-IoT Device Laptop1}
    \label{fig:outgoing_ports_laptop1}
\end{subfigure}

\caption{Daily Remote Network Domains and Ports in One Week.}
\label{fig:weekly_hosts_cam1_laptop1}
\end{figure}


Fig.~\ref{fig:weekly_hosts_cam1_laptop1} shows the number of remote network domains for a representative IoT (Cam1) and non-IoT device (Laptop1). In the figure, in addition to showing the total number of remote network domains in a day, we also show the number of new remote network domains, that is, the network domains that we have not seen in the previous days in the chosen week. From the figure we can see that, the IoT device Cam1 only communicates with one or two remote network domains (which is amazonaws.com, and additionally cloudfront.net, which is an integrated Content Delivery Network service with amazon aws~\cite{cloudfront}) throughout the week for TCP traffic. All IoT devices show similar behavior for TCP traffic. In contrast, for UDP traffic, Cam1 has communicated with a changing number of remote network domains throughout the week, with new remote network domains added each day. Most of the UDP traffic is related to NTP and DNS queries. Unlike TCP traffic, there are slightly different behaviors for UDP traffic among IoT devices. In particular, the IoT devices Plug1 and Bulb do not communicate with NTP or DNS servers.

Compared to the IoT devices, non-IoT devices behave quite differently. From the figure, we can see that Laptop1 communicates with a large number of remote network domains for both TCP and UDP traffic. In addition, it also communicates with new remote network domains each day. This is a reasonable observation, as the behavior of a non-IoT device is largely determined by the users who use this device. Users may access different remote servers and applications at will. On the other hand, the behavior of IoT devices is more likely to be pre-programmed and thus less likely to access random remote network domains and applications.

Figure~\ref{fig:outgoing_ports_cam1} shows the daily number of remote port numbers that Cam1 communicates with. From the figure we can see that, Cam1 communicates with a relatively small and stable set of remote port numbers. In particular, even though there are fluctuations in terms of number of remote port numbers, there are no new port numbers observed after the first day. In contrast, as we can see in Figure~\ref{fig:outgoing_ports_laptop1} that Laptop1 almost always communicates with some new remote port numbers for UDP traffic (and also one day for TCP traffic). Again, this observation can be similarly explained by the fact that, IoT devices are more likely pre-programmed with a fix set of applications, and consequently they only communicate with a relatively small and stable set of remote port numbers.



\subsection{Flow-Level Traffic Characteristics}\label{sec:flow}
In this subsection we investigate the {\em flow-level} traffic characteristics of IoT devices, including number of flows, flow duration, flow size, among others. These provide us with insights into the behaviors of IoT devices at the granularity of TCP and UDP flows. 



\subsubsection{Number of Flows and Flow Durations}
We first study the number of flows and flow durations of TCP and UDP flows for the deployed devices on the representative day (for 24 hours). The duration of a flow is defined as the time difference between the last packet and the first packet of the flow.




\begin{figure}[h!]
    \centering
    \includegraphics[width=0.37\textwidth]{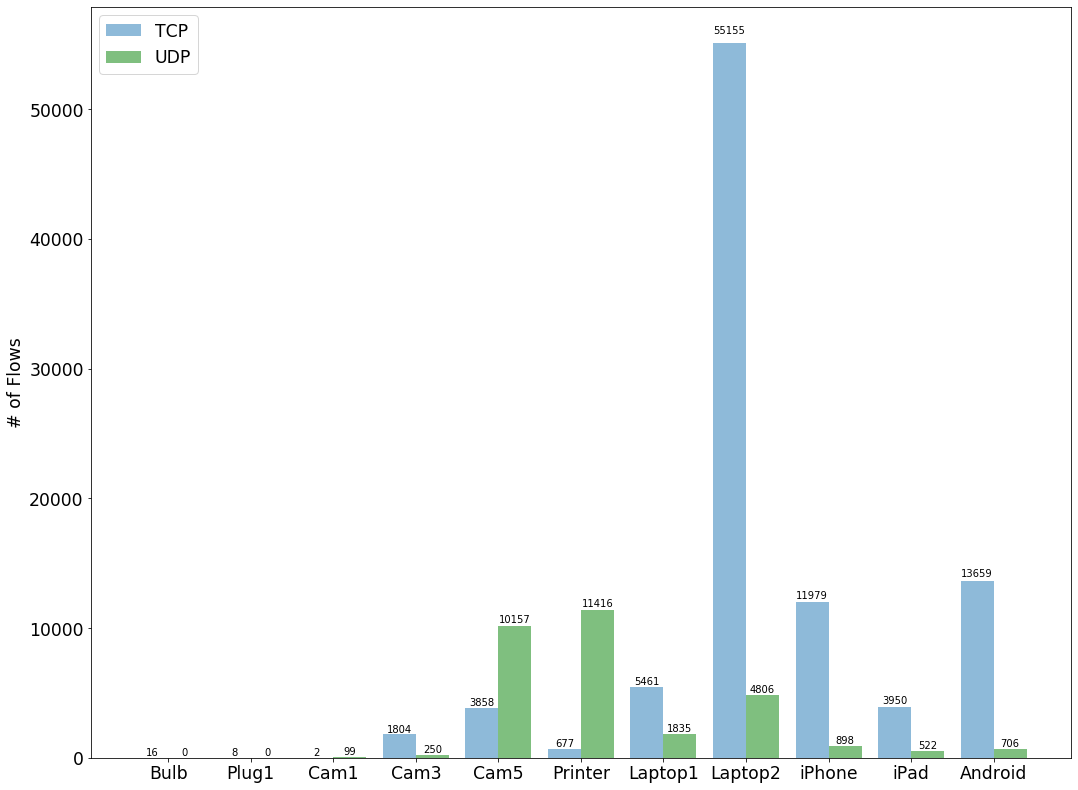}
    \caption{Number of TCP and UDP Flows.}
    \label{fig:number_of_flows}
\end{figure}


Figure~\ref{fig:number_of_flows} shows the number of flows of the representative IoT and non-IoT devices. From the figure we first note that non-IoT devices have more TCP flows than IoT devices. For example, the non-IoT device iPad generated $3950$ TCP flows, the lowest among the non-IoT devices, which is still higher than the highest number of TCP flows generated by IoT devices ($3858$ generated by Cam5). As we will see when we study the flow durations, this could be caused by the fact that IoT devices tend to generate long-lasting TCP flows. The situation of UDP flows is quite different. Some of the IoT devices do not generate UDP flows at all (including Bulb and Plug1); however, some other IoT devices have generated a large number of UDP flows (including Cam5 and Printer). After examining the traffic trace, we note that Cam5 has generated frequent UDP flows to remote port numbers such as $32100$ for its operational functionalities. Printer has some mDNS flows (with destination IP addresses $224.0.0.251$ and $224.0.0.252$, and port number $5353$), while all other UDP flows are to amazonaws servers with port $9930$.

\begin{figure}[th]
\centering  
\subcaptionbox{IoT devices \label{fig:hourly-num-flows-tcp-iot}}[0.48\linewidth]
    {\includegraphics[width=0.24\textwidth]{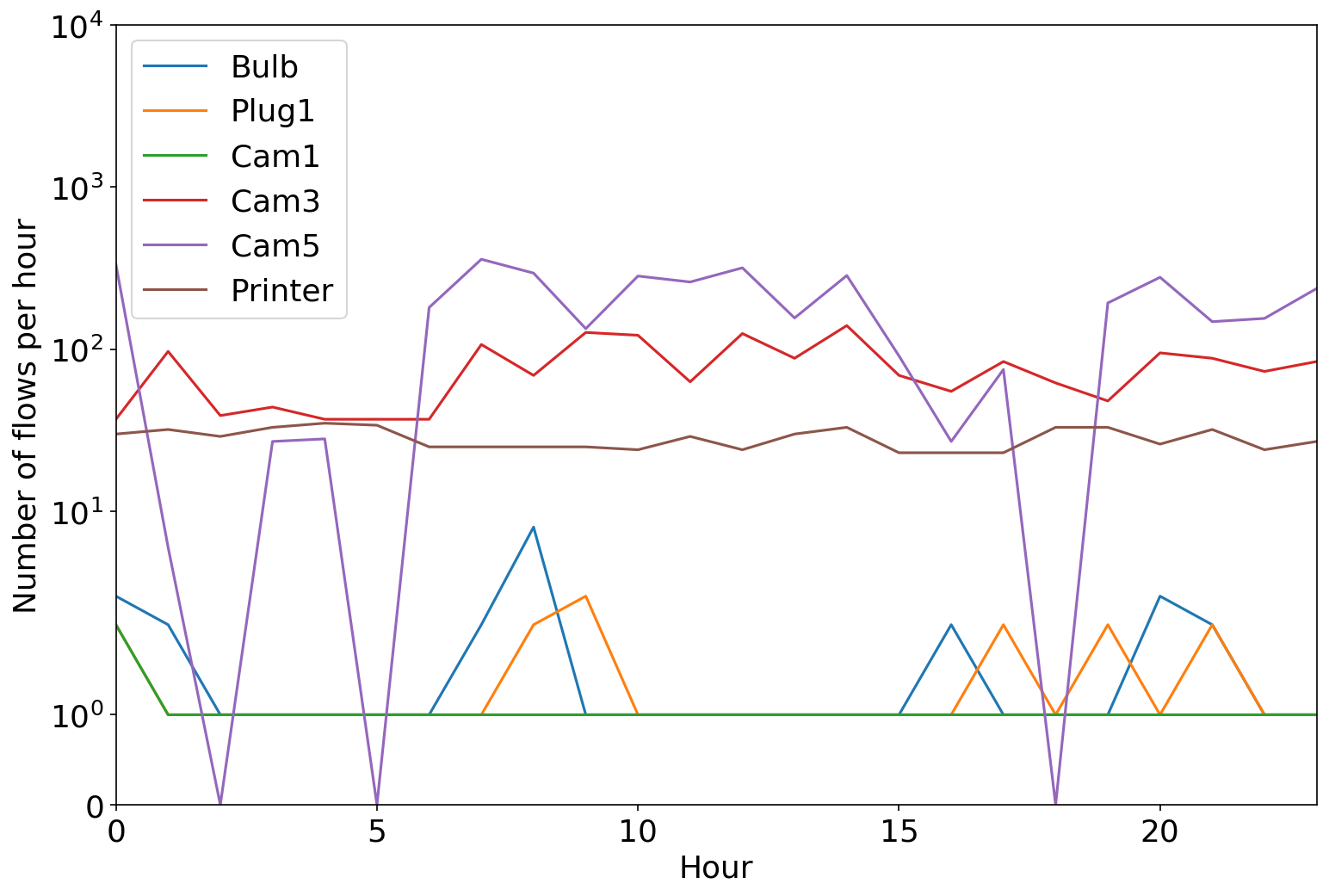}}
\hspace{0.1cm}
\subcaptionbox{Non-IoT devices \label{fig:hourly-num-flows-tcp-noniot}}[0.48\linewidth]
    {\includegraphics[width=0.24\textwidth]{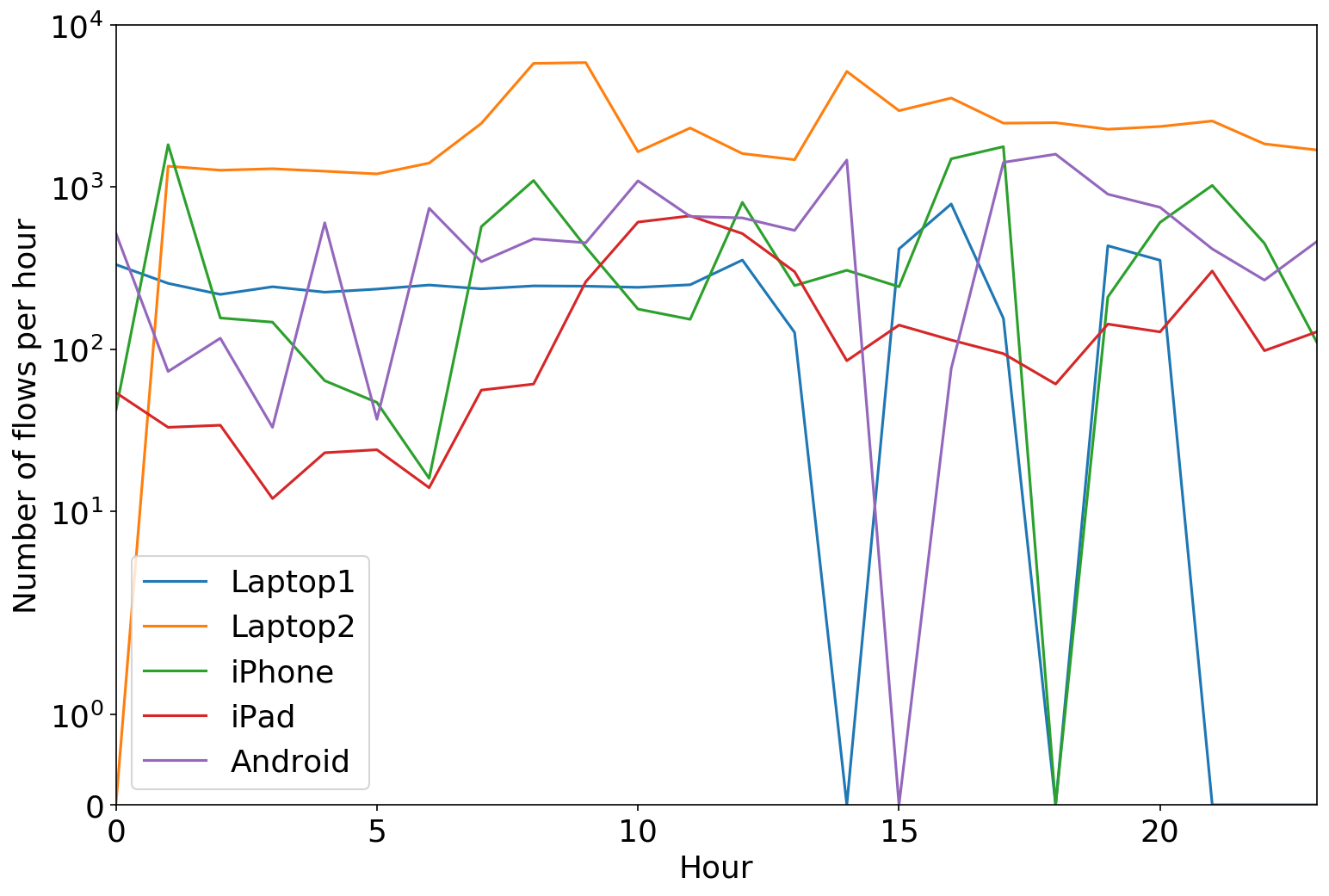}}
    
\caption{Hourly number of TCP flows.}
\label{fig:hourly-num-flows-tcp}
\end{figure}

Figures~\ref{fig:hourly-num-flows-tcp-iot} and~\ref{fig:hourly-num-flows-tcp-noniot} show the hourly number of TCP flows for representative IoT and non-IoT devices, respectively. From the figures, we note that most of the IoT devices have relatively stable number of hourly number of TCP flows, with some fluctuations, except Cam5. Overall, the hourly number of TCP flows of Cam5 is also stable; however, it has a number of instances where it has zero TCP flows. We manually examined the data trace for Cam5 and other IoT devices around the times when these instances occurred (for both TCP and UDP). Our conclusion is that, this was likely related to the specific application running on Cam5 for video delivery, which may crash or malfunction from time to time. 


Non-IoT devices also generate relatively stable hourly number of TCP flows, though with a relatively larger fluctuation (and also larger number of TCP flows). A number of non-IoT devices also generate zero number of TCP flows during a number of hourly time intervals. Based on our observation, this could be caused by either taking a device outside the home or accidentally turning off a device. 

\begin{figure}[th]
\subcaptionbox{IoT devices \label{fig:hourly-num-flows-udp-iot}}[0.48\linewidth]
    {\includegraphics[width=0.24\textwidth]{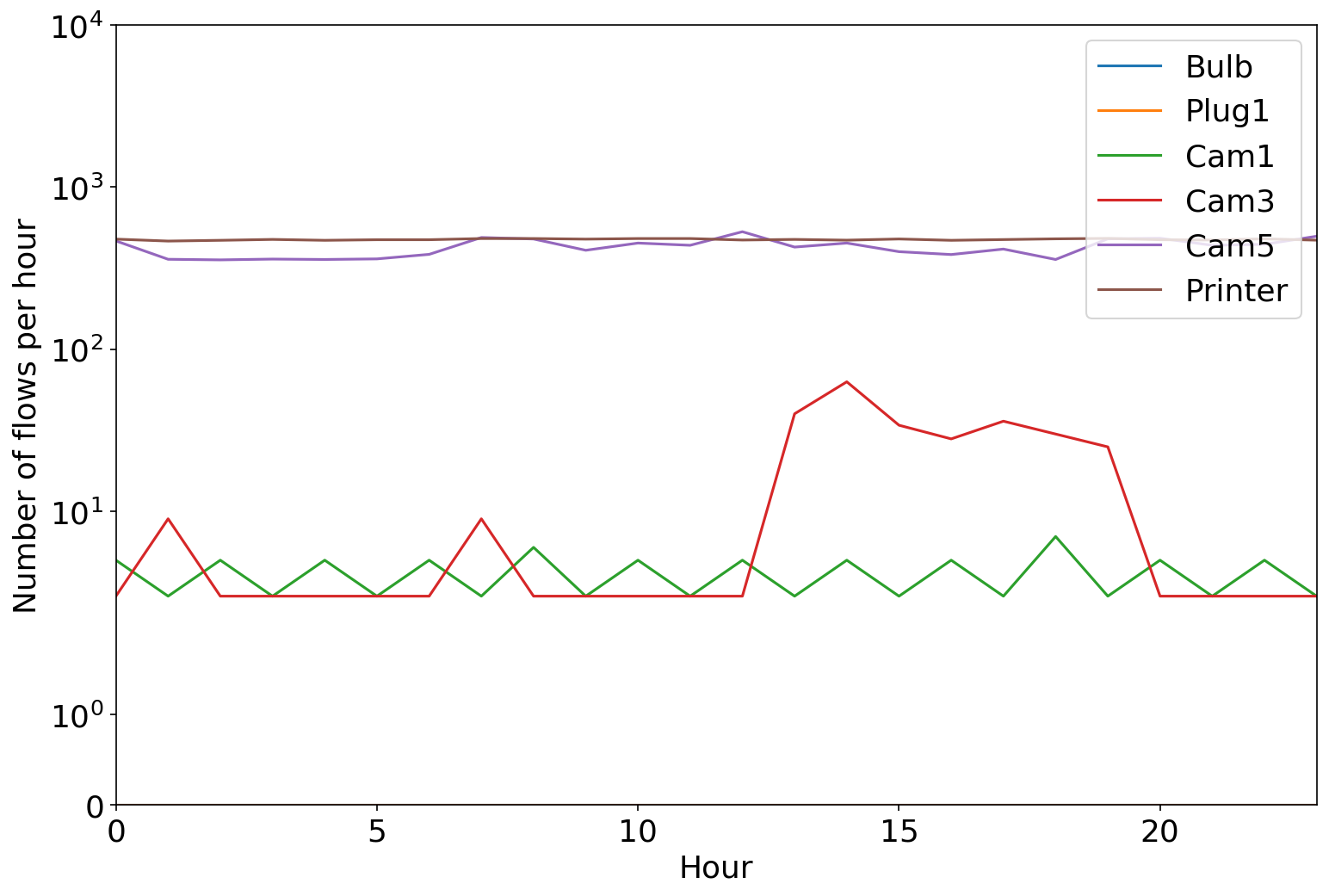}}
\hspace{0.1cm}
\subcaptionbox{Non-IoT devices \label{fig:hourly-num-flows-udp-noniot}}[0.48\linewidth]
    {\includegraphics[width=0.24\textwidth]{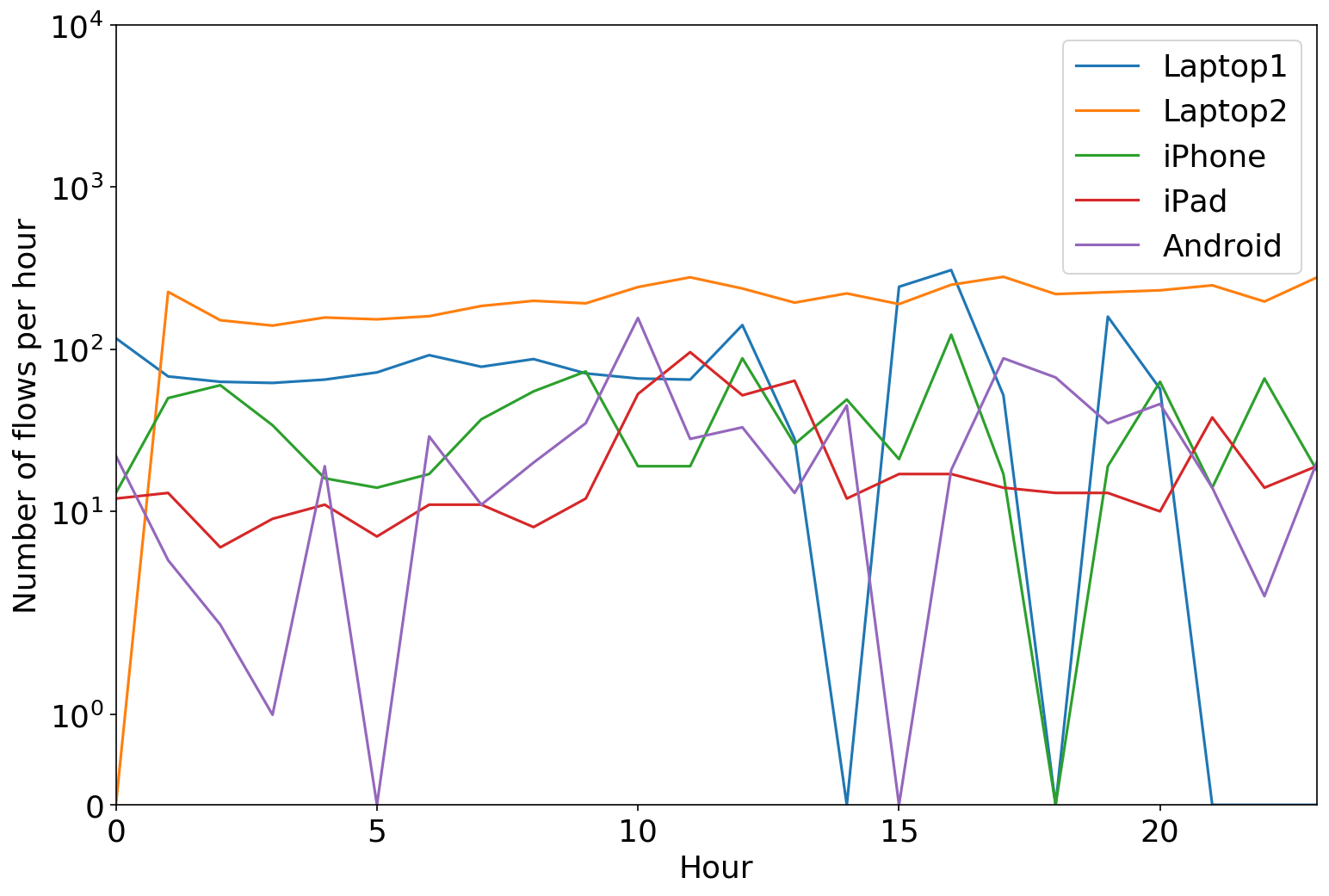}}
    
\caption{Hourly number of UDP flows.}
\label{fig:hourly-num-flows-udp}
\end{figure}

Figures~\ref{fig:hourly-num-flows-udp-iot} and~\ref{fig:hourly-num-flows-udp-noniot} show the hourly number of UDP flows of representative IoT and non-IoT devices, respectively. From the figures we can see that, the hourly number of UDP flows of IoT devices are very stable, except Cam3, which generates larger number of UDP flows from the noon time to the early night time. This could be caused by a larger amount of data traffic to be delivered during that time interval.

The hourly number of UDP flows of non-IoT devices has higher fluctuation than those of IoT devices. In addition, except one instance around $5$AM for Android, all the instances with zero hourly number of UDP flows are coincident with those of TCP flows of the same device. As we have discussed above, these should be caused by us either taking a device outside the home or accidentally turning it off.

\begin{figure}[th]
\centering
\subcaptionbox{IoT devices \label{fig:flow_duration_tcp_iot}}[0.48\linewidth]
    {\includegraphics[width=0.24\textwidth]{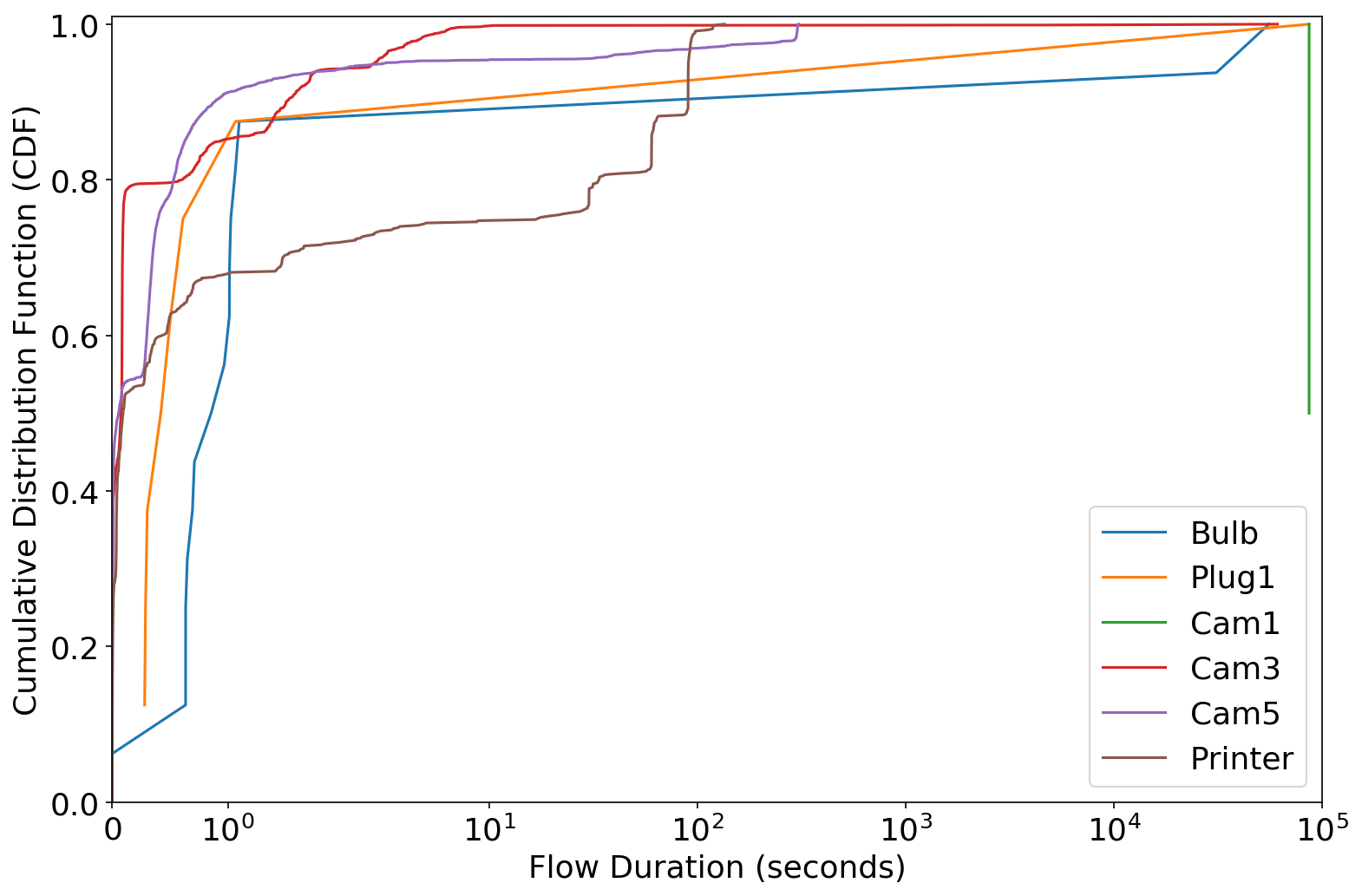}}
\hfill
\subcaptionbox{non-IoT devices \label{fig:flow_duration_tcp_noniot}}[0.48\linewidth]
    {\includegraphics[width=0.24\textwidth]{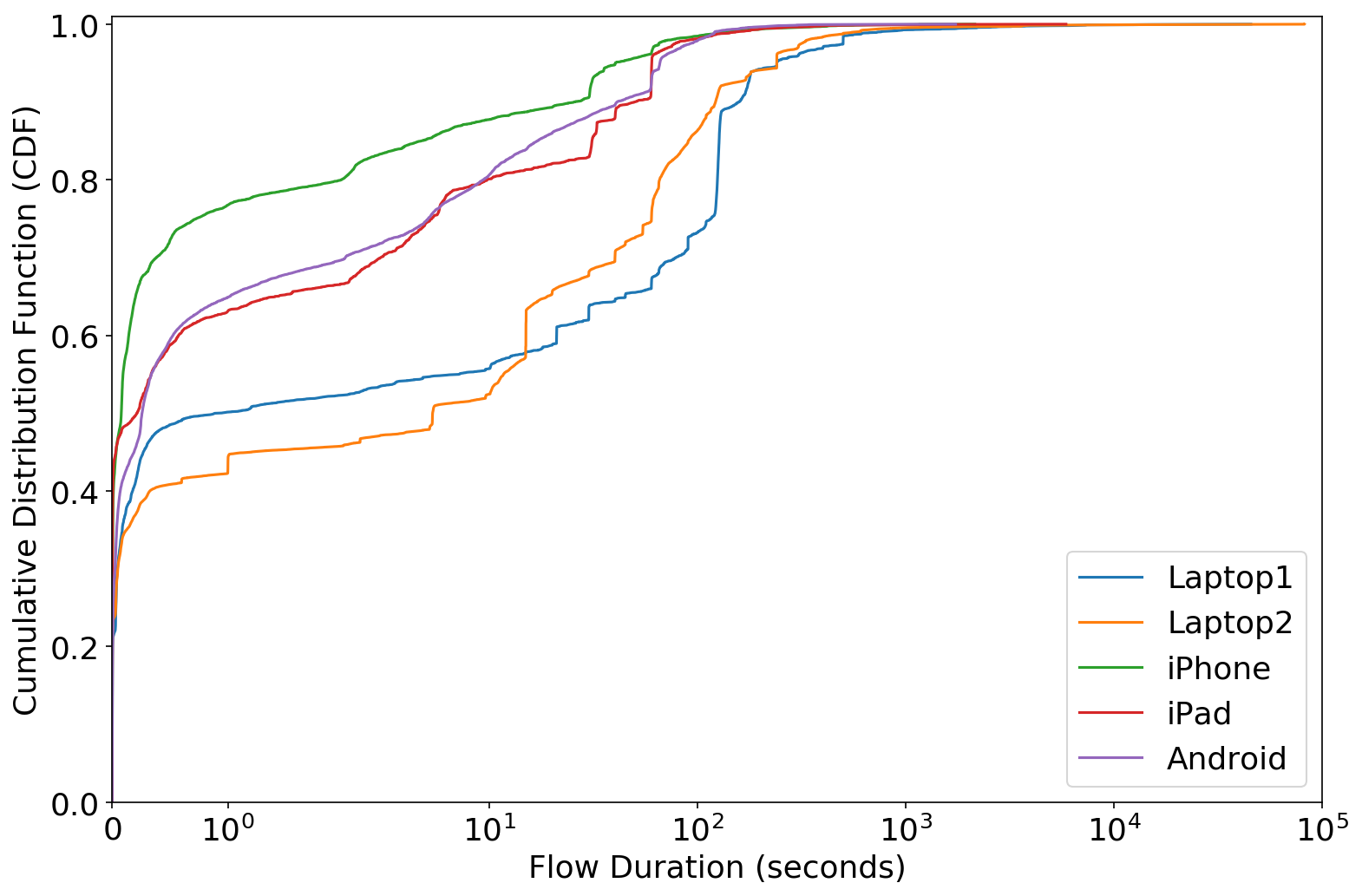}}
\caption{TCP flow duration.}
\label{fig:flow-durations-tcp}
\end{figure}

Figure~\ref{fig:flow_duration_tcp_iot} shows the TCP flow durations of the representative IoT devices. We first note that most IoT devices tend to generate either long-lasting flows or very short flows. For example, Cam1 only has two TCP flows and both of these flows last for close to $24$ hours (the length of our one-day data trace). Some flows of a few other IoT devices also last for close to $24$ hours for the same reason. Theses devices use long-lasting TCP flows to communicate with their corresponding servers. On the other hand, most TCP flows of IoT devices have short durations, except Printer. For example, about $85\%$ of TCP flows of most IoT devices have a duration less than $1$ second. Printer behaves slightly different from the other IoT devices;  about $30\%$ of its TCP flows have a duration between $1$ second and $100$ seconds.

Figure~\ref{fig:flow_duration_tcp_noniot} shows the TCP flow duration of the non-IoT devices. We note first that, most non-IoT devices do not have long-lasting TCP flows (except Laptop2). The majority of them last less than $1000$ seconds (about $16$ minutes). In addition, their flow durations are more spread out between $0$ second and $1000$ seconds. Like IoT devices, non-IoT devices also have a large portion of short TCP flows (for example, with flow duration less than $1$ second). However, the percentage of short TCP flows of non-IoT devices is in general lower than that of IoT devices.

\begin{figure}[th]
\centering  
\subcaptionbox{IoT devices \label{fig:flow_duration_udp_iot}}[0.48\linewidth]
    {\includegraphics[width=0.24\textwidth]{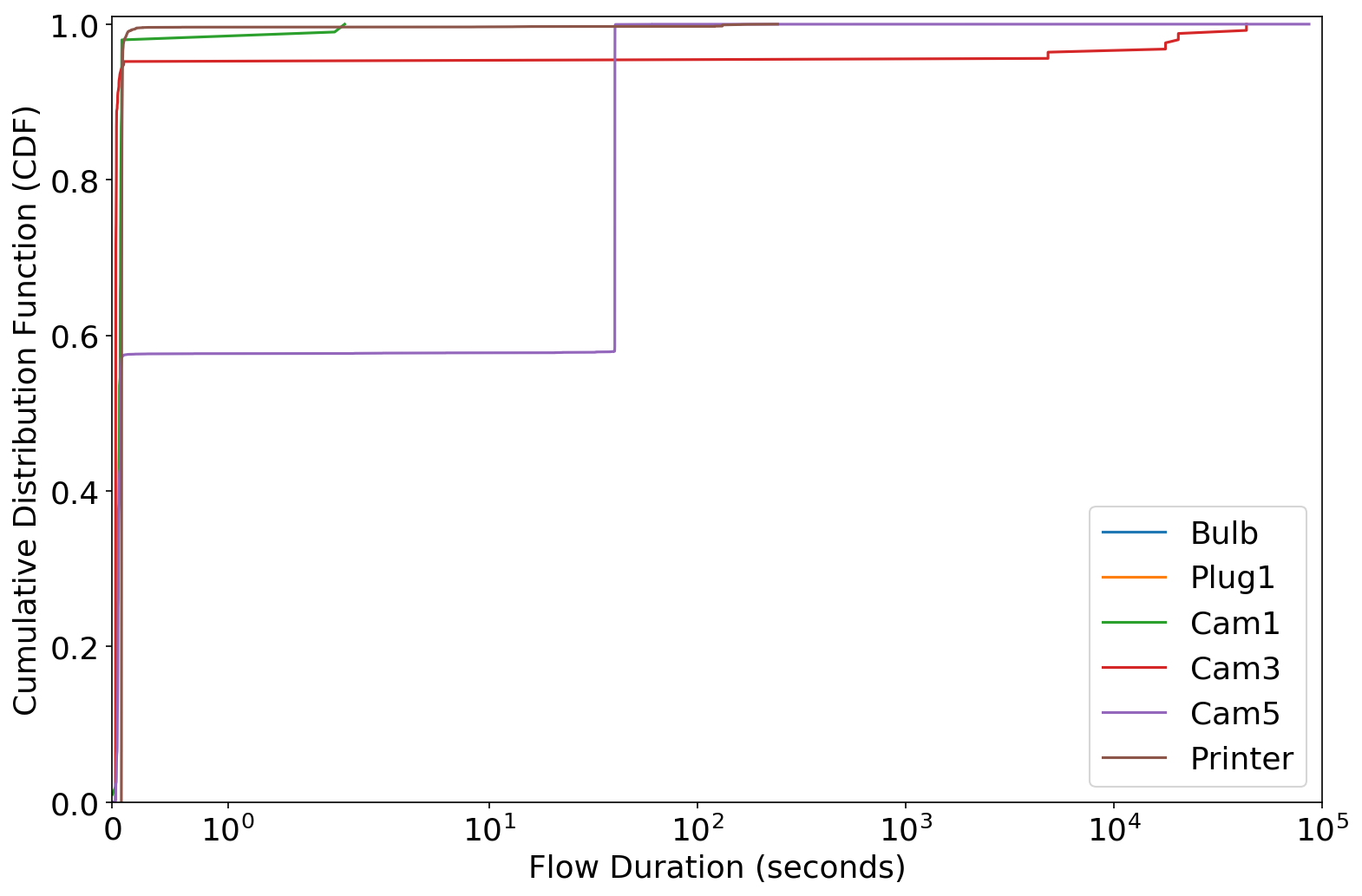}}
\hfill
\subcaptionbox{Non-IoT devices \label{fig:flow_duration_udp_noniot}}[0.48\linewidth]
    {\includegraphics[width=0.24\textwidth]{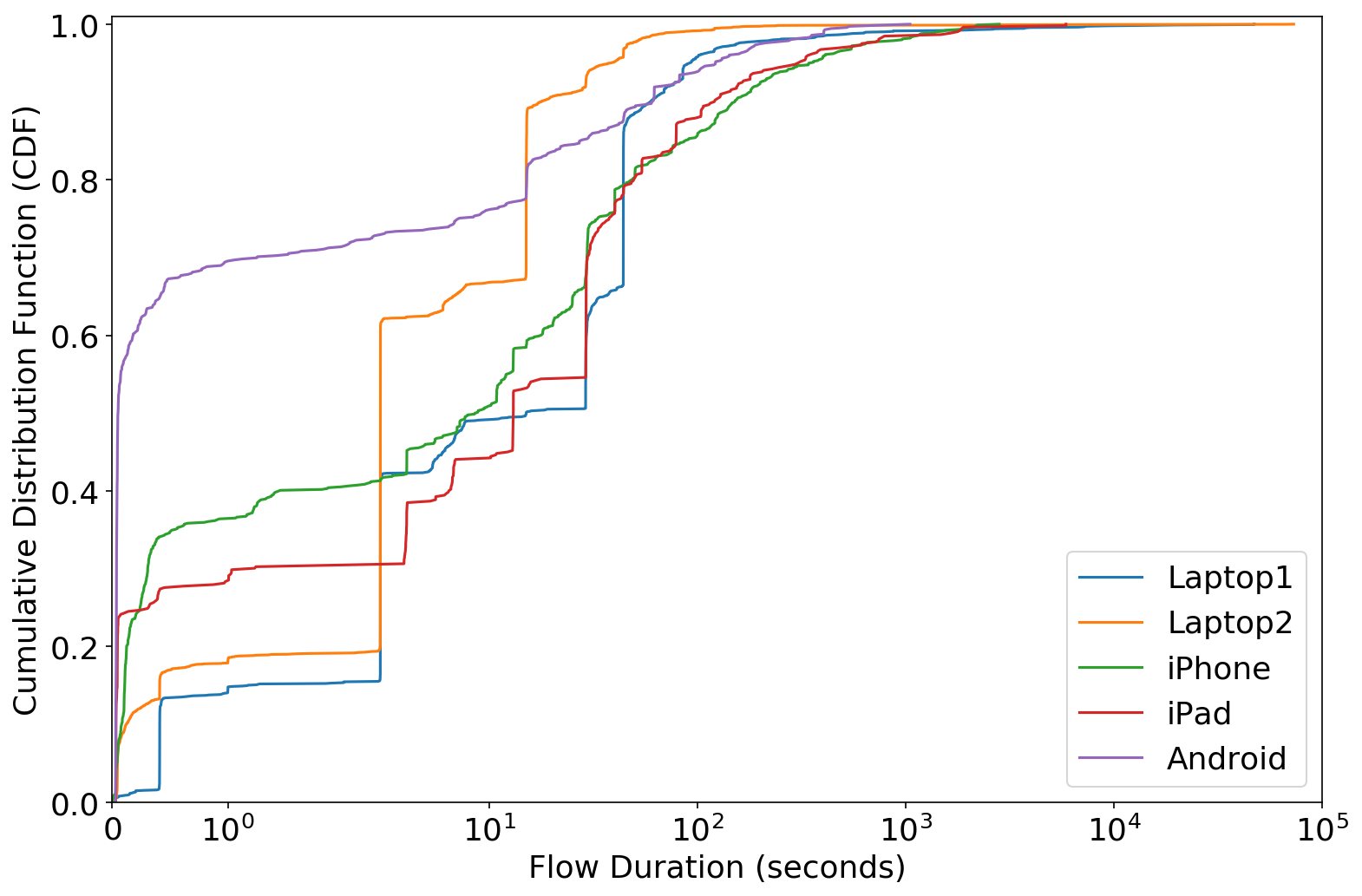}}
    
\caption{UDP flow duration.}
\label{fig:flow-durations-udp}
\end{figure}

Figures~\ref{fig:flow_duration_udp_iot} and~\ref{fig:flow_duration_udp_noniot} show the UDP flow durations of IoT and non-IoT devices, respectively. From the figure we can see that, IoT and non-IoT devices have different behaviors in terms of UDP flow durations. First we note that the majority of UDP flows of IoT devices have very short UDP flow durations (close to $0$ seconds). This is caused by the fact that IoT device only periodically communicate with certain remote UDP servers such as NTP servers. In contrast, UDP flows of non-IoT are in general much longer. For example, a large portion of UDP flows of all non-IoT devices have a duration between $1$ second and $1,000$ seconds. When a non-IoT device communicates with a remote UDP server, they tend to exchange traffic continuously, and therefore, resulting in longer UDP flows. 

Cam5 in Figure~\ref{fig:flow_duration_udp_iot} behaves differently from other (IoT) devices. It has a concentrated flow duration of around $40$ seconds for about $40\%$ of its UDP flows. Our manual examination of the data trace shows that, Cam5 uses remote port $32100$ to communicate with amazonaws servers to send video traffic. A large number of these flows have similar duration (around 40 seconds). We have similarly inspected the UDP traffic of Laptop2 (Figure~\ref{fig:flow_duration_udp_noniot}), which shows that it uses SSDP (with destination IP address $239.255.255.250$, and port $1900$) to facilitate UPnP (Universal Plug and Play). Most of the SSDP flows a duration around $3$ seconds. We see another jump around $15$ seconds, which are contributed by flows to a google service (google safe browsing).

\subsubsection{Flow Size}\label{sec:flow}
In this subsection we study the size of TCP and UDP flows in terms of both the number of packets and the amount of traffic (in bytes). We note that the amount of traffic in a packet is the number of bytes in the payload of a TCP or UDP packet; and therefore, the amount of traffic in a flow is the sum of the payload of all packets in the flow.

\begin{figure}[th]
\centering
\subcaptionbox{IoT devices \label{fig:flow-bytes-tcp-iot}}[0.48\linewidth]
    {\includegraphics[width=0.24\textwidth]{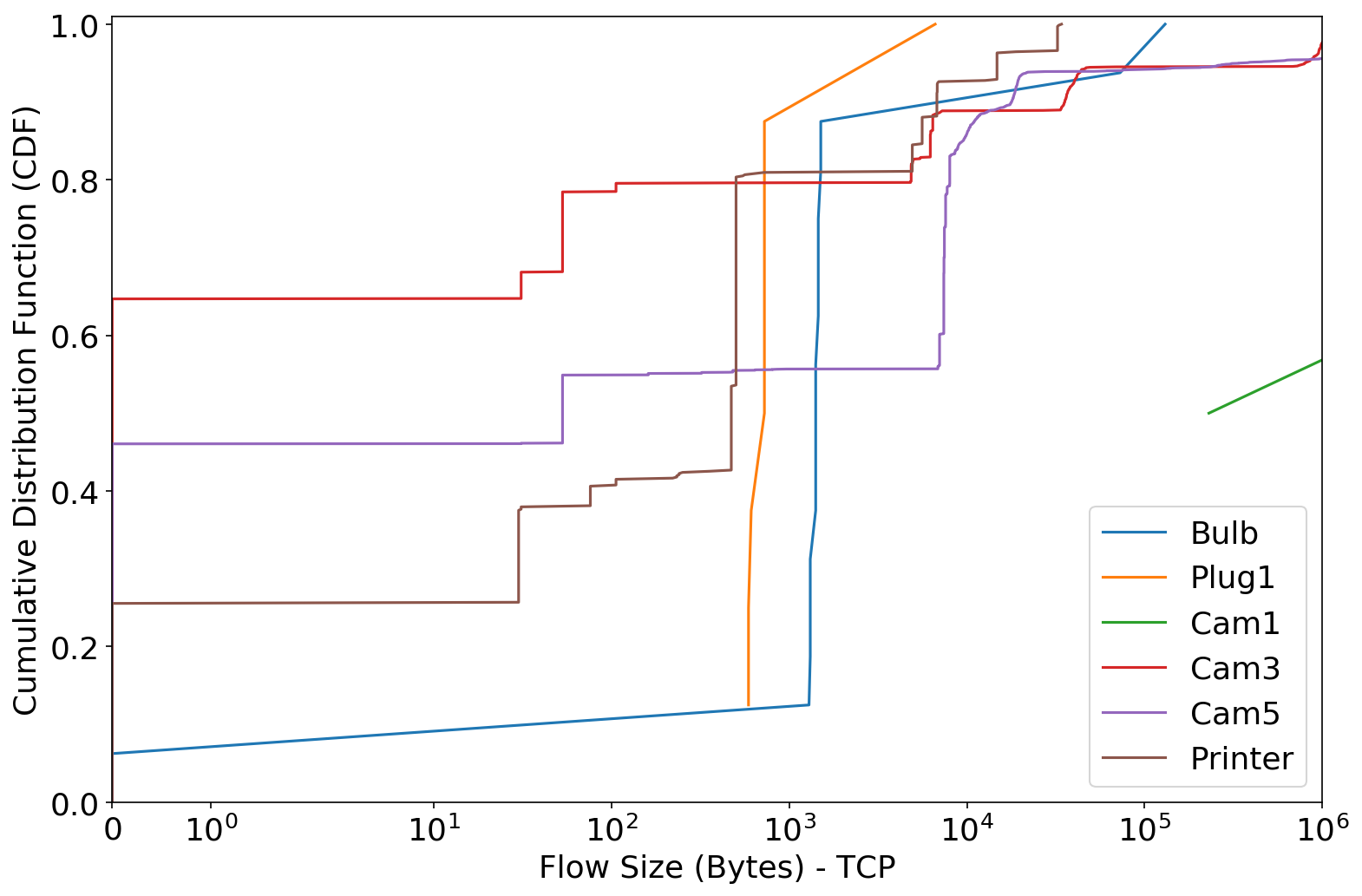}}
\hfill
\subcaptionbox{Non-IoT devices \label{fig:flow-bytes-tcp-noniot}}[0.48\linewidth]
    {\includegraphics[width=0.24\textwidth]{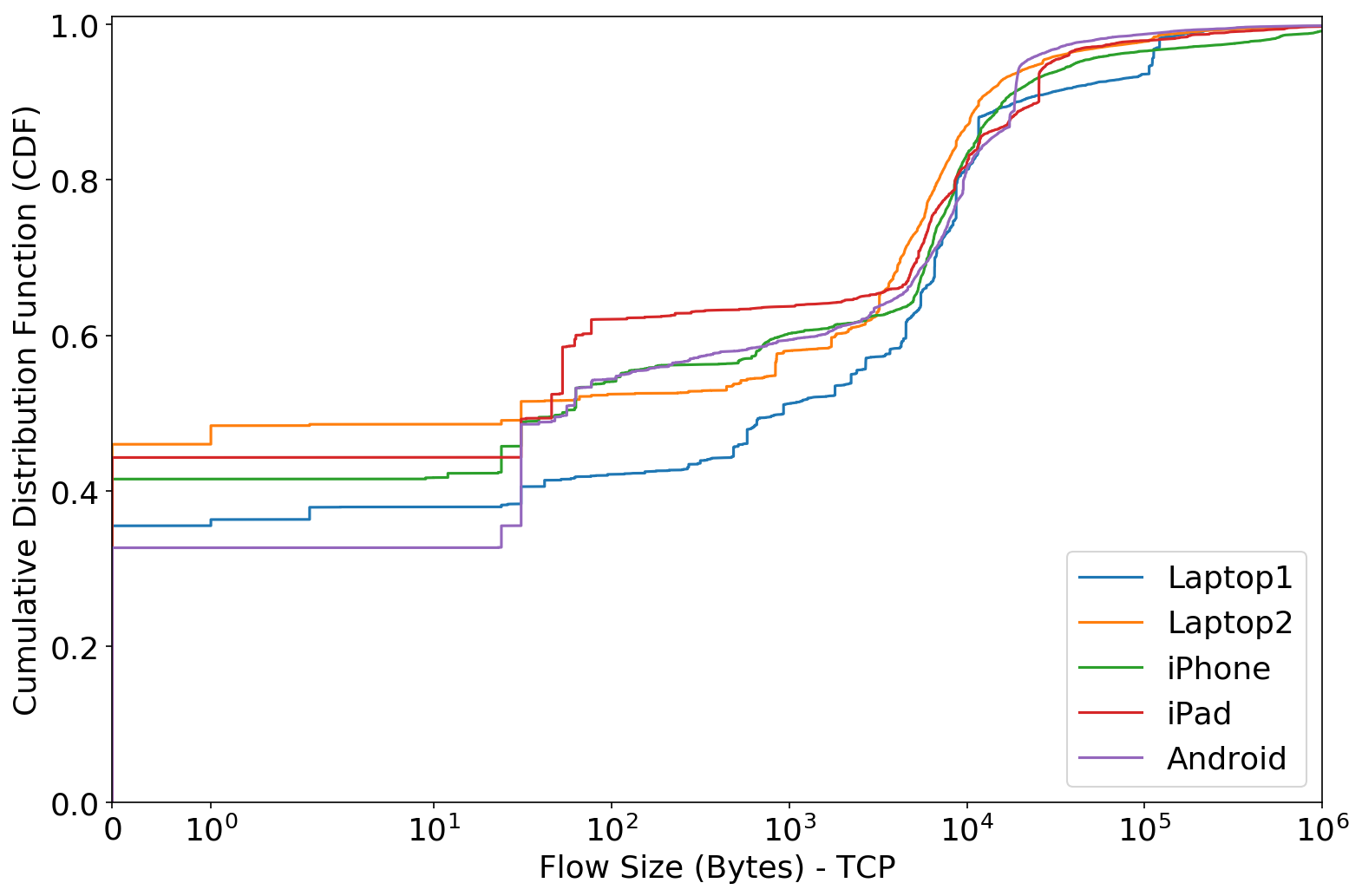}}
\caption{TCP flow size (bytes).}
\label{fig:flow-bytes-tcp}
\end{figure}

Figures~\ref{fig:flow-bytes-tcp-iot} and~\ref{fig:flow-bytes-tcp-noniot} show the flow sizes of TCP flows of IoT and non-IoT devices (in bytes), respectively. We note that all non-IoT devices have similar behavior in terms of flow sizes, for example, about $70\%$ of TCP flows of all non-IoT devices have a flow size less than or equal to $10^4$ bytes. In contrast, IoT devices have more diverse behaviors in terms of flow sizes. For example, both TCP flows of Cam1 are large flows. One generates $229,892$ bytes of traffic, while another generates more than $10$ GB of traffic (not shown in the figure). We note that 
Cam1 is positioned at the front porch facing a reasonably busy street, and as a consequence, Cam1 generates large volume of traffic. In addition, both TCP flows of Cam1 are long-lasting flows, which also contributes to the observation of large flow sizes. The other IoT devices in general generate smaller TCP flows compared to non-IoT devices.

\begin{figure}[th]
\centering
\subcaptionbox{IoT devices \label{fig:flow-bytes-udp-iot}}[0.48\linewidth]
    {\includegraphics[width=0.24\textwidth]{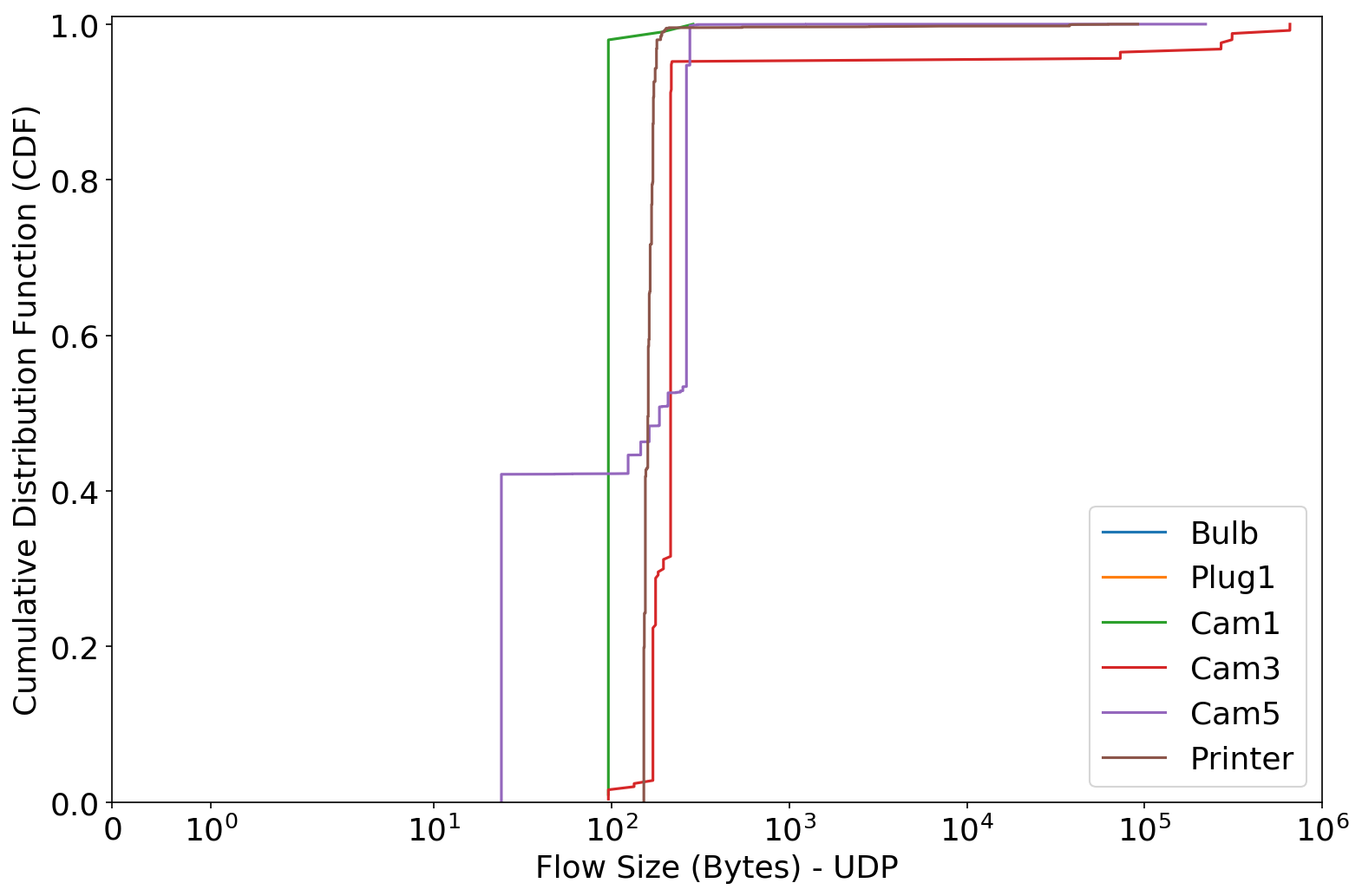}}
\hfill
\subcaptionbox{Non-IoT devices \label{fig:flow-bytes-udp-noniot}}[0.48\linewidth]
    {\includegraphics[width=0.24\textwidth]{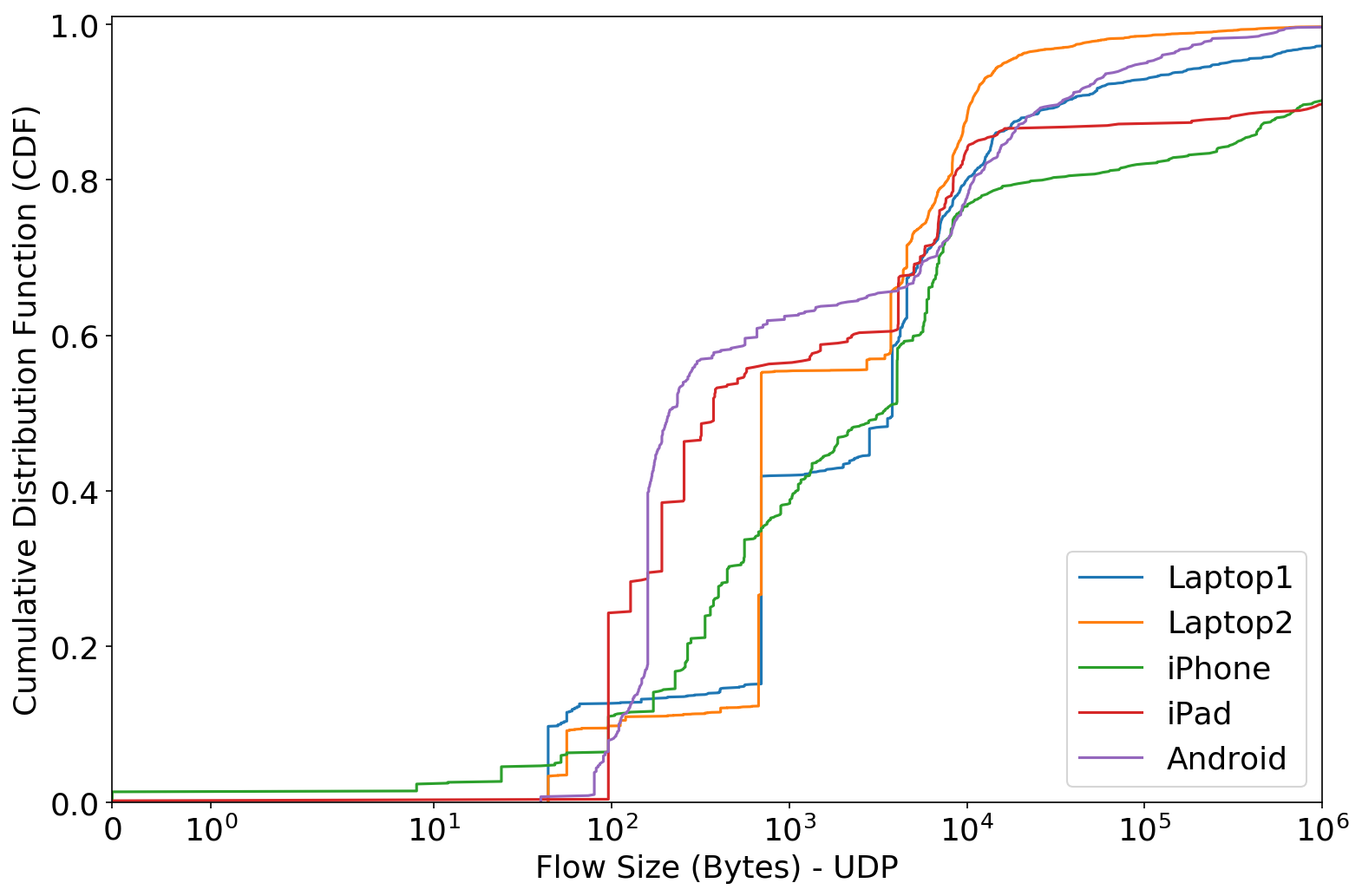}}
    
\caption{UDP flow size (bytes).}
\label{fig:flow-bytes-udp}
\end{figure}


Figures~\ref{fig:flow-bytes-udp-iot} and~\ref{fig:flow-bytes-udp-noniot} show the sizes of UDP flows of IoT and non-IoT devices (in bytes), respectively. We have similar observations with that of TCP flow sizes. In particular, UDP flows of non-IoT devices behave similarly in terms of flow sizes. They follow a similar shape and they in general generate more traffic than IoT flows. In contrast, most UDP flows of IoT devices have small sizes; for example, about $90\%$ of all UDP flows of IoT devices have a flow size less than $200$ bytes.  

\begin{figure}[th]
\centering
\subcaptionbox{IoT \label{fig:flowsize-pkts-tcp-iot}}[0.48\linewidth]
    {\includegraphics[width=0.24\textwidth]{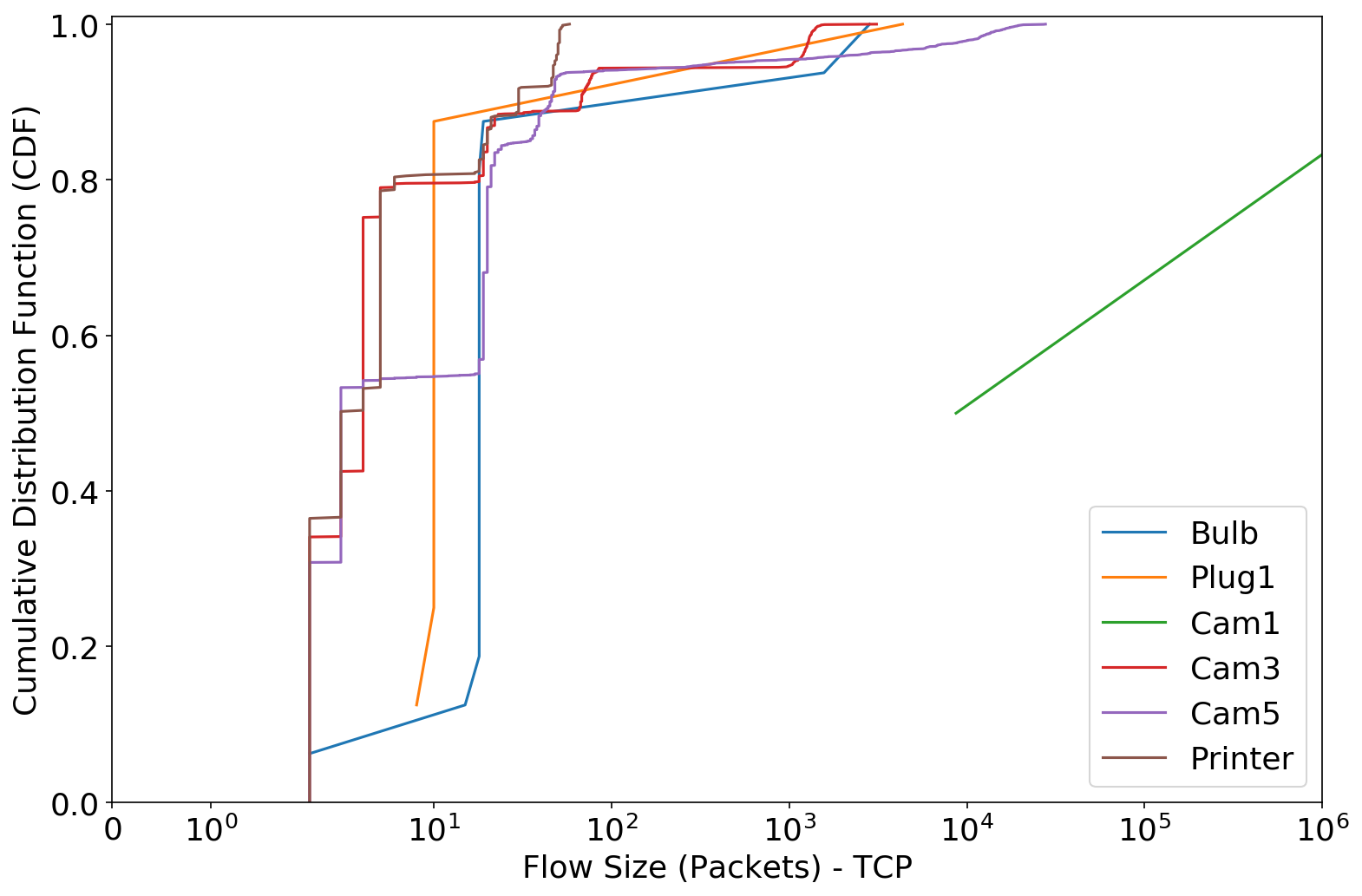}}
\hfill
\subcaptionbox{Non-IoT \label{fig:flowsize-pkts-tcp-noniot}}[0.48\linewidth]
    {\includegraphics[width=0.24\textwidth]{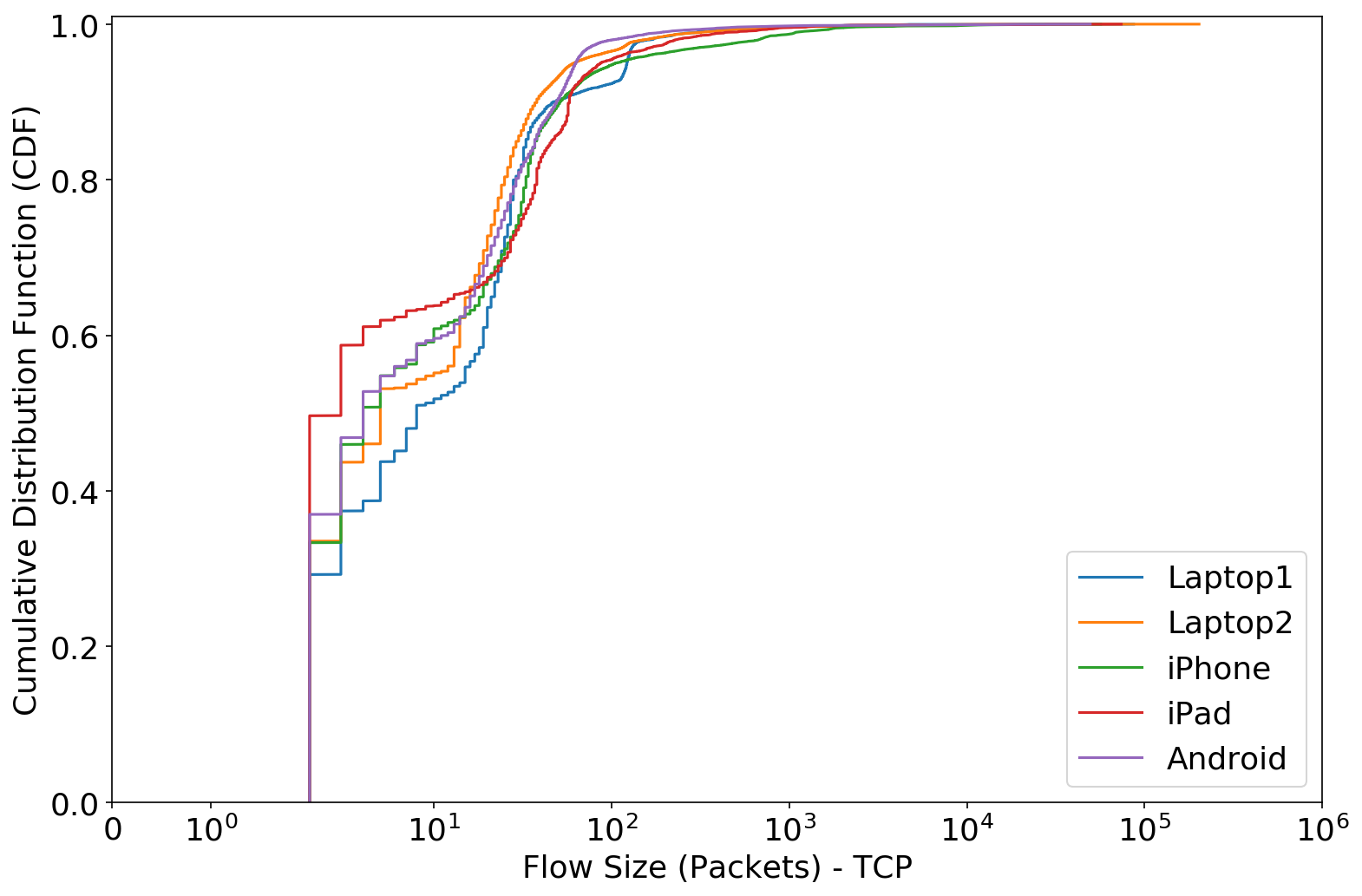}}
   \caption{TCP flow size (packets).}
\label{fig:flow-packets-tcp}
\end{figure}

Figures~\ref{fig:flowsize-pkts-tcp-iot} and~\ref{fig:flowsize-pkts-tcp-noniot} show the TCP flow sizes in terms of number of packets for IoT and non-IoT devices, respectively. From the figures we can see that in general non-IoT devices generate larger flows. For example, about $80\%$ of TCP flows of IoT devices generate less than $20$ packets. On the other hand, only about $60\%$ of TCP flows of non-IoT devices generate less than $20$ packets. Cam1 is an exception; both of its flows generate large number of packets ($8,681$ and $10,893,101$, respectively). Given that they are both long-lasting flows, this observation is not a surprise. 

\begin{figure}[th]
\centering
\subcaptionbox{IoT \label{fig:flowsize-pkts-udp-iot}}[0.48\linewidth]
    {\includegraphics[width=0.24\textwidth]{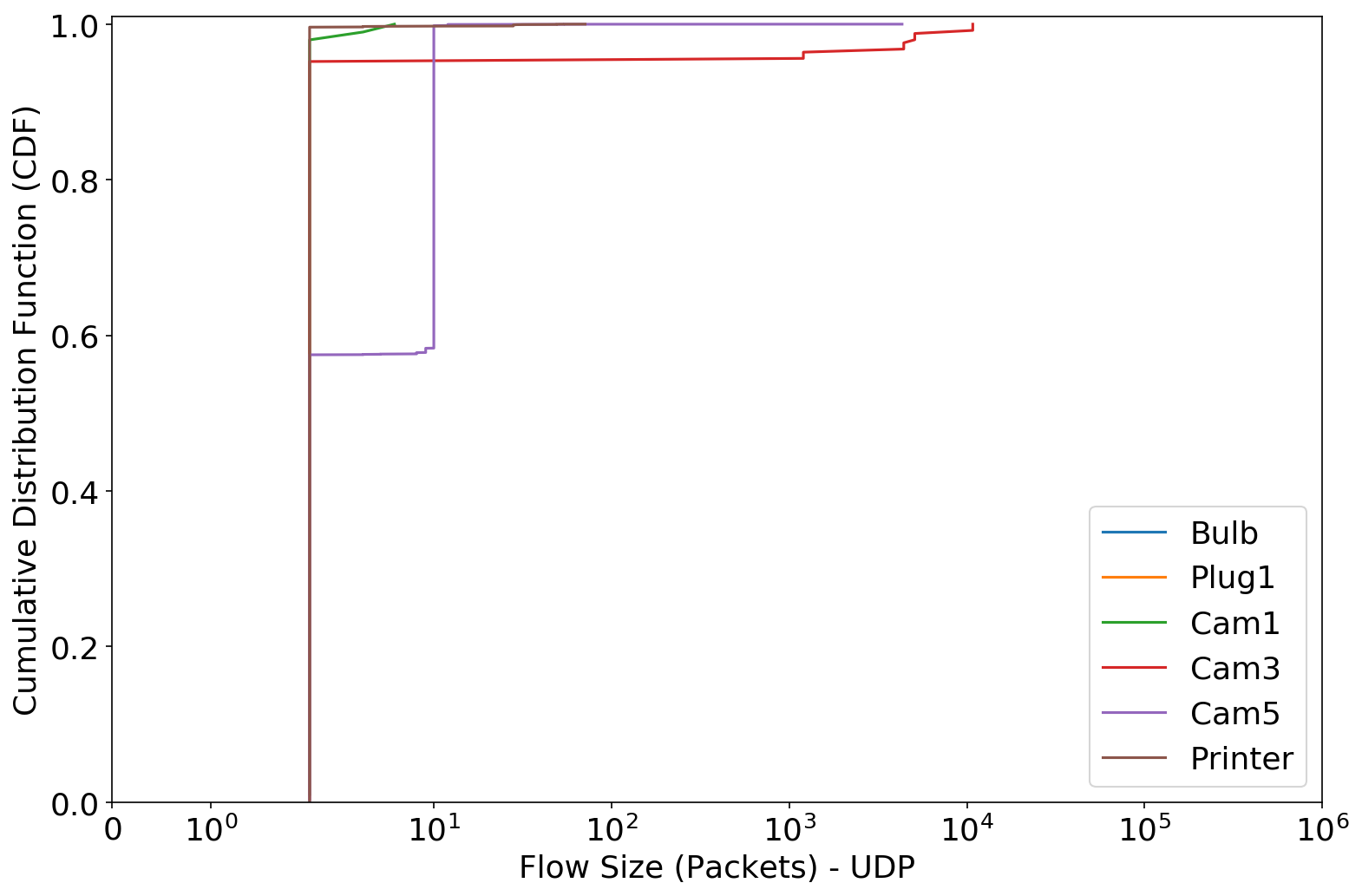}}
\hfill
\subcaptionbox{Non-IoT \label{fig:flowsize-pkts-udp-noniot}}[0.48\linewidth]
    {\includegraphics[width=0.24\textwidth]{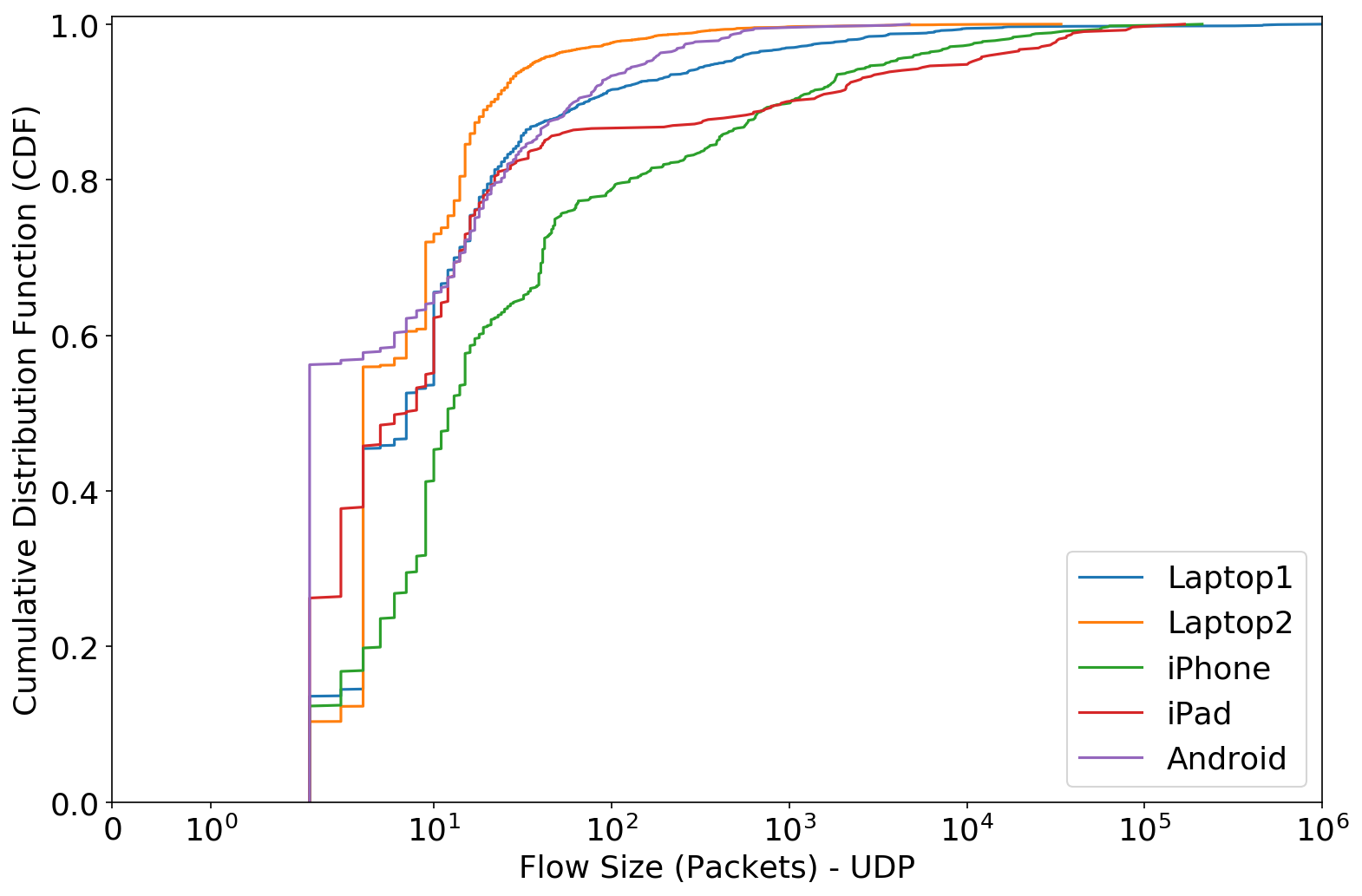}}
 \caption{UDP flow size (packets).}
\label{fig:flow-packets-udp}
\end{figure}

Figures~\ref{fig:flowsize-pkts-udp-iot} and~\ref{fig:flowsize-pkts-udp-noniot} show the UDP flow sizes in terms of packets for IoT and non-IoT devices, respectively. As we can see in the figures, IoT and non-IoT devices differ greater in UDP flow sizes than in TCP flow sizes. For example, above $90\%$ of UDP flows of most IoT devices generate $20$ packets; on the other hand, only less than $30\%$ of UDP flows of most non-IoT devices generate that amount of packets. In general, UDP flows of non-IoT devices generate greater amount of packets than IoT devices. This can be similarly understood based on different usages of UDP traffic in IoT and non-IoT devices, in particular, many IoT devices use NTP (on top of UDP) to synchronize their clocks.

\subsubsection{Flow Rate}
\begin{figure}[!h]
\begin{subfigure}[b]{0.24\textwidth}
    \includegraphics[width=\textwidth]{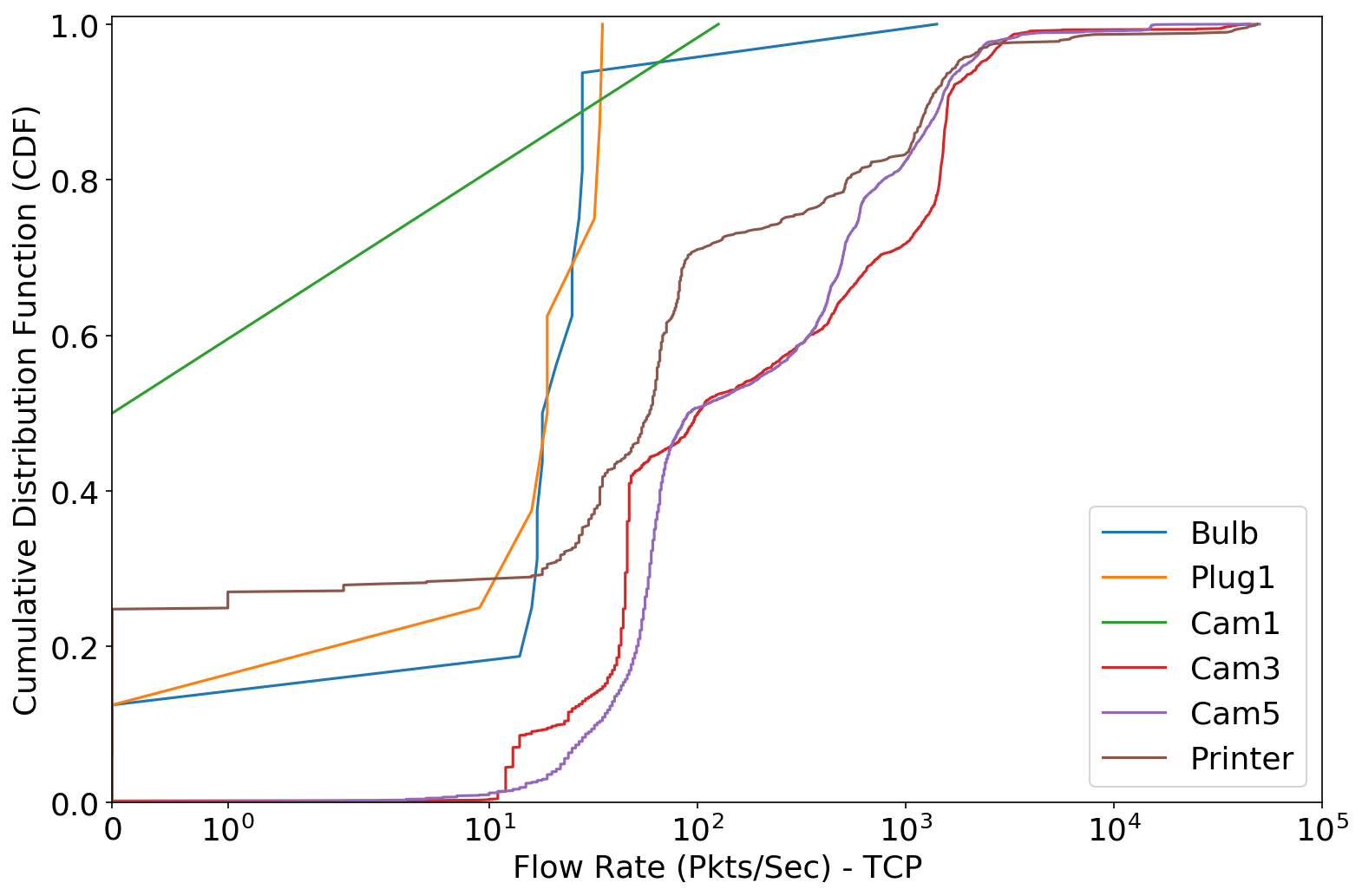}
    \caption{IoT devices}
    \label{fig:flow-rate-pkts-tcp-iot}
\end{subfigure}
\hfill
\begin{subfigure}[b]{0.24\textwidth}
    \includegraphics[width=\textwidth]{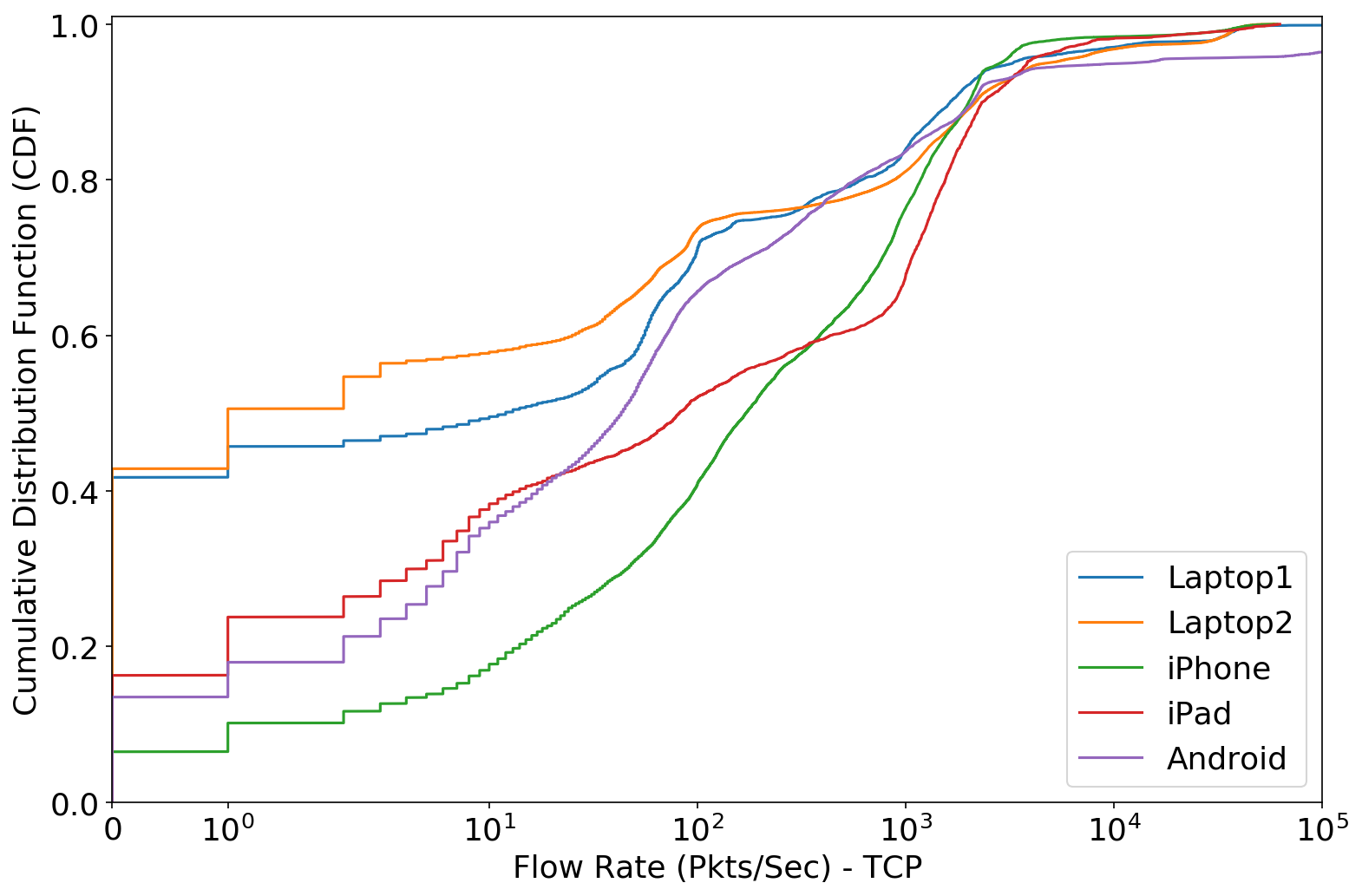}
    \caption{Non-IoT devices}
    \label{fig:flow-rate-pkts-tcp-noniot}
\end{subfigure}
\caption{TCP flow rate (packets/second).}
\label{fig:flow-rate-pkts-tcp}
\end{figure}

Figures~\ref{fig:flow-rate-pkts-tcp-iot} and~\ref{fig:flow-rate-pkts-tcp-noniot} show the TCP flow rates (in packets per second) of IoT and non-IoT devices, respectively. From the figures we can see that, the majority of TCP flows of IoT devices have a flow rate between $10$ and $3,000$ packets/second. In contrast, TCP flow rates of non-IoT devices are more spread out, with majority of them ranging from $1$ to $10,000$ packets/second. We also note that both IoT and non-IoT devices have a large portion of TCP flows with a flow rate close to zero.

\begin{figure}[!h]
\begin{subfigure}[b]{0.24\textwidth}
    \includegraphics[width=\textwidth]{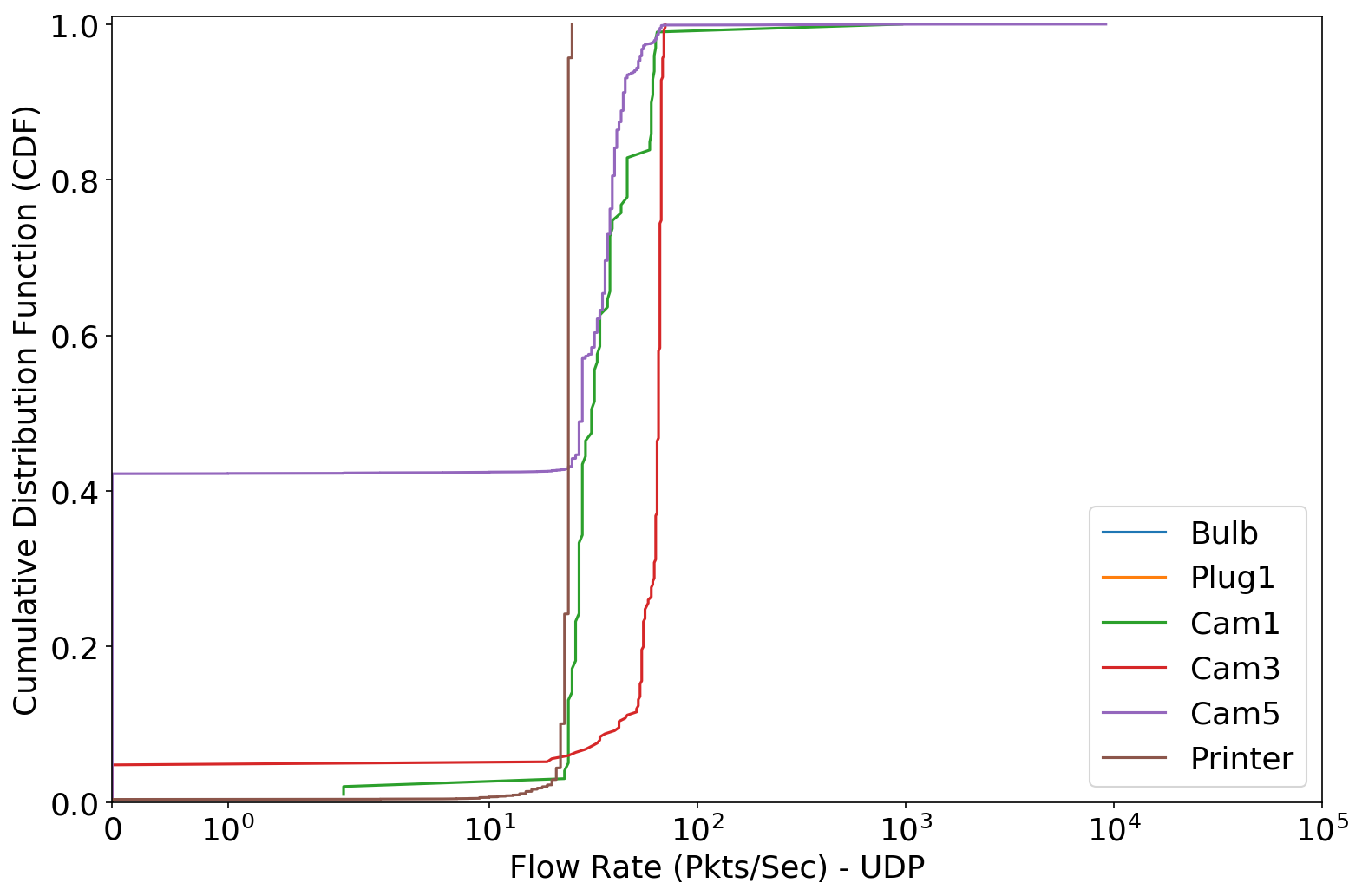}
    \caption{IoT devices}
    \label{fig:flow-rate-pkts-udp-iot}
\end{subfigure}
\hfill
\begin{subfigure}[b]{0.24\textwidth}
    \includegraphics[width=\textwidth]{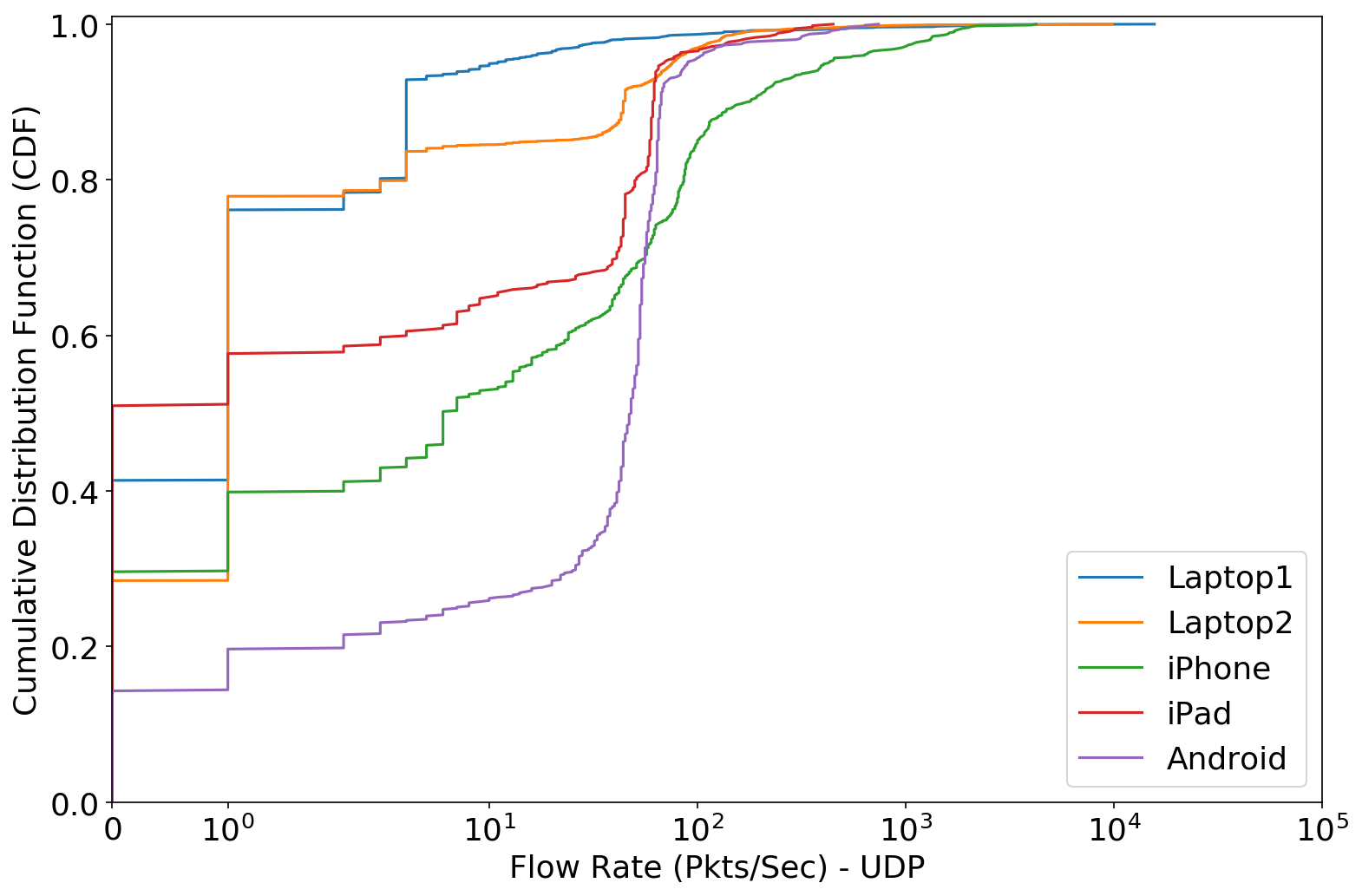}
    \caption{Non-IoT devices}
    \label{fig:flow-rate-pkts-udp-noniot}
\end{subfigure}
\caption{UDP flow rates (packets/second).}
\label{fig:flow-rate-pkts-udp}
\end{figure}

We can make some similar observations on UDP flow rates of both IoT and non-IoT devices, as shown in Figures~\ref{fig:flow-rate-pkts-udp-iot} and~\ref{fig:flow-rate-pkts-udp-noniot}. In particular, majority of UDP flows of IoT devices have a concentrated flow rate range between $20$ and $90$ packets/second. On the other hand, the UDP flows of non-IoT devices have a wider range of flow rates, ranging from $1$ to $1,000$ packets/second. Non-IoT devices have a large portion of UDP flows with a flow rate close to zero. The majority of IoT devices have a low percentage of flows with a flow rate close to zero, with an exception of Cam5, which has above $40\%$ of UDP flows with a close to zero flow rate.

\begin{figure}[!h]
\begin{subfigure}[b]{0.24\textwidth}
    \includegraphics[width=\textwidth]{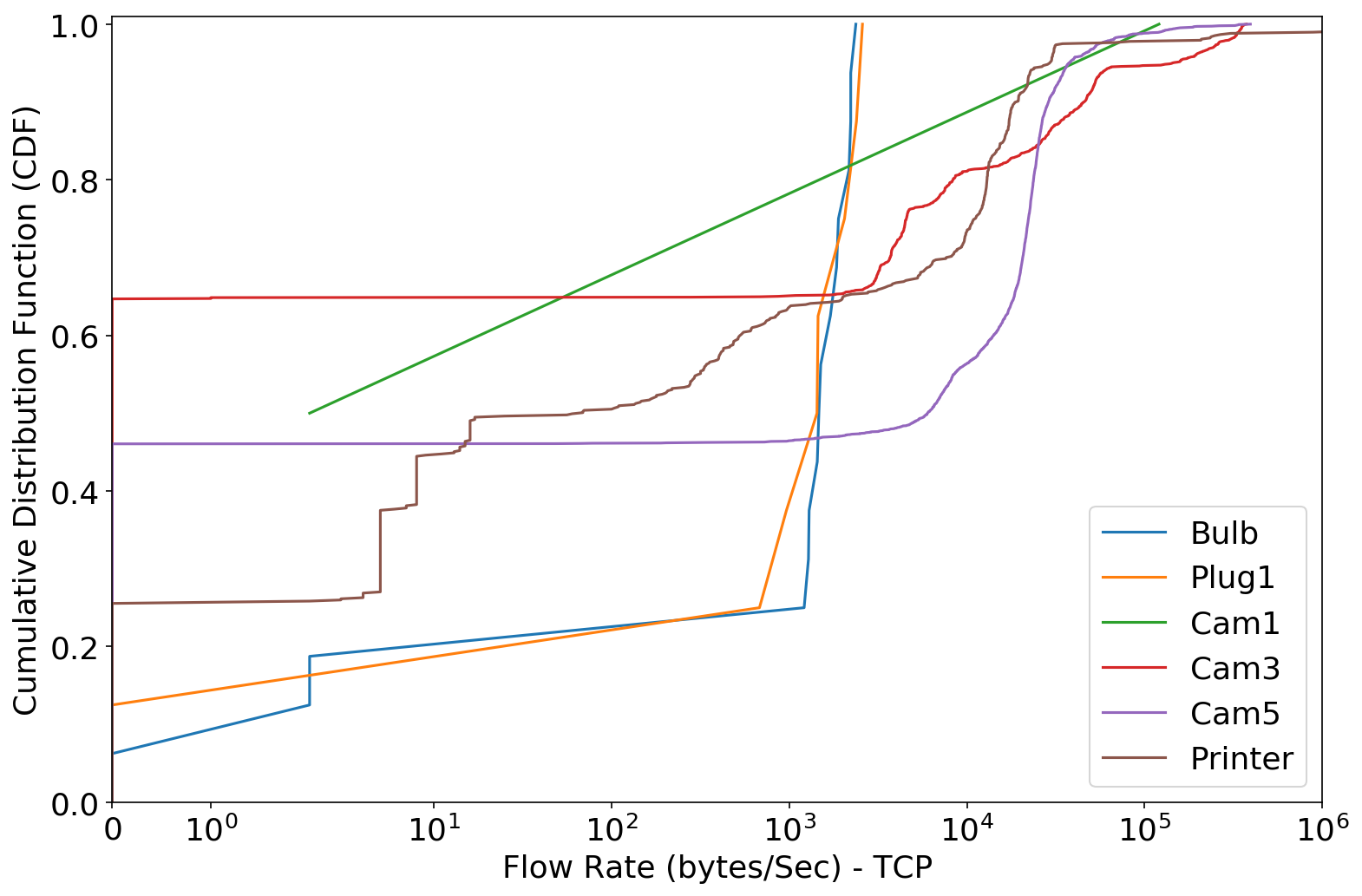}
    \caption{IoT devices}
    \label{fig:flow-rate-bytes-tcp-iot}
\end{subfigure}
\hfill
\begin{subfigure}[b]{0.24\textwidth}
    \includegraphics[width=\textwidth]{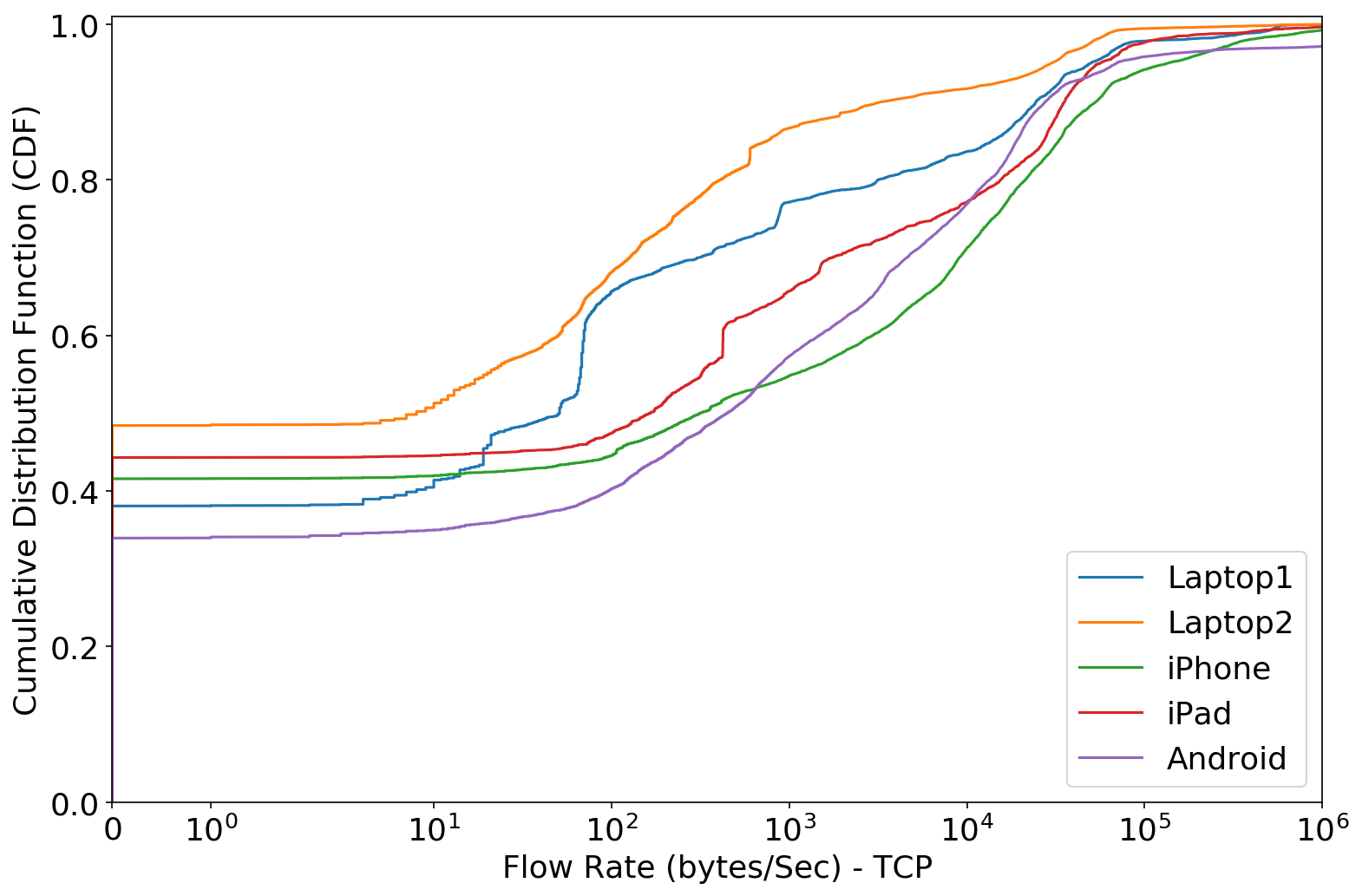}
    \caption{Non-IoT devices}
    \label{fig:flow-rate-bytes-tcp-noniot}
\end{subfigure}
\caption{TCP flow rates (bytes/second).}
\label{fig:flow-rate-bytes-tcp}
\end{figure}

Figures~\ref{fig:flow-rate-bytes-tcp-iot} and~\ref{fig:flow-rate-bytes-tcp-noniot} show the TCP flow rates (in bytes/second) for IoT and non-IoT devices, respectively. From the figures we can again observe that non-IoT devices have a similar behavior in terms of UDP flow rates; their CDFs show similar shapes. In contrast, IoT devices have more diverse TCP flow rates in terms of bytes/second.

\begin{figure}[!h]
\begin{subfigure}[b]{0.24\textwidth}
    \includegraphics[width=\textwidth]{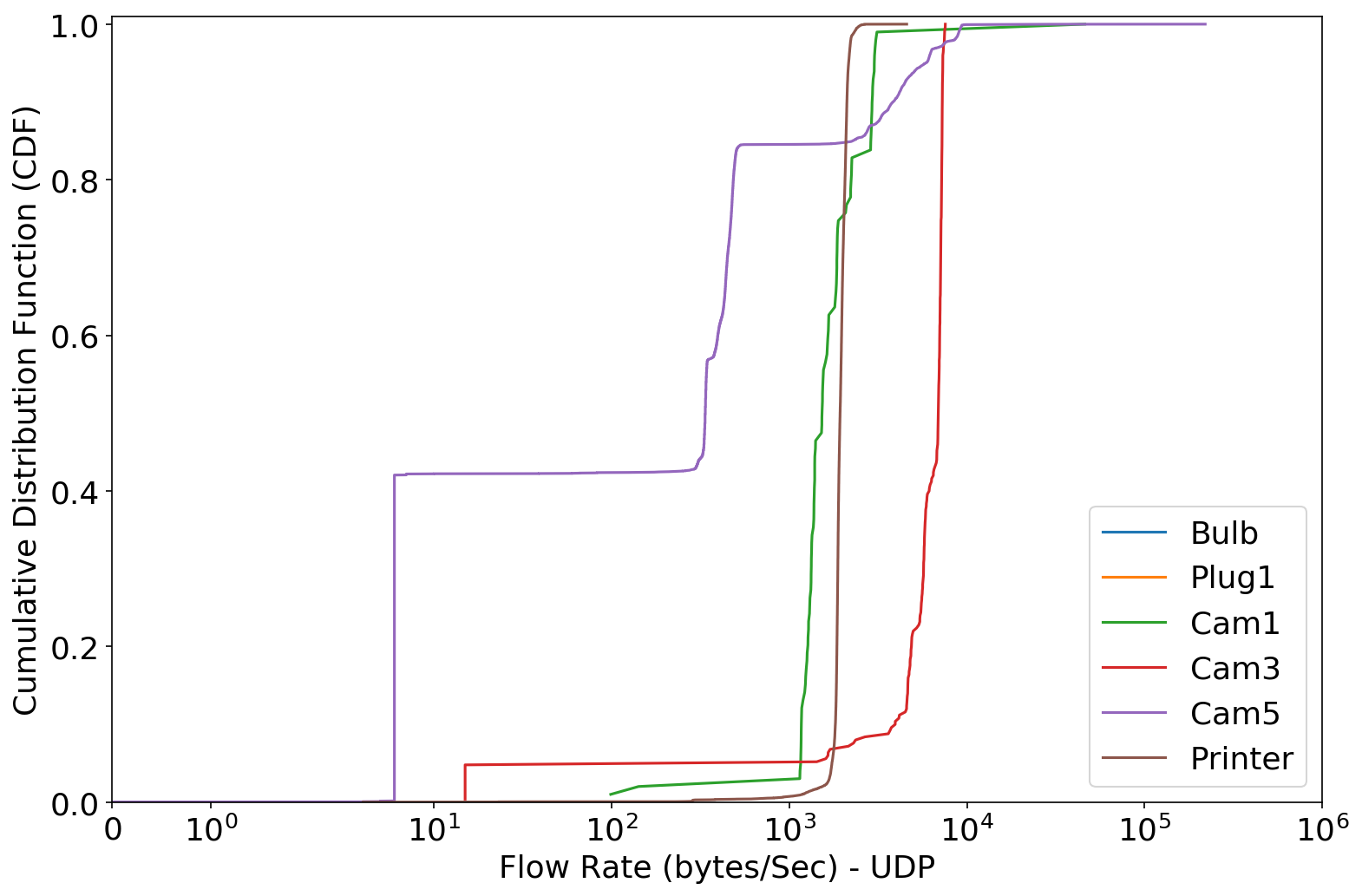}
    \caption{IoT devices}
    \label{fig:flow-rate-bytes-udp-iot}
\end{subfigure}
\hfill
\begin{subfigure}[b]{0.24\textwidth}
    \includegraphics[width=\textwidth]{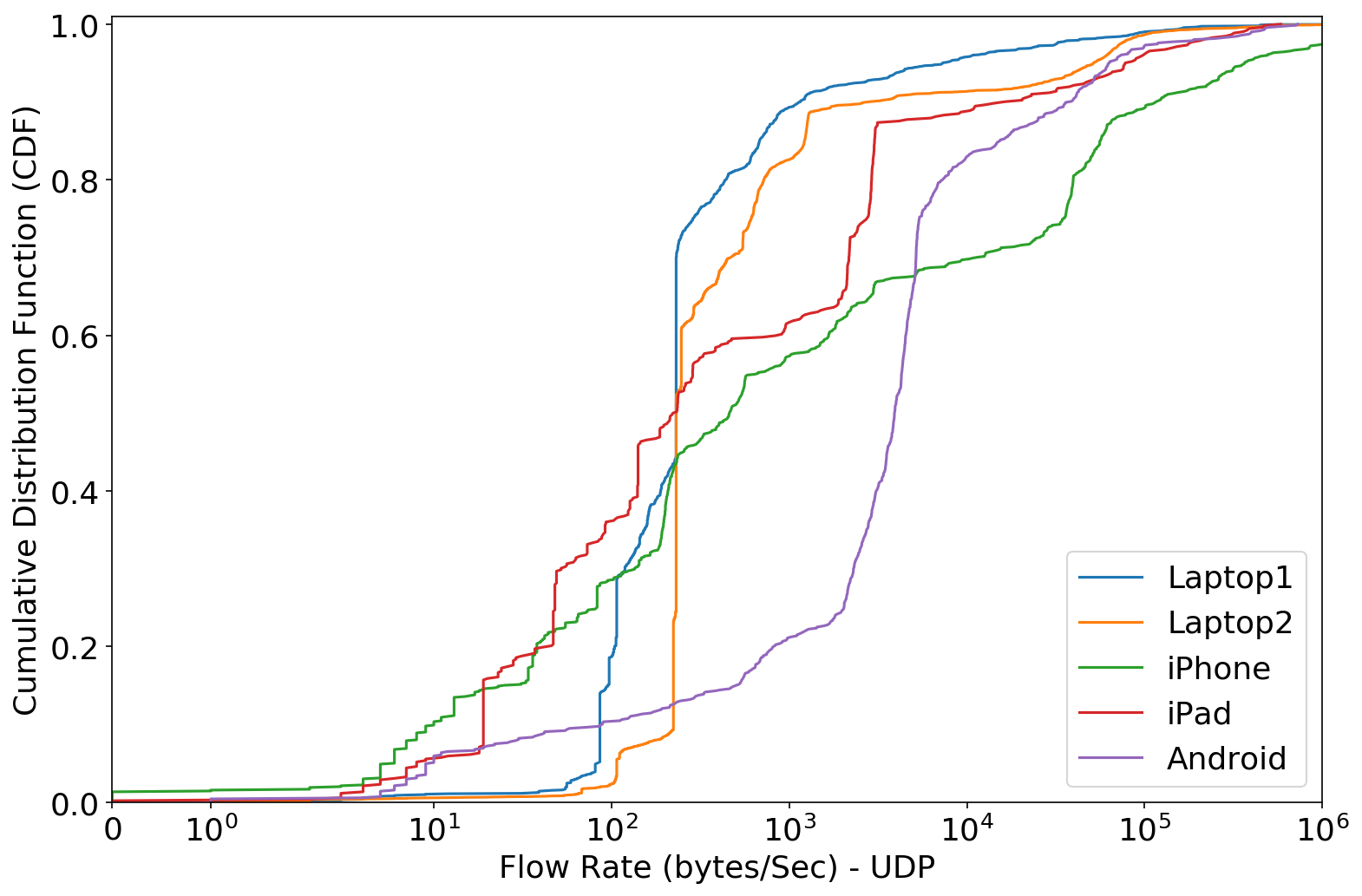}
    \caption{Non-IoT devices}
    \label{fig:flow-rate-bytes-udp-noniot}
\end{subfigure}
\caption{UDP flow rates (bytes/second).}
\label{fig:flow-rate-bytes-udp}
\end{figure}

Figures~\ref{fig:flow-rate-bytes-udp-iot} and~\ref{fig:flow-rate-bytes-udp-noniot} show the UDP flow rates (in bytes/second) for IoT and non-IoT devices, respectively. From the figure we can see that the majority of UDP flows of most of the IoT devices have flow rates between $1,000$ and $10,000$ bytes/second. Cam5 is an exception, with a few smaller UDP flow rate ranges. The UDP flows of non-IoT devices have more diverse flow rates, ranging from $10$ to $10^6$ bytes/second.

\subsubsection{Number of Flows}
In this subsection we study the number of flows in each time interval of $120$ seconds. We note that, a flow may be counted multiple times if it spans multiple time intervals.


\begin{figure}[th]
\centering
\subcaptionbox{IoT devices \label{fig:active-flows-tcp-iot}}[0.48\linewidth]
    {\includegraphics[width=0.24\textwidth]{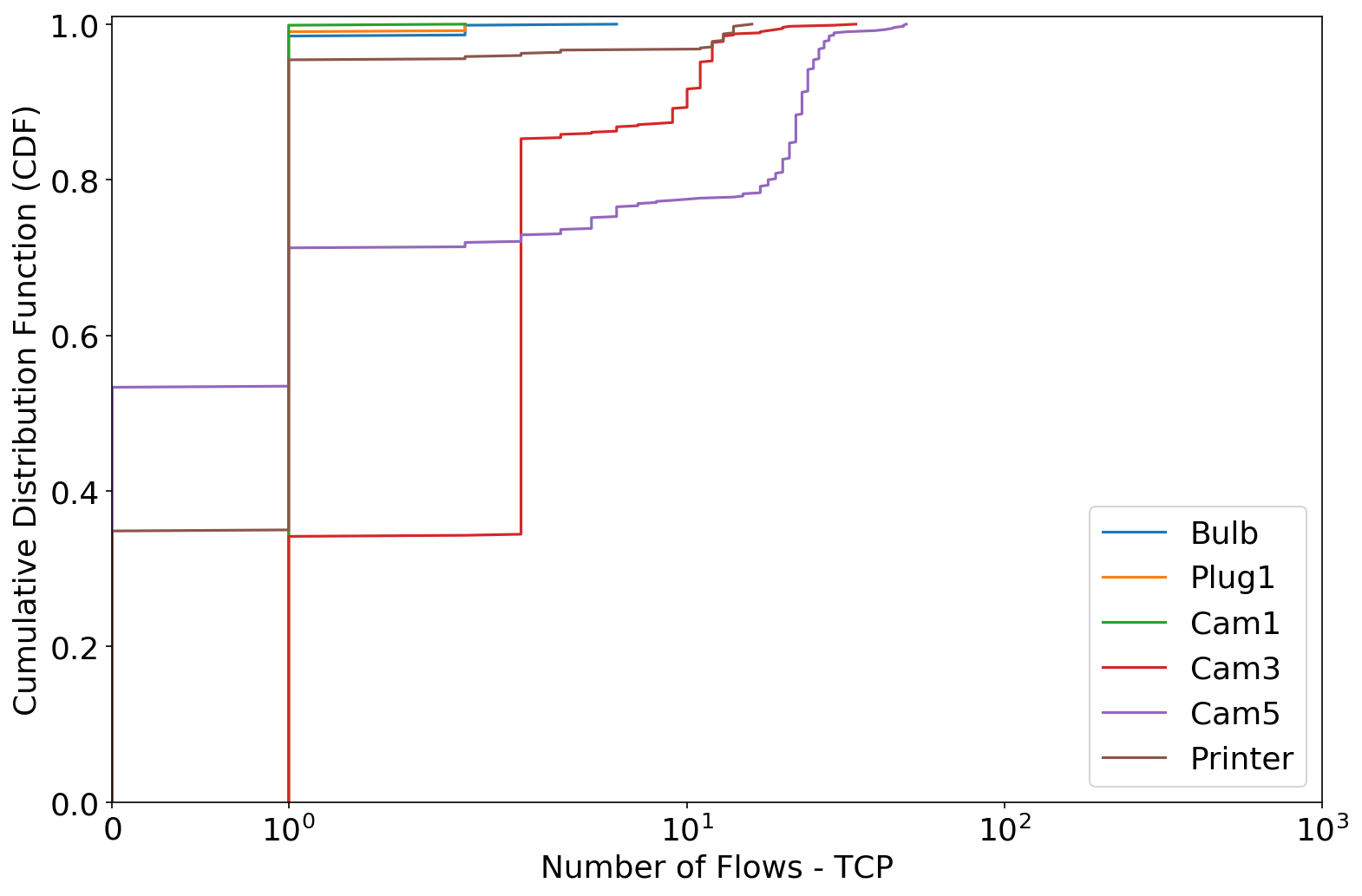}}
\hfill
\subcaptionbox{Non-IoT devices \label{fig:active-flows-tcp-noniot}}[0.48\linewidth]
    {\includegraphics[width=0.24\textwidth]{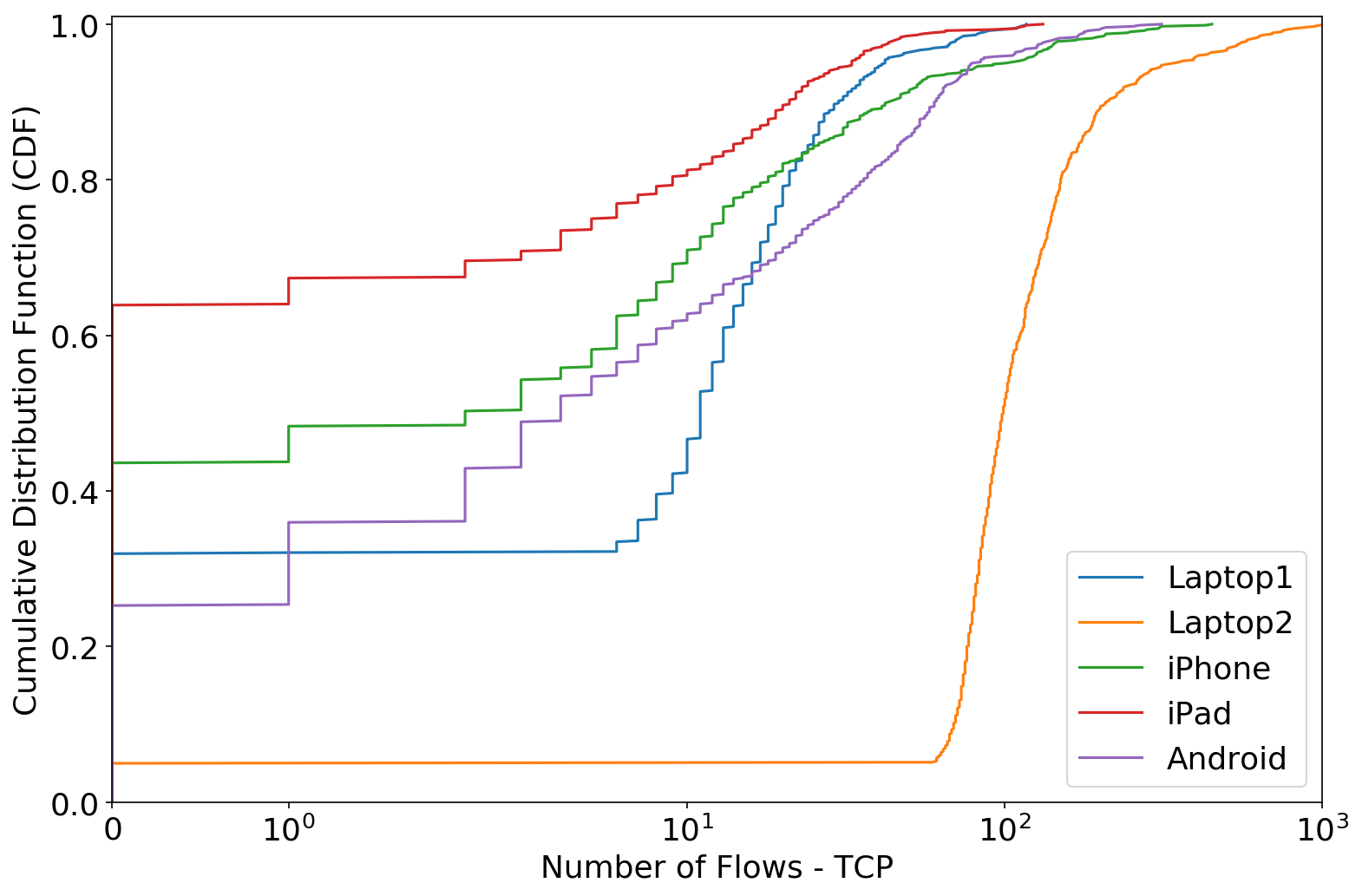}}
 \caption{Number of TCP flows.}
\label{fig:active-flows-tcp}
\end{figure}

Figures~\ref{fig:active-flows-tcp-iot} and~\ref{fig:active-flows-tcp-noniot} show the distribution of number of TCP flows in each time interval for IoT and non-IoT devices, respectively. From the figures we can see that, for all the IoT devices, a large portion of time intervals have no TCP flows at all. For example, Printer has no TCP flows in about $33\%$ of time intervals, and Plug1 has only one TCP flow in about $99\%$ of time intervals. Moreover, no time interval has more than $50$ TCP flows among all these IoT devices. In contrast, although non-IoT devices have a large portion of time intervals without any TCP flows, in general, the percentage of such time intervals is lower. In addition, the number of TCP flows in the time intervals is more diverse, ranging from $0$ to $1,039$ TCP flows.

\begin{figure}[th]
\centering
\subcaptionbox{IoT devices \label{fig:active-flows-udp-iot}}[0.48\linewidth]
    {\includegraphics[width=0.24\textwidth]{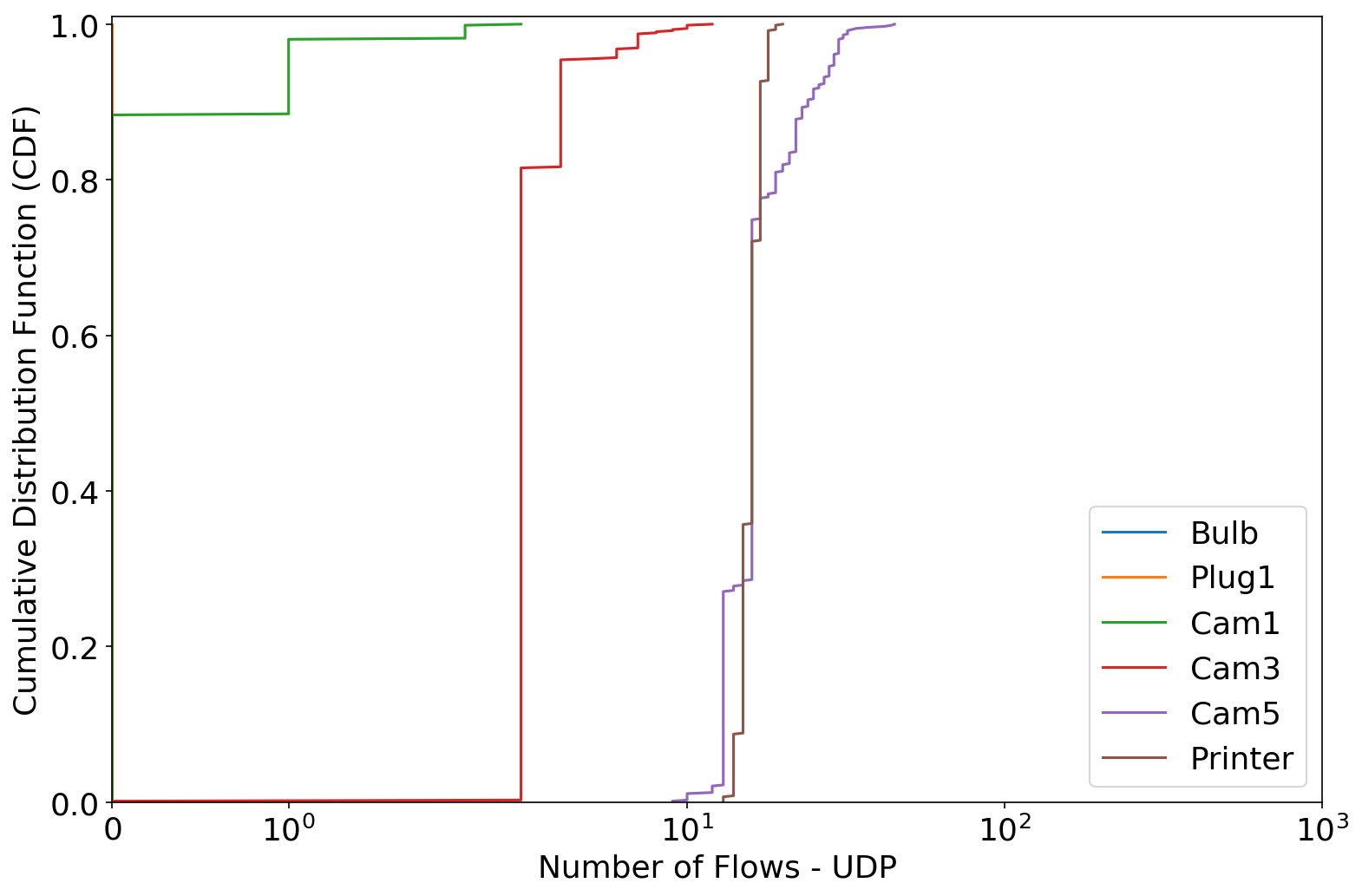}}
\hfill
\subcaptionbox{Non-IoT devices \label{fig:active-flows-udp-noniot}}[0.48\linewidth]
    {\includegraphics[width=0.24\textwidth]{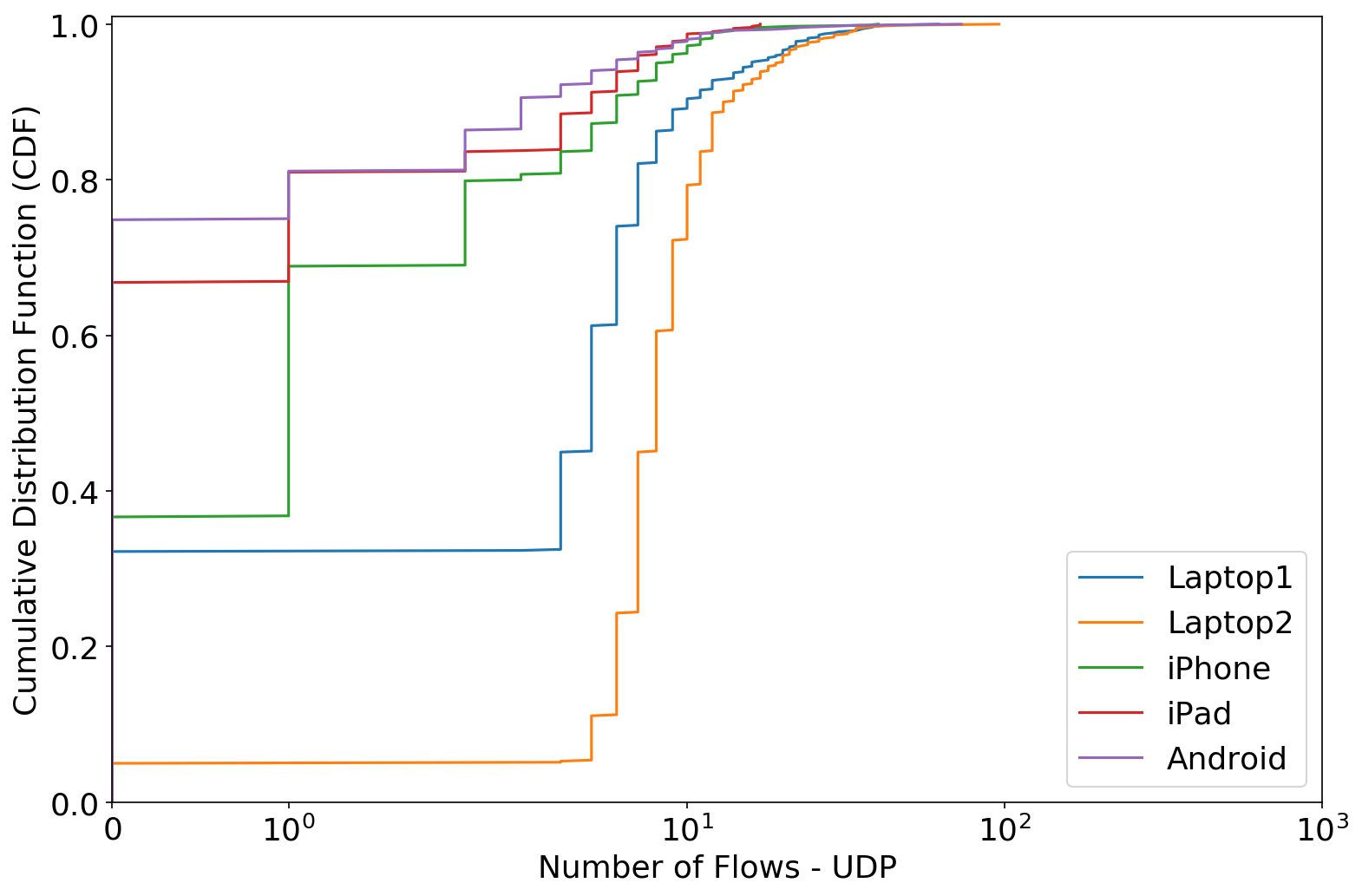}}
 \caption{Number of UDP flows.}
\label{fig:active-flows-udp}
\end{figure}

IoT and non-IoT devices also behave quite differently in terms of the number of UDP flows, as shown in Figures~\ref{fig:active-flows-udp-iot} and~\ref{fig:active-flows-udp-noniot}, respectively. First we note that Cam1 and Cam3 have smaller number UDP flows, with close to $80\%$ and $90\%$ of time intervals without UDP flows at all. In addition, they do not have more than $45$ UDP flows in any of time intervals. Cam5 and Printer have relatively larger number of UDP flows, with majority of them having about $16$ UDP flows. Overall, they also do not have more than $80$ UDP flows in any of the time intervals. Again, non-IoT devices have more diverse number of UDP flows, ranging from $0$ to about $100$ UDP flows.







\subsection{Packet Level Characteristics}\label{sec:packet}
In this subsection we study the packet-level characteristics of IoT devices.

\subsubsection{IATs of Packets in a Flow}
First we study the inter-arrival times (IATs) of outgoing packets of both TCP and UDP flows of IoT devices and contrast them with those of non-IoT devices. Note that although flows are bidirectional, in this study, we only investigate the IATs of consecutive outgoing packets of a flow.

\begin{figure}[th]
\centering
\subcaptionbox{IoT devices \label{fig:outgoing-iat-tcp-iot}}[0.48\linewidth]
    {\includegraphics[width=0.24\textwidth]{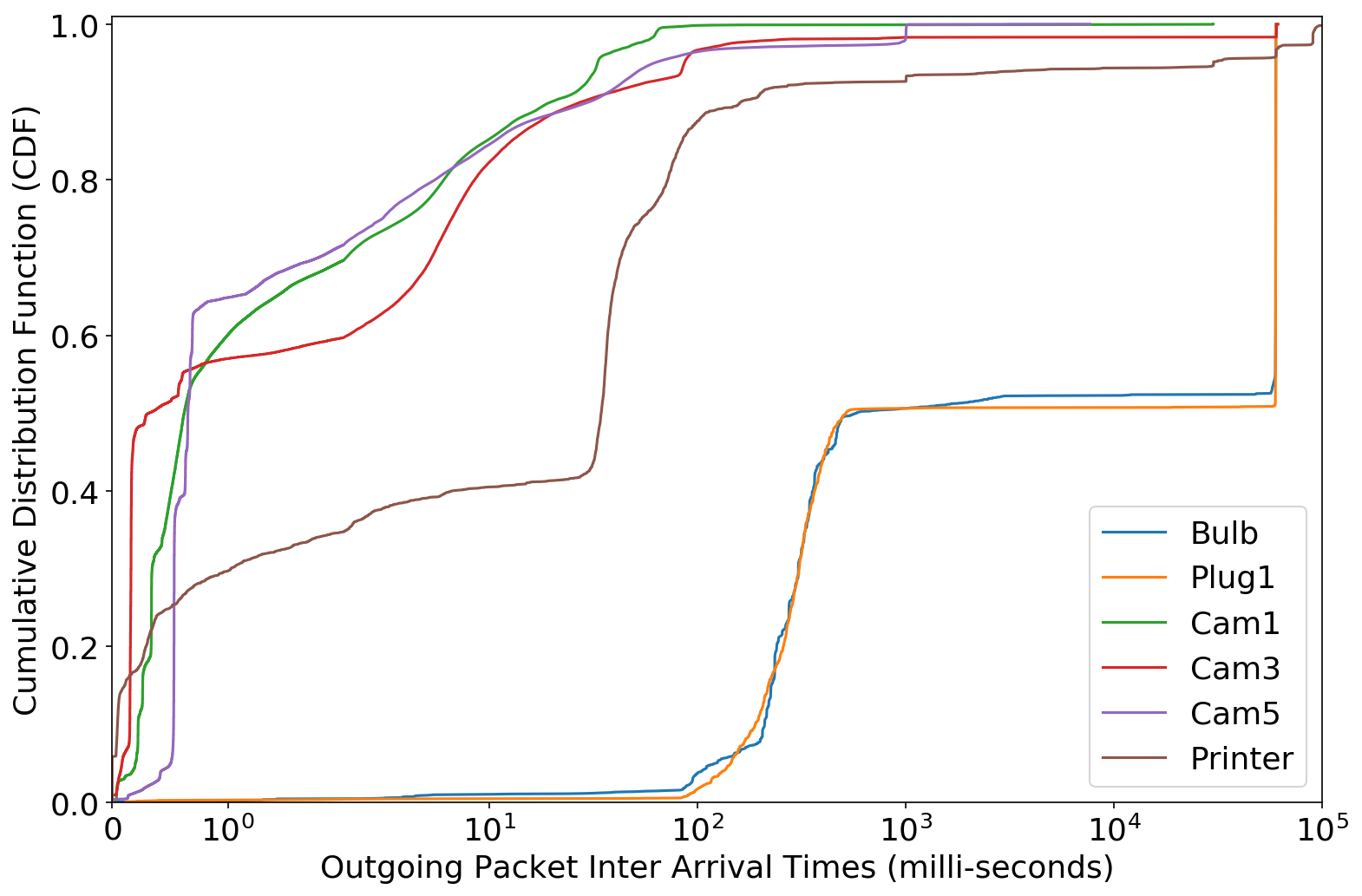}}
\hfill
\subcaptionbox{Non-IOT devices \label{fig:outgoing-iat-tcp-noniot}}[0.48\linewidth]
    {\includegraphics[width=0.24\textwidth]{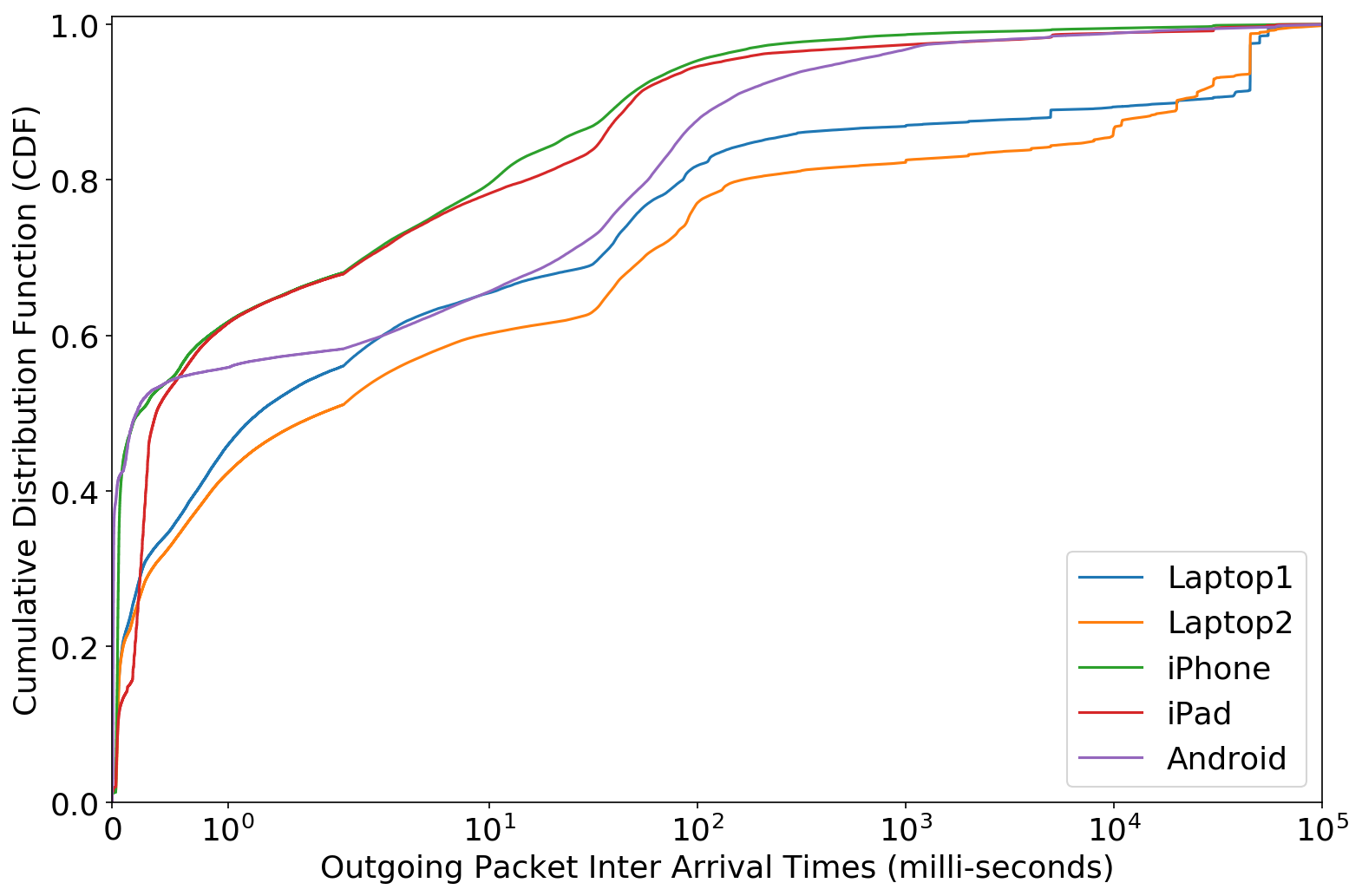}}
\caption{IATs of outgoing packets of TCP flows.}
\label{fig:outgoing-iat-tcp}
\end{figure}

Figures~\ref{fig:outgoing-iat-tcp-iot} and~\ref{fig:outgoing-iat-tcp-noniot} show the IATs of outgoing packets of TCP flows of IoT and non-IoT devices, respectively. From the figures we first note that the longest IATs of both IoT and non-IoT devices are close to $120$ seconds. This could be caused by the timeout threshold that we have used in extracting flows. We point out that, theoretically speaking, an IAT longer than $120$ seconds is possible, given that the flow timeout threshold is applied on bidirectional packets (instead of one way traffic). We also note that both IoT and non-IoT devices have a large portion of small IAT values. For example, all the security cameras (IoT devices) and phones and ipad (non-IoT devices) have around $60\%$ of IATs smaller than $1$ms. Two notable exceptions are Bulb and Plug1; they have in general larger IATs. We also note that, in general, IoT devices behave more differently from each other than non-IoT devices in terms of IATs of outgoing packets of TCP flows.

\begin{figure}[th]
\centering
\subcaptionbox{IoT devices \label{fig:outgoing-iat-udp-iot}}[0.48\linewidth]
    {\includegraphics[width=0.24\textwidth]{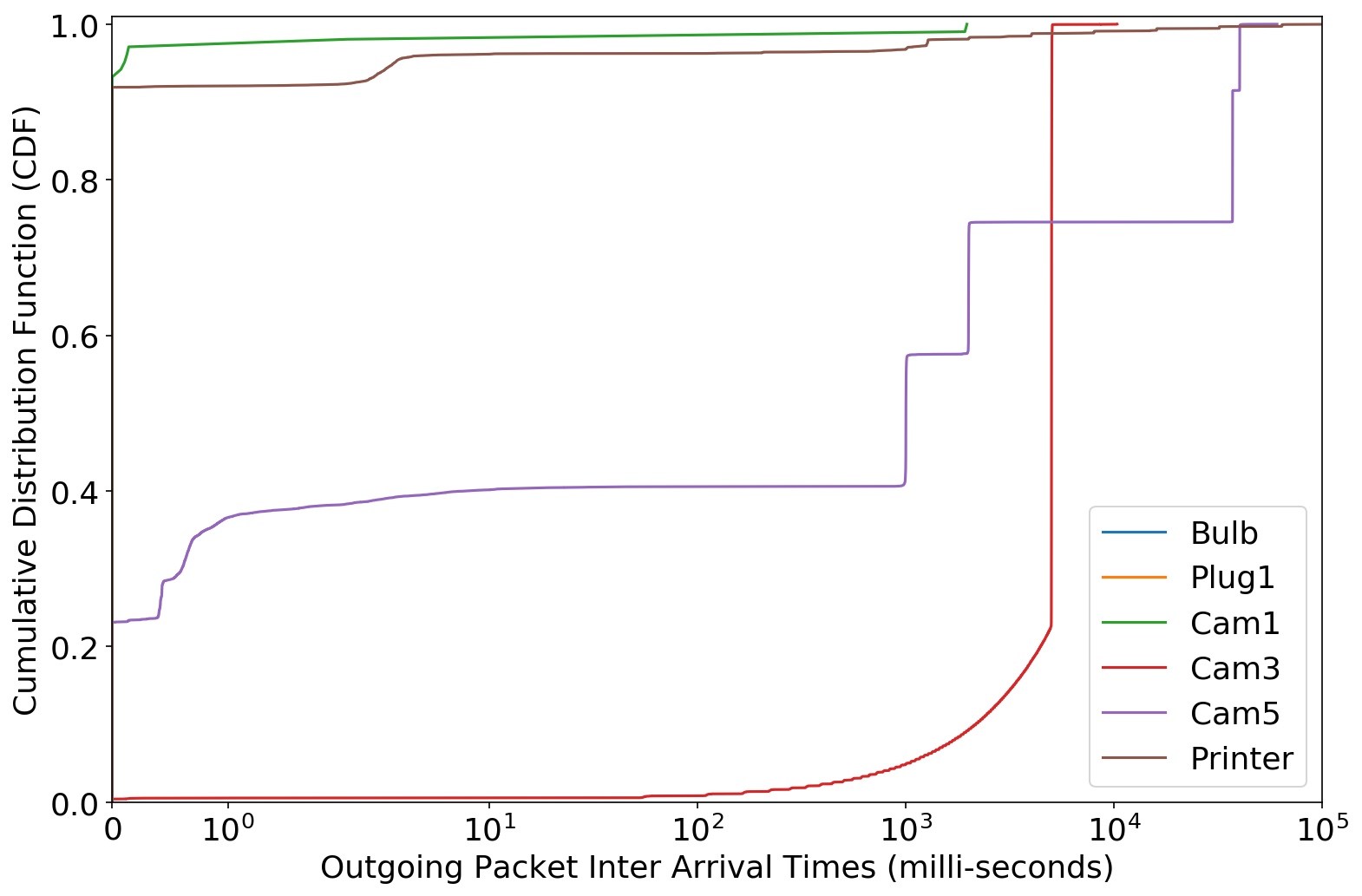}}
\hfill
\subcaptionbox{non-IOT devices \label{fig:outgoing-iat-udp-noniot}}[0.48\linewidth]
    {\includegraphics[width=0.24\textwidth]{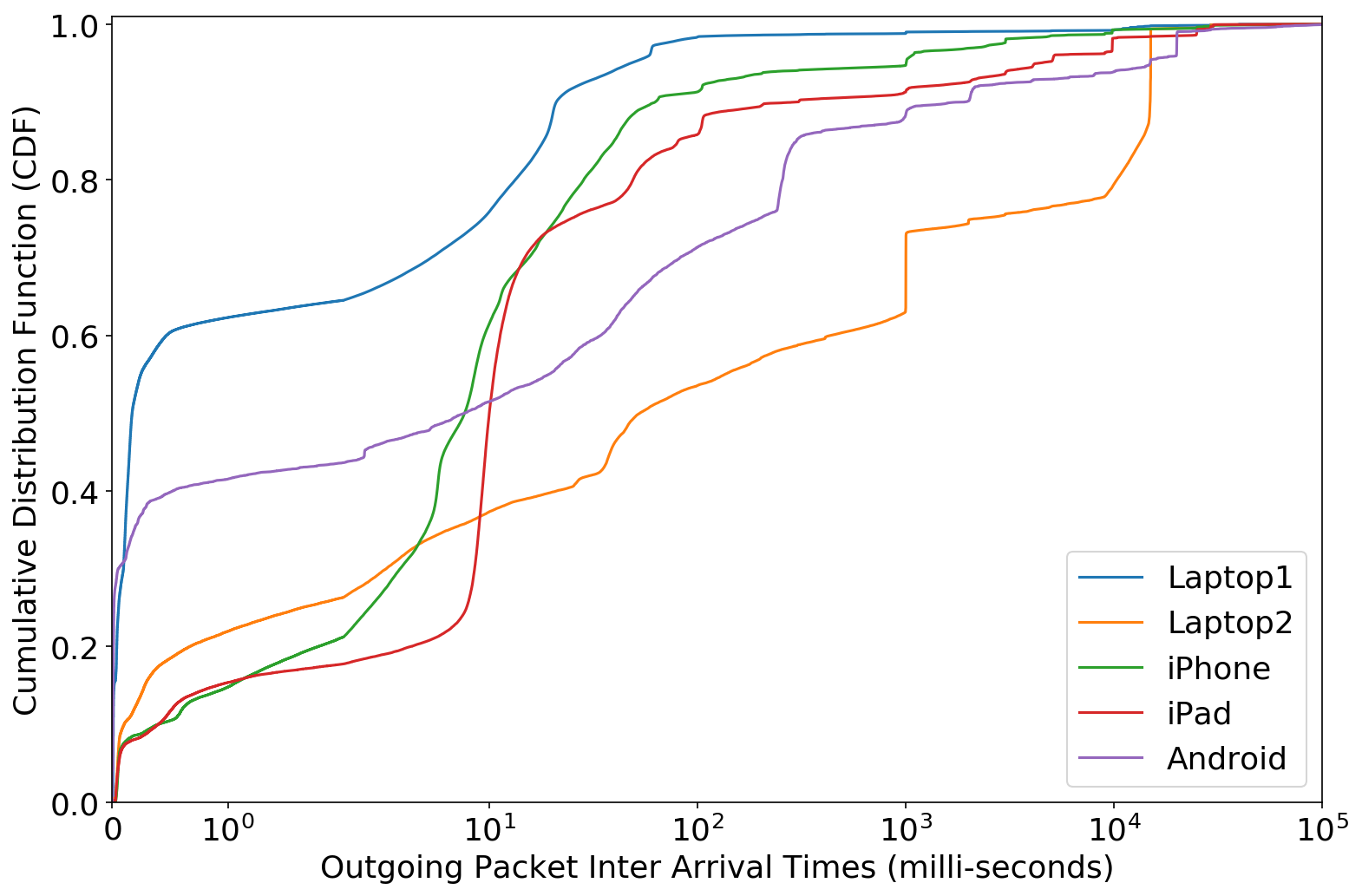}}
\caption{IATs of outgoing packets of UDP flows.}
\label{fig:outgoing-iat-udp}
\end{figure}

Figures~\ref{fig:outgoing-iat-udp-iot} and~\ref{fig:outgoing-iat-udp-noniot} show the IATs of outoging packets of UDP flows. From the figures we can similarly observe that, non-IoT devices behave closer to each other than IoT devices. Moreover, while IATs of IoT devices tend to concentrate at a few small ranges, the IATs of non-IoT tend to be more spread out.


\subsubsection{Hourly Number of Bytes}
In this subsection we study the hourly amount of all TCP and UDP traffic in terms of number of bytes. We note that the amount of traffic only consider the payload of a TCP or UDP packet.
\begin{figure}[!h]
\begin{subfigure}[b]{0.24\textwidth}
    \includegraphics[width=\textwidth]{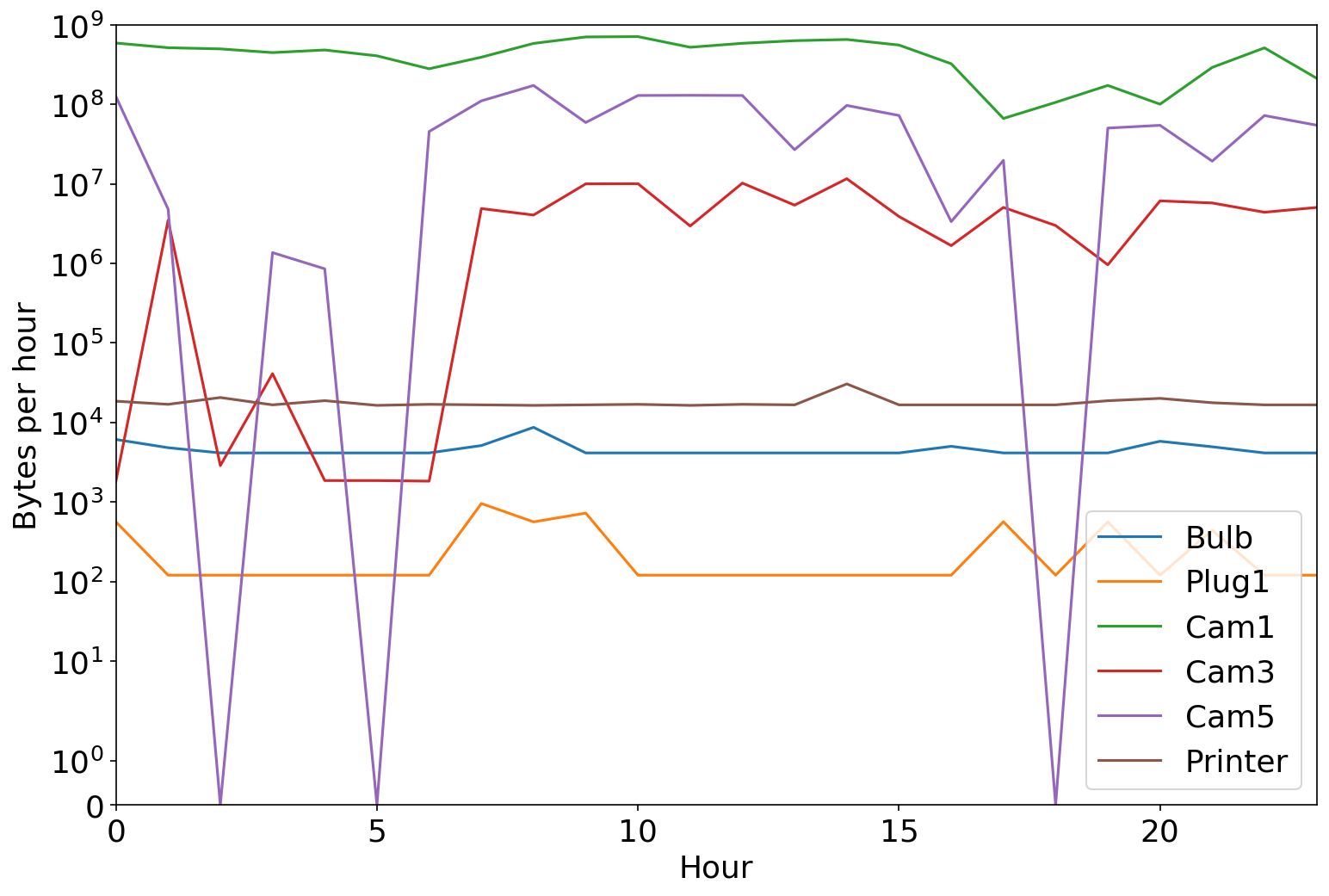}
    \caption{IoT devices}
    \label{fig:hourly-tcp-bytes-iot}
\end{subfigure}
\hfill
\begin{subfigure}[b]{0.24\textwidth}
    \includegraphics[width=\textwidth]{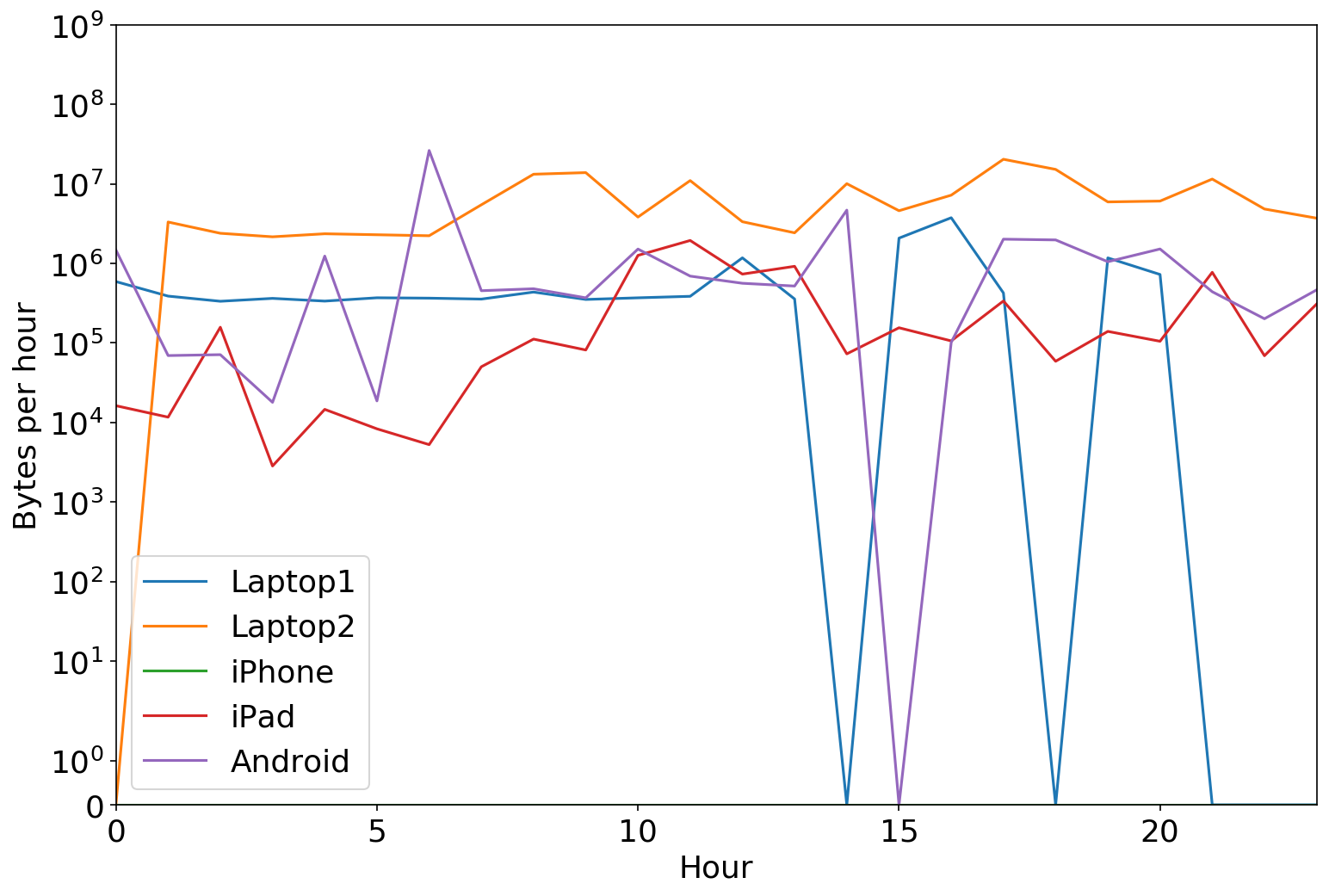}
    \caption{Non-IoT devices}
    \label{fig:hourly-tcp-bytes-noniot}
\end{subfigure}
\caption{Hourly number of bytes of TCP traffic.}
\label{fig:hourly-tcp-bytes}
\end{figure}

Figures~\ref{fig:hourly-tcp-bytes-iot} and~\ref{fig:hourly-tcp-bytes-noniot} show the hourly number of bytes of TCP traffic of IoT and non-IoT devices, respectively. From the figures we can see that, some of the IoT devices (Bulb, Plug1, and Printer) have very stable amount of hourly TCP traffic. On the other hand, the security cameras, in particular, Cam3 and Cam5, have more fluctuations in the amount of TCP traffic. Cam5 have a few hours without any TCP traffic, for the potential reason that we have discussed previously. In general. non-IoT devices generate TCP traffic with more fluctuations.

\begin{figure}[!h]
\begin{subfigure}[b]{0.24\textwidth}
    \includegraphics[width=\textwidth]{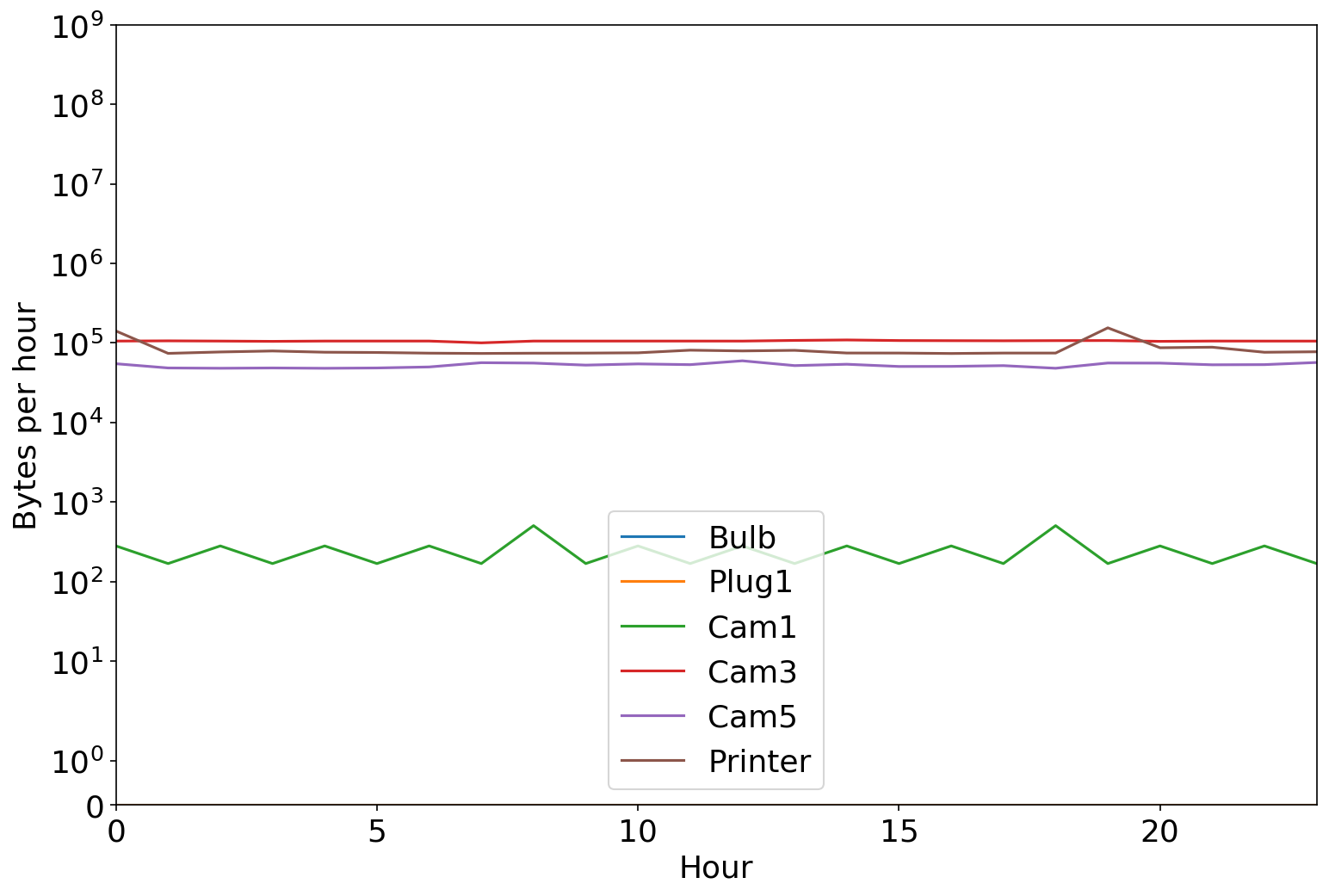}
    \caption{IoT}
    \label{fig:hourly-udp-bytes-iot}
\end{subfigure}
\hfill
\begin{subfigure}[b]{0.24\textwidth}
    \includegraphics[width=\textwidth]{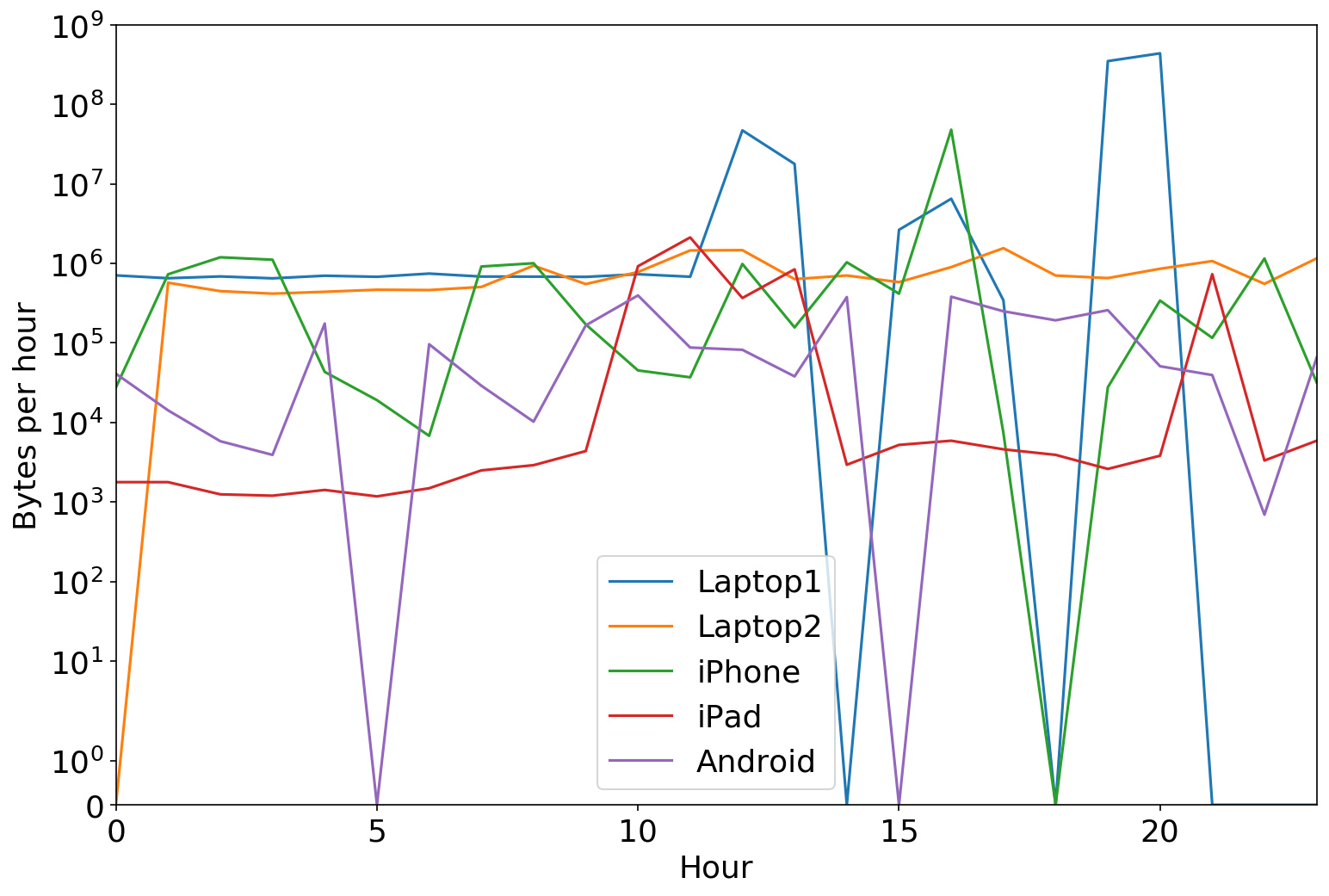}
    \caption{non-IoT}
    \label{fig:hourly-udp-bytes-noniot}
\end{subfigure}

\caption{Hourly number of bytes of UDP traffic.}
\label{fig:hourly-udp-bytes}
\end{figure}

The difference in terms of fluctuations in the number of bytes in UDP traffic is more notable between IoT and non-IoT devices (see Figures~\ref{fig:hourly-udp-bytes-iot} and~\ref{fig:hourly-udp-bytes-noniot}). While all the IoT devices generate relatively stable amount of UDP traffic, the amount of UDP traffic of non-IoT devices has more fluctuations.

\subsubsection{Hourly Number of Packets}
\begin{figure}[!h]
\begin{subfigure}[b]{0.24\textwidth}
    \includegraphics[width=\textwidth]{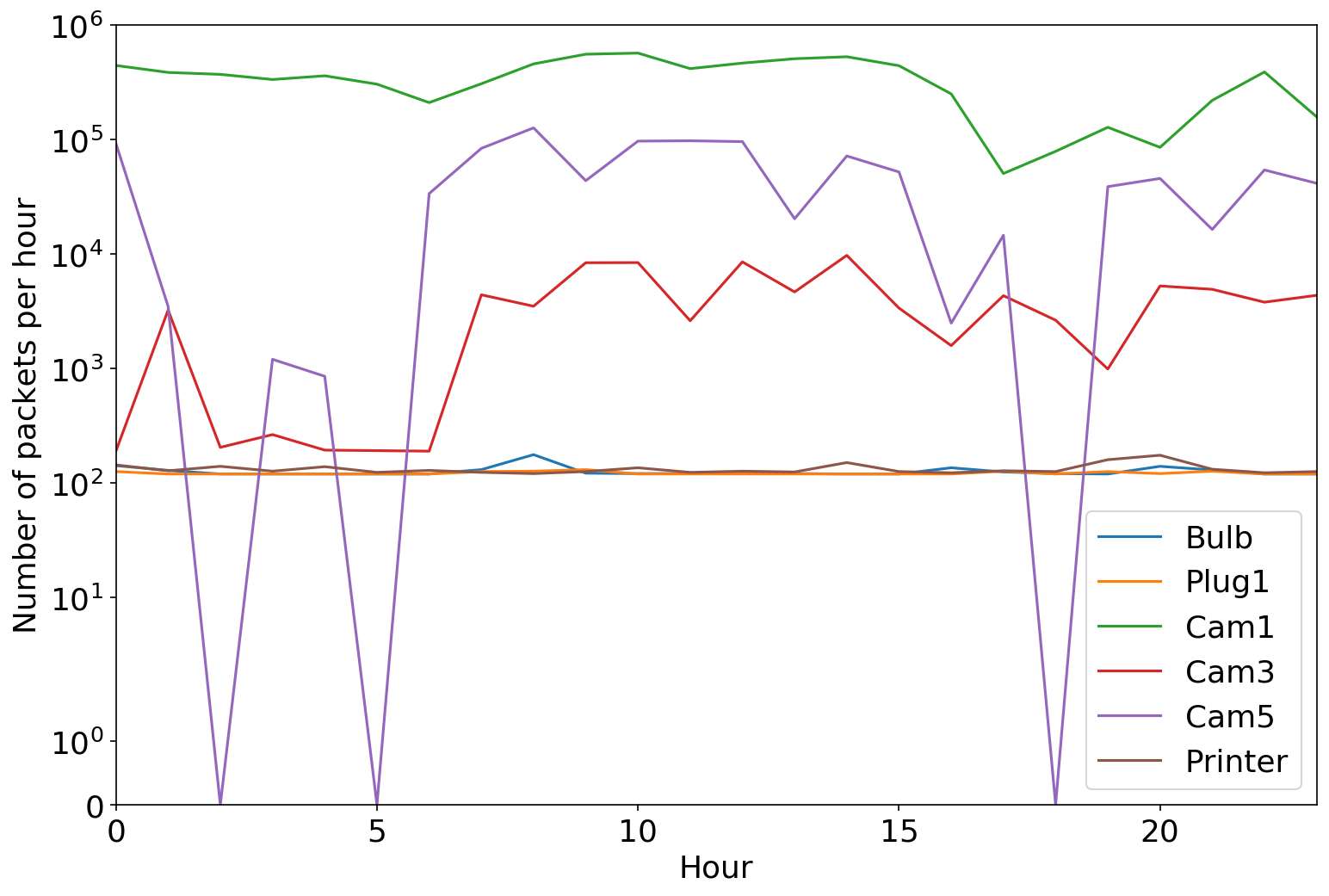}
    \caption{IoT devices}
    \label{fig:hourly-tcp-pkts-iot}
\end{subfigure}
\hfill
\begin{subfigure}[b]{0.24\textwidth}
    \includegraphics[width=\textwidth]{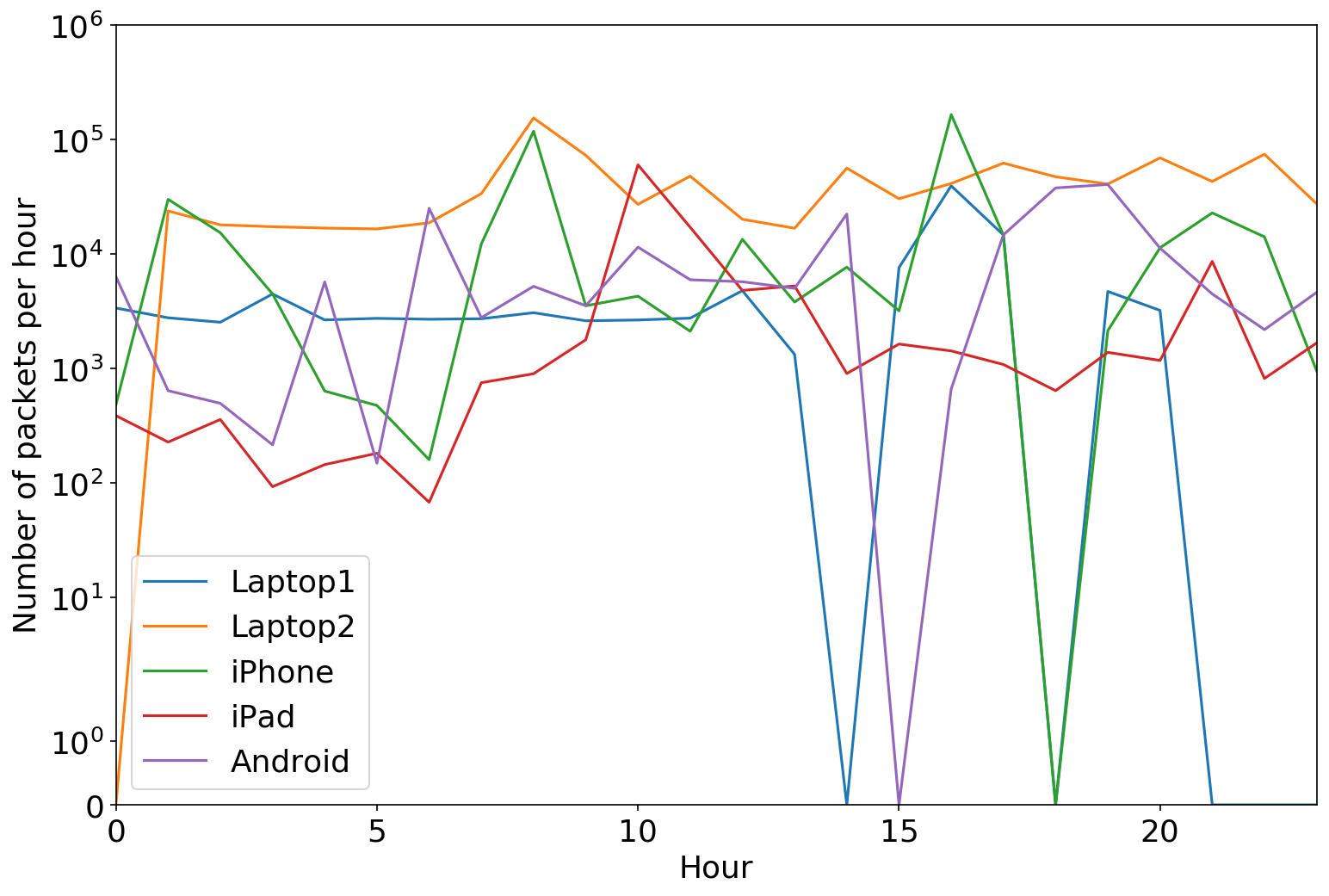}
    \caption{Non-IoT devices}
    \label{fig:hourly-tcp-pkts-noniot}
\end{subfigure}
\caption{Hourly number of packets of TCP traffic.}
\label{fig:hourly-tcp-pkts}
\end{figure}

\begin{figure}[!h]
\begin{subfigure}[b]{0.24\textwidth}
    \includegraphics[width=\textwidth]{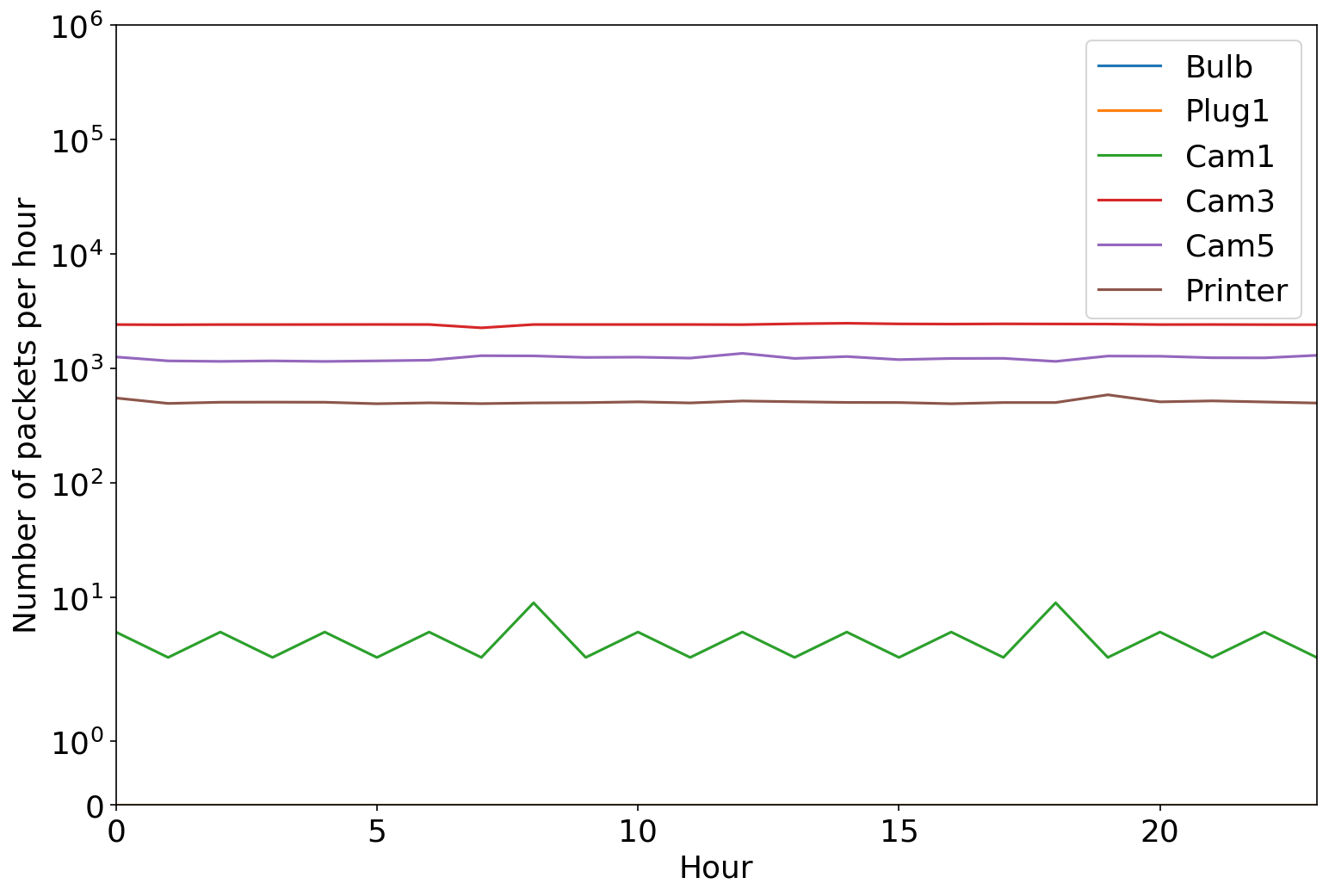}
    \caption{IoT devices}
    \label{fig:hourly-udp-pkts-iot}
\end{subfigure}
\hfill
\begin{subfigure}[b]{0.24\textwidth}
    \includegraphics[width=\textwidth]{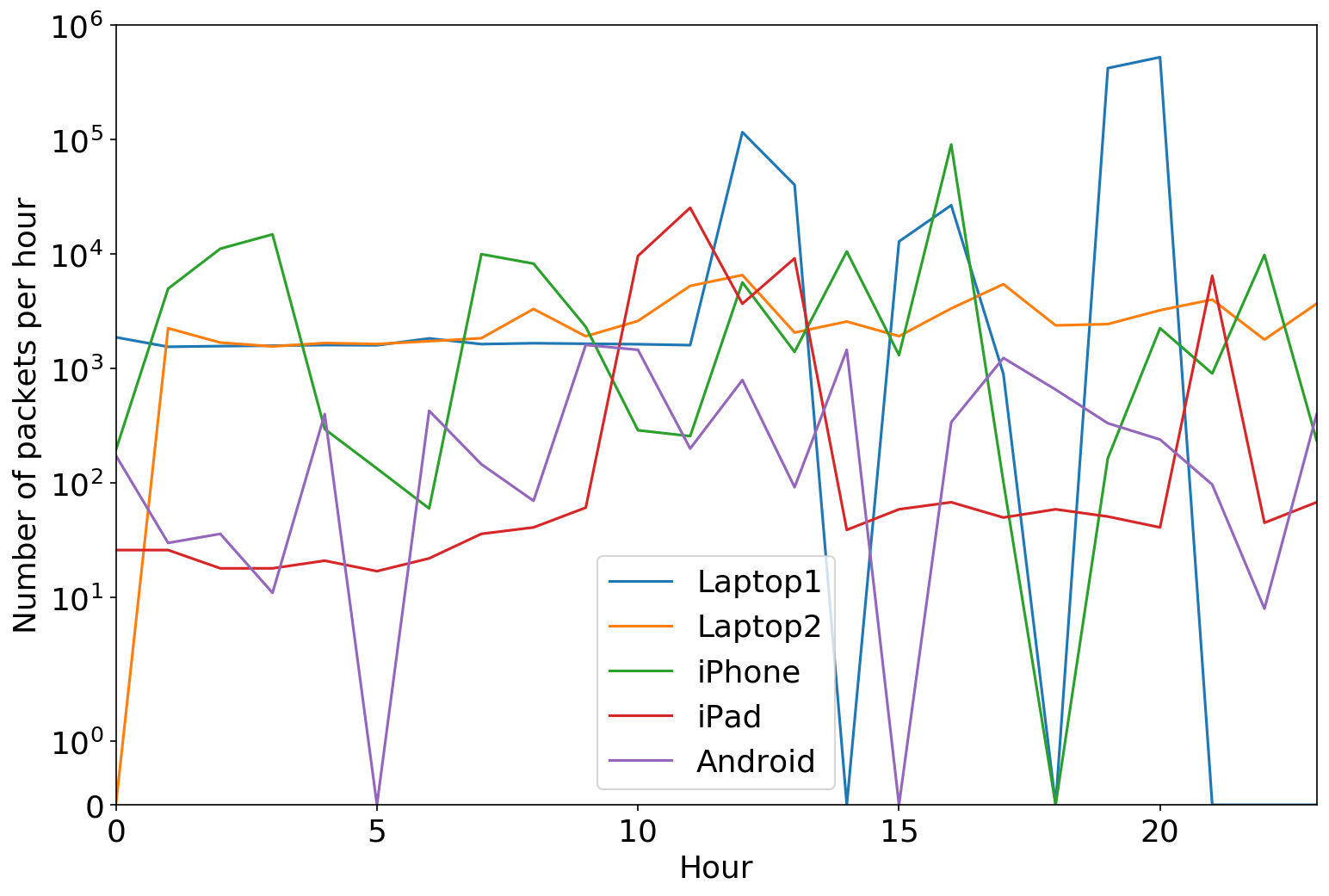}
    \caption{Non-IoT devices}
    \label{fig:hourly-udp-pkts-noniot}
\end{subfigure}

\caption{Hourly number of packets of UDP traffic.}
\label{fig:hourly-udp-pkts}
\end{figure}

Figures~\ref{fig:hourly-tcp-pkts} and~\ref{fig:hourly-udp-pkts} show the hourly amount of traffic in terms of the number of TCP and UDP packets, respectively. From the figures, we can make similar observations as we have seen in the hourly number of bytes (these two characteristics are closely related). In particular, we emphasize the notable difference between the UDP traffic of IoT and non-IoT devices as seen in Figures~\ref{fig:hourly-udp-pkts-iot} and~\ref{fig:hourly-udp-pkts-noniot}.

\section{Summary and Future Work}\label{sec:summary}
In this paper we analyzed the network traffic characteristics of IoT devices collected on a typical smart home environment consisting of a set of common IoT (and non-IoT) devices. We analyzed the network traffic characteristics of IoT devices from three complementary aspects: remote network servers, flow-level traffic characteristics, and packet-level traffic characteristics.  Our study provided critical insights into the operational and behavioral characteristics of IoT devices, which could have important implications on developing effective security and performance algorithms for IoT devices. 

In our future work, we plan to explore opportunities to include a larger set of different types of IoT devices in our study, including both adding additional IoT devices into our smart home testbed and utilizing IoT traffic traces collected by other researchers. Furthermore, we will apply the observations and insights made in the paper to develop effective algorithms to identify and classing IoT devices and to detect compromised IoT devices.

\section*{Acknowledgment}
This material is based upon work supported by the National Science
Foundation under Grant No. 1662487, Office of Naval Research Contract No.
N000142012049, and the Florida Center for Cybersecurity (FC2) Collaborative Seed Award Program (No. 24108-1106-00-I). 
Any opinions, findings, and conclusions or recommendations expressed in this
material are those of the authors and do not necessarily reflect the views
of the NSF, ONR, or FC2.



\bibliographystyle{unsrt}
\bibliography{main}

\vspace{10pt}

\end{document}